\documentclass[11pt,a4paper]{article}
\usepackage{jheppub}
\usepackage{bm}
\usepackage{feynmf}
\usepackage{mathrsfs}
\usepackage{slashed}

\allowdisplaybreaks

\DeclareMathOperator\tr{tr}
\newcommand\vek[1]{\bm{#1}}
\newcommand\he[1]{#1^\dagger}
\newcommand\adj[1]{\overline{#1}}
\newcommand\imag{\text{i}}
\newcommand\gr[1]{\mathrm{#1}}
\newcommand\La{\mathscr{L}}
\newcommand\dd{\mathop{\text d}\nolimits}
\newcommand\sumint[1]{\int\kern-1.5em\sum\nolimits_{#1}}
\newcommand\braced[1]{\{#1\}}
\newcommand\mufourpiT[1]{\left(\frac\Lambda{4\pi T}\right)^{#1\epsilon}}
\newcommand\OO{\mathcal{O}}
\newcommand{\ms}{\mu_\sigma}
\newcommand{\MSbar}{$\overline{\text{MS}}$}
\newcommand\td{\text{3d}}
\newcommand\fd{\text{4d}}

\graphicspath{
  {./}
  {./scans/}
}

\newcommand\selfzero[2]{\parbox{20mm}{%
\fmfframe(0,1)(0,1){
\begin{fmfgraph*}(20,20)
\fmfleft{l}
\fmfright{r}
\fmf{#1,label=$#2$,l.si=right,l.di=2pt}{r,l}
\end{fmfgraph*}}}}

\newcommand\selfone[3]{\parbox{20mm}{%
\fmfframe(0,1)(0,1){
\begin{fmfgraph}(20,20)
\fmfleft{l}
\fmfright{r}
\fmf{#1}{v,l}
\fmf{#3}{r,v}
\fmf{#2,right,tension=0.5}{v,v}
\fmfdot{v}
\end{fmfgraph}}}}

\newcommand\selftwo[4]{\parbox{25mm}{%
\fmfframe(0,1)(0,1){
\begin{fmfgraph}(25,20)
\fmfleft{l}
\fmfright{r}
\fmf{#1,tension=2}{vl,l}
\fmf{#4,tension=2}{r,vr}
\fmf{#2,right,tension=0.5}{vr,vl}
\fmf{#3,right,tension=0.5}{vl,vr}
\fmfdot{vl,vr}
\end{fmfgraph}}}}

\newcommand\vertone[9]{\parbox{20mm}{%
\fmfframe(0,1)(0,1){
\begin{fmfgraph*}(20,20)
\fmfleftn{l}{2}
\fmfright{r}
\fmf{#1,label=$#7$,l.si=left,l.di=2pt}{l1,v}
\fmf{#2,label=$#8$,l.si=left,l.di=2pt}{v,l2}
\fmf{#3,label=$#9$,l.si=right,l.di=2pt}{r,v}
\fmfdot{v}
\fmfv{label=$#4$,l.di=2pt}{l1}
\fmfv{label=$#5$,l.di=2pt}{l2}
\fmfv{label=$#6$,l.di=2pt}{r}
\end{fmfgraph*}}}}

\newcommand\vertoneinv[9]{\parbox{20mm}{%
\fmfframe(0,1)(0,1){
\begin{fmfgraph*}(20,20)
\fmfleftn{l}{2}
\fmfright{r}
\fmf{#1,label=$#7$,l.si=left,l.di=2pt}{l1,v}
\fmf{#2,label=$#8$,l.si=left,l.di=2pt}{v,l2}
\fmf{#3,label=$#9$,l.si=left,l.di=2pt}{v,r}
\fmfdot{v}
\fmfv{label=$#4$,l.di=2pt}{l1}
\fmfv{label=$#5$,l.di=2pt}{l2}
\fmfv{label=$#6$,l.di=2pt}{r}
\end{fmfgraph*}}}}

\newcommand\scatone[8]{\parbox{20mm}{%
\fmfframe(0,1)(0,1){
\begin{fmfgraph*}(20,20)
\fmfleftn{l}{2}
\fmfrightn{r}{2}
\fmf{#1}{l1,v}
\fmf{#2}{v,l2}
\fmf{#3}{r1,v}
\fmf{#4}{v,r2}
\fmfdot{v}
\fmfv{label=$#5$,l.di=2pt}{l1}
\fmfv{label=$#6$,l.di=2pt}{l2}
\fmfv{label=$#7$,l.di=2pt}{r1}
\fmfv{label=$#8$,l.di=2pt}{r2}
\end{fmfgraph*}}}}

\newcommand\scattwo[6]{\parbox{25mm}{%
\fmfframe(0,1)(0,1){
\begin{fmfgraph}(25,20)
\fmfleftn{l}{2}
\fmfrightn{r}{2}
\fmf{#1,tension=2}{l1,vl}
\fmf{#2,tension=2}{vl,l2}
\fmf{#5,tension=2}{r1,vr}
\fmf{#6,tension=2}{vr,r2}
\fmf{#4,right,tension=0.5}{vl,vr}
\fmf{#3,right,tension=0.5}{vr,vl}
\fmfdot{vl,vr}
\end{fmfgraph}}}}

\newcommand\scatthree[7]{\parbox{25mm}{%
\fmfframe(0,1)(0,1){
\begin{fmfgraph}(25,20)
\fmfleftn{l}{2}
\fmfrightn{r}{2}
\fmf{#1,tension=1.5}{l1,vl}
\fmf{#2,tension=1.5}{vl,l2}
\fmf{#6,tension=1.5}{r1,vr1}
\fmf{#7,tension=1.5}{vr2,r2}
\fmf{#3,tension=0.5}{vr2,vl}
\fmf{#4,tension=0.5}{vr1,vr2}
\fmf{#5,tension=0.5}{vl,vr1}
\fmfdot{vl,vr1,vr2}
\end{fmfgraph}}}}

\newcommand\scatthreeinv[7]{\parbox{25mm}{%
\fmfframe(0,1)(0,1){
\begin{fmfgraph}(25,20)
\fmfleftn{l}{2}
\fmfrightn{r}{2}
\fmf{#1,tension=1.5}{l1,vl}
\fmf{#2,tension=1.5}{vl,l2}
\fmf{#6,tension=1.5}{r1,vr1}
\fmf{#7,tension=1.5}{vr2,r2}
\fmf{#3,tension=0.5}{vl,vr2}
\fmf{#4,tension=0.5}{vr2,vr1}
\fmf{#5,tension=0.5}{vr1,vl}
\fmfdot{vl,vr1,vr2}
\end{fmfgraph}}}}

\newcommand\scatfour[8]{\parbox{25mm}{%
\fmfframe(0,1)(0,1){
\begin{fmfgraph}(25,20)
\fmfleftn{l}{2}
\fmfrightn{r}{2}
\fmf{#1,tension=1.5}{l1,vl1}
\fmf{#2,tension=1.5}{vl2,l2}
\fmf{#7,tension=1.5}{r1,vr1}
\fmf{#8,tension=1.5}{vr2,r2}
\fmf{#3,tension=0.5}{vl2,vl1}
\fmf{#4,tension=0.5}{vr2,vl2}
\fmf{#5,tension=0.5}{vr1,vr2}
\fmf{#6,tension=0.5}{vl1,vr1}
\fmfdot{vl1,vl2,vr1,vr2}
\end{fmfgraph}}}}

\newcommand\scatextwo[5]{\parbox{25mm}{%
\fmfframe(0,1)(0,1){
\begin{fmfgraph}(25,20)
\fmfleftn{l}{2}
\fmfrightn{r}{2}
\fmf{#1,tension=1}{l1,vl}
\fmf{#2,tension=1}{vl,l2}
\fmf{#4,tension=1}{r1,vr}
\fmf{#5,tension=1}{vr,r2}
\fmf{#3,tension=1}{vl,vr}
\fmfdot{vl,vr}
\end{fmfgraph}}}}

\newcommand\scatexthree[7]{\parbox{30mm}{%
\fmfframe(0,1)(0,1){
\begin{fmfgraph}(30,20)
\fmfleftn{l}{2}
\fmfrightn{r}{2}
\fmf{#1,tension=2}{l1,vl}
\fmf{#2,tension=2}{vl,l2}
\fmf{#6,tension=2}{r1,vr}
\fmf{#7,tension=2}{vr,r2}
\fmf{#3,tension=1.5}{vm,vl}
\fmf{#5,right,tension=0.5}{vm,vr}
\fmf{#4,right,tension=0.5}{vr,vm}
\fmfdot{vl,vr,vm}
\end{fmfgraph}}}}

\newcommand\scatexfour[8]{\parbox{30mm}{%
\fmfframe(0,1)(0,1){
\begin{fmfgraph}(30,20)
\fmfleftn{l}{2}
\fmfrightn{r}{2}
\fmf{#1,tension=1.5}{l1,vl}
\fmf{#2,tension=1.5}{vl,l2}
\fmf{#7,tension=1.5}{r1,vr1}
\fmf{#8,tension=1.5}{vr2,r2}
\fmf{#3,tension=1.5}{vl,vm}
\fmf{#4,tension=0.5}{vr2,vm}
\fmf{#6,tension=0.5}{vr1,vr2}
\fmf{#5,tension=0.5}{vm,vr1}
\fmfdot{vl,vr1,vr2,vm}
\end{fmfgraph}}}}

\begin{document}
\title{Dimensional reduction of the Standard Model coupled to a new
  singlet scalar field}

\preprint{HIP-2016-27/TH }

\author[a]{Tom\'a\v{s} Brauner,}
\author[b]{Tuomas V.I.~Tenkanen,}
\author[a]{Anders Tranberg,}
\author[b]{Aleksi Vuorinen}
\author[a,b]{and David J.~Weir}
\affiliation[a]{Faculty of Science and Technology,
  University of Stavanger,\\
  N-4036 Stavanger,
  Norway}
\affiliation[b]{Department of Physics and Helsinki Institute of Physics,\\
  P.O.~Box 64,
  FI-00014 University of Helsinki,
  Finland}
\emailAdd{tomas.brauner@uis.no}
\emailAdd{tuomas.tenkanen@helsinki.fi}
\emailAdd{anders.tranberg@uis.no}
\emailAdd{aleksi.vuorinen@helsinki.fi}
\emailAdd{david.weir@helsinki.fi}

\abstract{We derive an effective dimensionally reduced theory for the
  Standard Model augmented by a real singlet scalar. We treat the
  singlet as a superheavy field and integrate it out, leaving an
  effective theory involving only the Higgs and
  $\gr{SU(2)}_L\times\gr{U(1)}_Y$ gauge fields, identical to the one
  studied previously for the Standard Model. This opens up the
  possibility of efficiently computing the order and strength of the
  electroweak phase transition, numerically and nonperturbatively, in
  this extension of the Standard Model. Understanding the phase
  diagram is crucial for models of electroweak baryogenesis and for
  studying the production of gravitational waves at thermal phase
  transitions.}

\begin{fmffile}{fig5}

\fmfset{arrow_len}{3mm}
\fmfset{zigzag_width}{1mm}
\fmfset{curly_len}{2mm}
 
\maketitle


\section{Introduction}
\label{sec:introduction}

\subsection{Background}
\label{sec:background}

Quantitatively understanding the origin of the observed
matter-antimatter asymmetry in the present-day universe is one of the
major open challenges in cosmology. A widely studied scenario is that
of electroweak baryogenesis \cite{Kuzmin:1985mm,Cohen:1993nk} (see
Refs.~\cite{Rubakov:1996vz,Morrissey:2012db} for reviews).  This
assumes that the excess in baryon density was generated during the
electroweak phase transition in the early universe, when the Higgs
field obtained its nonzero vacuum expectation value (VEV). While all
of the main ingredients for generating a baryon asymmetry can be found
in the Standard Model (SM) --- an electroweak phase transition as well
as the breaking of charge conjugation (C), parity (P), CP and baryon
number symmetries --- it unfortunately turns out that purely SM
electroweak baryogenesis fails to live up to its promise.

The problem originates on the one hand from the severe suppression of
CP violation at high
temperatures~\cite{Shaposhnikov:1987pf,Farrar:1993sp,Farrar:1993hn,Gavela:1993ts,Gavela:1994ds,Gavela:1994dt,Brauner:2011vb,Brauner:2012gu},
and perhaps even more importantly from the fact that the electroweak
phase transition within the Standard Model is not of first order, but
merely of the crossover type. This conclusion was reached in the
mid-1990s after extensive efforts to build a dimensionally reduced
effective theory to describe the long-distance dynamics of the SM
close to the phase transition~\cite{Kajantie:1995dw}, and to
subsequently study it via nonperturbative lattice
simulations~\cite{Kajantie:1995kf,Kajantie:1996mn,Kajantie:1996qd}.
Later studies also confirmed this result with four-dimensional
simulations \cite{Csikor:1998ge,Csikor:1998eu,Aoki:1999fi}.

As a result of these studies, alternative scenarios such as
leptogenesis~\cite{Fukugita:1986hr,Luty:1992un} (see
Refs.~\cite{Buchmuller:2004nz,Davidson:2008bu} for comprehensive
reviews) and cold electroweak
baryogenesis~\cite{GarciaBellido:1999sv,Krauss:1999ng,Copeland:2001qw,Tranberg:2003gi}
have been suggested to explain the observed baryon asymmetry. What is
common to these scenarios is that they involve degrees of freedom
beyond the Standard Model, albeit sometimes at much higher energy
scales. There are, however, many sound reasons to expect new physics
around the TeV scale, and a plethora of different scenarios have been
proposed to describe this new physics. It is clearly reasonable to
investigate whether electroweak baryogenesis might be viable within
these models.

Given such a model of new physics at the TeV scale, the only degrees
of freedom requiring nonperturbative treatment at high temperatures
are known to be the static modes of the bosonic fields. Following the
strategy taken in the original SM
works~\cite{Kajantie:1995dw,Kajantie:1995kf,Kajantie:1996mn}, the task
therefore becomes to first derive three-dimensional effective theories
for these modes, and subsequently perform lattice studies of these
dimensionally reduced theories.

The recent direct observation of gravitational waves
\cite{Abbott:2016blz} further strengthens the interest in
investigating high-energy phase transitions in the early universe. The
gravitational waves sourced by bubble collisions and the subsequent
nonequilibrium dynamics of a first-order electroweak-scale phase
transition may be within the sensitivity range of the space-based
detector eLISA
\cite{Grojean:2006bp,No:2011fi,Hindmarsh:2013xza,Caprini:2015zlo}, due
for launch in 2034. Understanding the strength of such a phase
transition in extensions of the Standard Model makes the detection or
absence of such primordial gravitational waves a valuable source of
information about particle physics; information that is complementary
to collider experiments (see e.g. Ref.~\cite{Hashino:2016xoj} for a related discussion). Indeed, eLISA may be able to probe new
physics at temperatures above 10 TeV, a region beyond the reach of
proposed colliders~\cite{Caprini:2015zlo}.

In this paper, we shall focus on one such model, the singlet-extended
Standard Model (SSM)~\cite{Barger:2007im,Ashoorioon:2009nf,Robens:2015gla,Kanemura:2016lkz,Kanemura:2015fra,Beniwal:2017eik}, which has been studied in various different contexts including even inflationary physics \cite{Enqvist:2014zqa, Tenkanen:2016idg}. This has,
in its most general form, seven parameters in the scalar sector, of
which two are fixed by the experimental values of the Higgs mass and
the Higgs VEV. The remaining five-dimensional parameter space is a
challenge to scan, which explains why no comprehensive attempt at a
nonperturbative study has been made. The common approach has been a
semi-analytic daisy-resummed one-loop effective potential
treatment~\cite{Huber:2000mg,O'Connell:2006wi,Ahriche:2007jp,Profumo:2007wc,Espinosa:2011ax,Cline:2012hg,Damgaard:2013kva,Profumo:2014opa,Kozaczuk:2015owa,Damgaard:2015con},
which allows for a complete sampling of the parameter space and direct
comparison with experimental constraints. However, it is known that
perturbative treatments tend to over-estimate the strength of the
phase transition~\cite{Kajantie:1995kf,Kajantie:1996mn}. Hence we
expect that the region of (strongly) first order phase transitions is
smaller than what has so far been identified.  In the present work, we
derive the dimensionally reduced effective theory for the SSM. This
will be used in simulations, to be detailed in a companion
paper~\cite{TBD}. Our main result here is a set of explicit matching
relations, which allow us to relate a given set of four-dimensional
SSM parameters (and temperature) to the (fewer) parameters of the
three-dimensional theory.

We shall present our computation in a highly explicit manner,
displaying most of the intermediate results and presenting the final
results in a such a form that the Standard Model limit is simple to
take. There are two reasons for this. First, the original derivation
of the dimensionally reduced effective theory of SM, carried out in
the seminal paper~\cite{Kajantie:1995dw}, was presented in a rather
compact way, suppressing many calculational details. Second, apart
from the SSM, it is naturally very interesting and well-motivated to
study baryogenesis in a number of other beyond-SM models, which could
be subjected to the same procedure presented here. We hope that by
providing more details of the calculations, our work will be useful
for a broader audience interested in the derivation or use of
dimensionally reduced effective theories either in the SM or in
different beyond-SM scenarios.

A note of caution is, however, necessary. In our derivation of the
dimensionally reduced effective field theory, we work to one loop
order for all the parameters of the effective theory and only perform
the matching to physical parameters at tree level. While this does not
match the accuracy of the original Standard Model calculation
performed in Ref.~\cite{Kajantie:1995dw}, we do not expect this to
affect the phenomenological implications of our
calculation. Nevertheless we shall revisit this issue in
Ref.~\cite{TBD}.

This paper is organized as follows. In the remainder of this
introductory section, we explain the basic principles of dimensional
reduction on a very general level. In Section~\ref{sec:SSM_euclidean},
we then introduce the SSM, including the forms of its four- and
three-dimensional Lagrangians as well as the associated
parameters. The actual dimensional reduction of the model is performed
in Section~\ref{sec:dimred_superheavy}. 
In Section~\ref{sec:discussion}, we discuss our findings and
  investigate the extent to which the inclusion of a scalar singlet
  improves the prospects for a first order phase transition.  Many
details of the calculations, ranging from Feynman rules to the results
for individual graphs, are deferred to the appendices.

\subsection{Dimensional reduction framework}
\label{sec:dimredframework}

Dimensional reduction is a generic physical principle governing the
properties of quantum field theories at high temperatures,
stating that the low-energy behavior of static Green's functions can
be determined through a lower-dimensional effective theory. In short,
it follows from the fact that in thermal equilibrium, four-dimensional
fields can be reduced to infinite towers of three-dimensional field
modes -- termed Matsubara modes -- by means of a Fourier
series expansion in the imaginary time variable $\tau$. The effective
masses of the three-dimensional fields become
\begin{equation}
M_\text{boson}^2=M_0^2 + (2n \pi T)^2,\qquad
M_\text{fermion}^2=M_0^2 + [(2n+1) \pi T]^2,
\end{equation}
where $M_0$ denotes the field mass at zero temperature and $n$ takes integer values. Consequently, at high temperatures, that is for $T\gtrsim M_0$ for all fields, all modes except for the bosonic zero modes ($n=0$) obtain thermal masses at least of order $\pi T$, and thus decouple from physics at length scales parametrically larger than $1/T$.

Let us now follow the discussion of Ref.~\cite{Kajantie:1995dw} and specialize to a model whose bosonic sector can be described via a Euclidean Lagrangian density of the generic form
\begin{equation}
 \La= \frac14F_{\mu\nu}F_{\mu\nu}+D_\mu\he\phi D_\mu\phi+\mu^2\he\phi\phi+\lambda(\he\phi\phi)^2+ g_Y \bar{\psi}\phi\psi + \delta\La,
\end{equation}
where $A_\mu$ (appearing inside $F_{\mu\nu}$ and $D_\mu$) is a gauge
field, $\phi$ a complex scalar, $\psi$ a fermion, and $\delta\La$
corresponds to counterterms. We further assume that the Yukawa
coupling $g_Y$ and the scalar self-coupling $\lambda$ scale as
$g_Y\sim g$ and $\lambda\sim g^2$ in terms of the gauge coupling
$g$. Then it can be verified that at one-loop order,
interactions contribute to the masses of the zero Matsubara modes of
the $\phi$, $A_0$ and $A_r$ fields\footnote{In order to avoid
  confusion with the isospin doublet index $i,j,\dotsc$, employed for
  the Higgs field and the SM fermions, we use the letters $r,s,\dotsc$
  to label spatial vectors.} as
\begin{equation}
M_{\phi}^2-M_0^2\sim g^2 T^2,\qquad
M_{A_0}^2\sim g^2T^2,\qquad
M_{A_r}^2=0,
\end{equation}
where the last of the relations is consistent with the fact that the dimensionally reduced effective theory possesses three-dimensional gauge invariance.

From the above considerations, we see the emergence of a scale
hierarchy in the system. The thermal scale $\pi T$ is canonically
dubbed \emph{superheavy}, while the mass scale of the $A_0$ field,
$gT$, is referred to as \emph{heavy}. Finally, the mass of the $\phi$
field depends on the value of the mass parameter $M_0$: should $M_0$
be comparable to $\pi T$, the corresponding field mode is treated as
superheavy, whereas for $M_0$ of order $gT$, it is heavy. An exception
may, however, occur near a phase transition, where the ${\mathcal
  O}(g^2T^2)$ one-loop correction to $M_{\phi}^2$ exactly cancels the
(negative) tree-level $M_0^2$. In this case, the mass of the $n=0$
mode of the scalar field becomes of order $g^2T$ and the field is
referred to as \emph{light}. The $n=0$ component of the spatial gauge
fields $A_r$, which is protected by gauge invariance, is naturally
light as well.

The formal procedure of dimensional reduction consists of successively
integrating out the superheavy and heavy energy scales from the
system. This implies deriving effective Lagrangians for the relevant
field modes, which is most easily done with the following recipe (see
e.g.~Ref.~\cite{Appelquist:1974tg,Braaten:1995cm}):
\begin{enumerate}
\item Determine the relevant light degrees of freedom of the effective theory.
\item Write down the most general local Lagrangian consistent with the symmetries of the theory, including three-dimensional gauge invariance.
\item Order the operators in the Lagrangian in terms of their dimensions and discard terms beyond a given order.
\end{enumerate}
The essence of dimensional reduction is that the three-dimensional
effective theory obtained with the above procedure is capable of
reproducing the long-distance --- length scales $1/(gT)$ and above ---
Green's functions of the full four-dimensional theory. This can be
done to arbitrary accuracy, provided that operators of high enough
dimension are included in the corresponding Lagrangian density. In
practice, this implies matching various Green's functions for the two
theories, and deriving from them expressions for the parameters of the
effective theory.

Let us now specialize to the case of a high-temperature phase
transition, and assume that the thermal correction to the mass of the
$n=0$ scalar field mode exactly cancels its negative zero-temperature
mass parameter, so that the field becomes light. In this case,
dimensional reduction proceeds in two successive stages. In the first
step, we integrate out only the superheavy modes, leaving behind a
three-dimensional superrenormalizable effective theory for the spatial
gauge field $A_r$, the massive temporal gauge field $A_0$, and the
scalar $\phi$. This theory is capable of describing physics at length
scales $1/(gT)$, but still contains two distinct scales: the
${\mathcal O}(gT)$ mass of $A_0$ and the ${\mathcal O}(g^2T)$ mass of
$\phi$. The former can then also be integrated out, leaving a theory
for the light modes only, i.e.~the fields $A_r$ and $\phi$. The
construction of the Lagrangians and the matching calculations needed
for the determination of the corresponding parameters are discussed at
length in Ref.~\cite{Kajantie:1995dw}.

For the remainder of this paper, we take the basic principles of
dimensional reduction as given, referring the interested reader to
Refs.~\cite{Kajantie:1995dw,Appelquist:1974tg,Braaten:1995cm}. These
principles will be applied to the study of the SSM, which is
introduced in the next section. There, we shall also write down the
explicit forms of the effective Lagrangians corresponding to
two different scenarios where the new singlet scalar is treated as
superheavy and heavy, respectively, even though we shall only
  carry out the dimensional reduction in the superheavy case.  The
matching calculations are then presented in the following section,
which is dedicated to the case where the extra singlet is superheavy.

\section{Standard Model with singlet scalar in Euclidean space}
\label{sec:SSM_euclidean}

In this section, we review the Standard Model coupled to a
singlet scalar field. In addition, we present the form of the
three-dimensional effective Lagrangians for two scenarios, in
which the singlet is treated as superheavy and heavy,
respectively. Throughout the discussion, we shall work in a Euclidean
spacetime of $D=d+1=4-2\epsilon$ dimensions.

\subsection{Full four-dimensional theory}
\label{sec:full4D}

The classical Euclidean Lagrangian of our four-dimensional theory reads 
\begin{equation}
\La=\La_\text{gauge}+\La_\text{ghost}+\La_\text{fermion}+\La_\text{scalar}+\La_\text{Yukawa}+\delta\La,
\label{classlag}
\end{equation}
where the gauge field, ghost, fermion, scalar and Yukawa sector Lagrangians are defined as follows (the counterterm part $\delta\La$ will be discussed later): 
\begin{align}
\La_\text{gauge}={}&\frac14G^a_{\mu\nu}G^a_{\mu\nu}+\frac14F_{\mu\nu}F_{\mu\nu}+\frac14H^\alpha_{\mu\nu}H^\alpha_{\mu\nu},\\
\La_\text{ghost}={}&\partial_\mu\adj\eta^a D_\mu\eta^a+\partial_\mu\adj\xi\partial_\mu\xi+\partial_\mu\adj\zeta^\alpha D_\mu\zeta^\alpha,\\
\La_\text{fermion}={}&\sum_A\left(\adj\ell_A\slashed D\ell_A+\adj e_A\slashed De_A+\adj q_A\slashed Dq_A+\adj u_A\slashed Du_A+\adj d_A\slashed Dd_A\right),\\
\La_\text{scalar}={}&D_\mu\he\phi D_\mu\phi-\mu_h^2\he\phi\phi+\lambda_h(\he\phi\phi)^2+\frac12(\partial_\mu\sigma)^2+\frac12\mu_\sigma^2\sigma^2\\
\notag
&+\mu_1\sigma+\frac13\mu_3\sigma^3+\frac14\lambda_\sigma\sigma^4+\frac12\mu_m\sigma\he\phi\phi+\frac12\lambda_m\sigma^2\he\phi\phi,\\
\La_\text{Yukawa}={}&\sum_{A,B}\left[h^{(e)}_{AB}\adj\ell_Ae_B\phi+h^{(d)}_{AB}\adj q_Ad_B\phi+h^{(u)}_{AB}\adj q_Au_B\tilde\phi\right]+\text{h.c.}
\end{align}
We shall work in the Landau gauge. The theory includes the following fields:
\begin{itemize}
\item The $\gr{SU(2)}_L$, $\gr{U(1)}_Y$ and $\gr{SU(3)}_c$ gauge
  fields $A^a_\mu$, $B_\mu$, and $C^\alpha_\mu$ appearing inside the
  field strength tensors $G^a_{\mu\nu}$, $F_{\mu\nu}$ and
  $H^\alpha_{\mu\nu}$. The associated gauge couplings are $g$,
  $g'$, and $g_s$, and the corresponding ghost fields $\eta^a$, $\xi$,
  and $\zeta^\alpha$.
\item The left-handed doublet and right-handed singlet lepton fields with a flavor index, $\ell_A$ and $e_A$, as well as the left-handed doublet quark fields $q_A$ and right-handed singlet up- and down-type quark fields $u_A$ and $d_A$.
\item The Higgs field $\phi^i$, with the charge-conjugated Higgs doublet  $\tilde\phi\equiv\imag\tau_2\phi^*$, where $\tau_2$ is the second Pauli matrix.
\item The extra real singlet scalar field $\sigma$.
\end{itemize}

The relation $Q=I_3+\frac Y2$ between electric charge $Q$ and isospin
$I_3$ defines the hypercharge of the fields as follows: $Y_\ell=-1$,
$Y_e=-2$, $Y_q=\frac13$, $Y_u=\frac43$, $Y_d=-\frac23$, $Y_\phi=1$,
$Y_\sigma=0$. Finally, we shall for completeness write down explicit
expressions for the covariant derivatives and field strength
tensors. The covariant derivatives read in different cases
\begin{align}
D_\mu\psi&=\biggl(\partial_\mu-\imag g\frac{\vec\tau}2\cdot\vec A_\mu-\imag g_s\frac{\vec\lambda}2\cdot\vec C_\mu-\imag g'\frac Y2B_\mu\biggr)\psi
&&\text{(for $q_A$)},\\
D_\mu\psi&=\biggl(\partial_\mu-\imag g\frac{\vec\tau}2\cdot\vec A_\mu-\imag g'\frac Y2B_\mu\biggr)\psi
&&\text{(for $\ell_A,\phi$)},\\
D_\mu\psi&=\biggl(\partial_\mu-\imag g_s\frac{\vec\lambda}2\cdot\vec C_\mu-\imag g'\frac Y2B_\mu\biggr)\psi
&&\text{(for $u_A,d_A$)},\\
D_\mu\psi&=\biggl(\partial_\mu-\imag g'\frac Y2B_\mu\biggr)\psi
&&\text{(for $e_A,\sigma$)},
\end{align}
where $\vec\tau$ and $\vec\lambda$ denotes the vector of Pauli and Gell-Mann matrices, respectively. Finally, the field strength tensors take the forms 
\begin{align}
G^a_{\mu\nu}&=\partial_\mu A^a_\nu-\partial_\nu A^a_\mu+g\epsilon^a_{\phantom abc}A^b_\mu A^c_\nu, \\
F_{\mu\nu}&=\partial_\mu B_\nu-\partial_\nu B_\mu,\\
H^\alpha_{\mu\nu}&=\partial_\mu C^\alpha_\nu-\partial_\nu C^\alpha_\mu+g_sf^\alpha_{\phantom \alpha\beta\gamma}C^\beta_\mu C^\gamma_\nu.
\end{align}
In the Yukawa part $\La_\text{Yukawa}$, $h^{(e)}, h^{(d)}$ and $h^{(u)}$ stand for the flavor-mixing matrices, while h.c.~represents hermitian conjugate. In the final stages of our calculation, we shall use an approximation where only the top quark Yukawa coupling $g_Y$ is nonzero. The Yukawa sector then simplifies to 
\begin{equation}
\La_\text{Yukawa} = g_Y (\bar{q}_t \tilde\phi t + \bar{t} \tilde\phi^\dagger q_t)\qquad\text{(if top-quark only)}.
\end{equation}
For the sake of convenience, the Feynman rules in the unbroken phase
of this theory are listed in Appendix~\ref{sec:4d_feynrules_unbroken}.

Since $\sigma$ is a real singlet, we can choose\footnote{A
  similar shift is not permitted for the Higgs field because of gauge
  invariance.} the zero-temperature VEV, around which we perturb, to
be at $\sigma=0$. This shift amounts to a redefinition of the
parameters of the potential, and since $\sigma=0$ is defined to be a
minimum, we have that $\mu_\sigma^2\geq 0$. Our choice also imposes a
relation between $\mu_1$ and $\mu_m$ in the vacuum where the Higgs
field has a VEV, given by $\langle\phi^\dagger\phi\rangle =v^2/2$,
\begin{equation}
\label{eq:fix_mu_1}
\mu_1= -\frac{\mu_m v^2}{4}.
\end{equation}
To start with, however, we will not impose this constraint, treating
$\mu_1$ and $\mu_m$ as independent parameters. Keeping the parameter
$\mu_1$ explicit will allow us to see in
Section~\ref{sec:matching_1loop} that the matching relations for the
three-dimensional parameters are independent of the renormalization
scale of the four-dimensional theory; including the running of $\mu_1$
is essential to ensure this property. Later on, in
Section~\ref{sec:rel_physical}, we shall impose the
condition \eqref{eq:fix_mu_1} when we relate the \MSbar{} scheme
parameters to physical observables in the vacuum. We will assume
throughout that $\mu_\sigma^2>0$. As argued above, this represents no
loss of generality.

\subsubsection{Renormalization}
\label{sec:renormalization}

All fields and couplings appearing in the above Lagrangian are the
renormalized ones, while the counterterms, given explicitly in
Section~\ref{sec:counterterms_beta_functions}, are included in
$\delta\La$. We use the following conventions for the relations
between the renormalized fields and couplings and their bare
counterparts, denoted by the subscript $(b)$:
\begin{align}
\vec A_{\mu(b)} &\equiv Z^{1/2}_A \vec A_\mu = (1+\delta Z_A)^{1/2} \vec A_\mu,\\
B_{\mu(b)} &\equiv Z^{1/2}_B B_\mu = (1+\delta Z_B)^{1/2} B_\mu,\\
\phi_{(b)} &\equiv Z^{1/2}_\phi \phi = (1+\delta Z_\phi)^{1/2} \phi,\\
\sigma_{(b)} &\equiv Z^{1/2}_\sigma \sigma = (1+\delta Z_\sigma)^{1/2} \sigma,
\end{align}
for the fields, and
\begin{align}
 \label{eq:bare_couplings} 
 g_{(b)} &\equiv g + \delta g, \qquad g'_{(b)} \equiv g' + \delta g', 
& g_{Y(b)} &\equiv g_{Y} + \delta g_Y, \\
\mu^2_{h(b)} &\equiv Z^{-1}_\phi (\mu^2_{h} + \delta \mu^2_h),
&\lambda_{h(b)} &\equiv Z^{-2}_\phi (\lambda_{h} + \delta \lambda_h),  \\
\mu_{1(b)} &\equiv Z^{-1/2}_\sigma (\mu_{1} + \delta \mu_1),
&\mu^2_{\sigma(b)} &\equiv Z^{-1}_\sigma (\mu^2_{\sigma} + \delta \mu^2_\sigma), \\
\mu_{3(b)} &\equiv Z^{-3/2}_\sigma (\mu_{3} + \delta \mu_3),
&\mu_{m(b)} &\equiv Z^{-1}_\phi Z^{-1/2}_\sigma (\mu_{m} + \delta \mu_m),\\
\lambda_{\sigma(b)} &\equiv Z^{-2}_\sigma (\lambda_{\sigma} + \delta \lambda_\sigma),
&\lambda_{m(b)}&\equiv Z^{-1}_\phi Z^{-1}_\sigma (\lambda_{m} + \delta \lambda_m),
\end{align}
for the couplings. It is worth pointing out that at the one-loop level
at which we work, the singlet scalar does not receive any wavefunction
renormalization, that is, $Z_\sigma=1$.

\subsubsection{Scaling of parameters}
\label{sec:param_relations}

We assume that the parameters of the theory obey the following
parametric scaling relations in terms of the $\gr{SU(2)}_L$ coupling
$g$:
\begin{itemize}
\item $g', g_s,g_Y \sim g$, 
\item $\lambda_h, \lambda_m, \lambda_\sigma \sim g^2$,
\item $\mu_h, \mu_3 \sim g T$,
\item $\mu_1 \sim g T^3$,
\item $\mu_m \sim g^n T$,  and $\mu_\sigma \sim g^m T$,
\end{itemize}
where we keep some freedom in the choice of the scaling power for the mass and cubic interaction of the singlet scalar. To find a suitable choice for $m$ and $n$, consider schematically the tree-level contribution to the Higgs four-point function originating from a $\sigma$ exchange at vanishing external momenta,
\begin{equation}
\scatextwo{scalar}{scalar}{double}{scalar}{scalar} \simeq \frac{\mu^2_m}{\mu^2_\sigma} \sim g^{2(n-m)}.
\label{sigma_exchange}
\end{equation}
We require this contribution to be at least of order $g^2$, so that it
does not exceed the value of the Higgs self-coupling $\lambda_h$. We
therefore have two interesting and very distinctive options:
superheavy $\sigma$ (corresponding to $m=0$) combined with $n=1$, and
heavy $\sigma$ (corresponding to $m=1$), combined with $n=2$.

In the first case, even the zero mode of $\sigma$ is superheavy
and will therefore be integrated out, together with the
non-vanishing Matsubara modes. The three-dimensional effective
  theory is then, up to operators of order six and higher in the
fields, the same as in the Standard Model. However, the
dimensional reduction step contains new technical aspects compared to
the Standard Model case considered in
Ref.~\cite{Kajantie:1995dw}, as one cannot expand the superheavy
$\sigma$ mass term in the denominator of sum-integrals, but has to
consider massive sum-integrals instead,
cf.~Section~\ref{sec:master_integrals}. Furthermore, in addition to the
one-particle-irreducible (1PI) diagrams usually sufficient for
matching of the four-dimensional and three-dimensional theories, one
needs to include graphs which are one-$\sigma$-reducible.

When $\sigma$ is itself heavy, it remains in the dimensionally reduced
theory for the heavy scale. Sum-integrals with $\sigma$ propagators
can then be expanded in the mass parameter, which generates
higher-order corrections analogous to those stemming from the
Higgs mass parameter. Moreover, in this case the contributions
originating from the coupling $\mu_m$ are highly suppressed.

We emphasize that our scaling relations above differ from those of
Ref.~\cite{Kajantie:1995dw}; we do not assume $g'$ to be
parametrically smaller than $g$. As a result, we have to retain
  the $\gr{U(1)}_Y$ gauge field, treating it on the same footing as
the $\gr{SU(2)}_L$ gauge field.

\subsection{Effective three-dimensional theories}
\label{sec:eff3D}

Unless explicitly stated otherwise, we choose to denote the fields of
the effective theories with the same symbols as those of the
four-dimensional theory. However, the effective theory gauge couplings
are denoted by $g_3$, $g'_3$ and $g_{s,3}$. The classical Lagrangian
density of the effective theory (again in the Landau gauge) then has
the schematic form
\begin{equation}
\La^{(3)}=\La^{(3)}_\text{gauge}+\La^{(3)}_\text{ghost}+\La^{(3)}_\text{scalar}+\La^{(3)}_\text{temporal}+\delta \La^{(3)}.
\end{equation}
We include the $\gr{SU(2)}_L$ and $\gr{U(1)}_Y$ gauge fields in the gauge sector part
\begin{equation}
\La^{(3)}_\text{gauge}=\frac14G^a_{rs}G^a_{rs}+\frac14F_{rs}F_{rs},
\end{equation}
where only spatial Lorentz indices are summed over. The explicit forms
of $\La^{(3)}_\text{ghost}$ and $\delta \La^{(3)}$ are not relevant
for the present discussion. The scalar and temporal gauge field
sectors are discussed below for our two different cases.

\subsubsection{The superheavy $\sigma$ case}
\label{sec:superheavy_sigma}

As explained above, in this case the neutral scalar is completely
integrated out in the dimensional reduction step. To the order we are
working, the three-dimensional Lagrangian therefore coincides with
that of SM, with the temporal gauge field part reading
\begin{align}
\La^{(3)}_\text{temporal}={}&\frac12(D_rA^a_0)^2+\frac12m_D^2A^a_0A^a_0+\frac12(\partial_rB_0)^2+\frac12m_D'^2B_0^2
+\frac12(D_rC^\alpha_0)^2 \\ \notag
&+\frac12m_D''^2C^\alpha_0C^\alpha_0
+\frac14\lambda_3(A^a_0A^a_0)^2
+\frac14\lambda_3'B_0^4+\frac14\lambda_3''A^a_0A^a_0B_0^2+h_3\he\phi\phi A^a_0A^a_0 \\ \notag
&+h_3'\he\phi\phi B_0^2+h_3''B_0\he\phi\vec A_0\cdot\vec\tau\phi +\delta_3\he\phi\phi C^\alpha_0C^\alpha_0,
\end{align}
with the covariant derivatives of the adjoint fields reading $D_rA^a_0 = \partial_r A^a_0 + g_3 \epsilon^a_{\phantom abc}A^b_rA^c_0$ and $D_rC^\alpha_0 = \partial_r C^\alpha_0 + g_s f^\alpha_{\phantom \alpha\beta\rho}C^\beta_rC^\rho_0$.

Finally, the scalar part of the Lagrangian is
\begin{equation}
\La^{(3)}_\text{scalar}=D_r\he\phi D_r\phi-\mu_{h,3}^2\he\phi\phi+\lambda_{h,3}(\he\phi\phi)^2.
\end{equation}
In this case, the second step of dimensional reduction from the heavy
to the light scale is identical to that of SM, with the heavy temporal
gauge fields $A^a_0,B_0$ and $C^\alpha_0$ integrated out. The results for the
parameters of the effective theory for the light scale, denoted by
$\bar{g}_3$, $\bar{g}_3'$, $\bar{\mu}_{h,3}$, $\bar{\lambda}_{h,3}$,
can be taken from Ref.~\cite{Kajantie:1995dw}
(apart from the contribution of temporal gluon fields $C^\alpha_0$), and are therefore only briefly reviewed in Section~\ref{sec:int_heavyscale}. In Ref.~\cite{Kajantie:1995dw}, gluons were completely neglected from the three-dimensional theory, expecting the effect of this omission to be subdominant in the final conclusions regarding the order and properties of the electroweak phase transition. We have, however, included the leading order contribution from temporal gluons for completeness.

\subsubsection{The heavy $\sigma$ case}
\label{sec:heavy_sigma}
When the $\sigma$ field is heavy, the static (zero Matsubara) mode of
the $\sigma$ field appears in the effective theory for the heavy
scale, resulting in additional terms in the Lagrangian. The scalar
part $\La^{(3)}_\text{scalar}$ now includes the operators
\begin{equation}
\frac12(\partial_r\sigma)^2+\mu_{1,3}\sigma+\frac12\mu_{\sigma,3}^2\sigma^2+\frac13\mu_{3,3}\sigma^3+\frac14\lambda_{\sigma,3}\sigma^4+\frac12\mu_{m,3}\sigma\he\phi\phi+\frac12\lambda_{m,3}\sigma^2
\he\phi\phi,
\end{equation}
while $\La^{(3)}_\text{temporal}$ acquires the new terms 
\begin{equation}
x_3\sigma A^a_0A^a_0+x_3'\sigma B_0^2+y_3\sigma^2 A^a_0A^a_0+y_3'\sigma^2B_0^2+y_3''\sigma\he\phi\vec A_0\cdot\vec\tau\phi.
\end{equation}
The derivation of the effective theory for the light scale differs
from the SM computation in that one needs to integrate out the zero
mode of $\sigma$. Although in principle straightforward, this
calculation is left for future work. For the remainder of this paper,
we focus exclusively on the superheavy $\sigma$ case, where the
singlet scalar is completely integrated out already in the first
dimensional reduction step.

\subsubsection{Terms neglected from $\La^{(3)}$}  
\label{sec:terms_neglected}

Before we close this section, we will briefly list and discuss
examples of operators that have been discarded from the
three-dimensional effective theory for various reasons:
\begin{itemize}
\item The effects of the $\gr{SU(3)}_c$ gauge fields, i.e.~gluons, are partially neglected, as we discard the operators $H^\alpha_{rs}H^\alpha_{rs}$, $(C^\alpha_0C^\alpha_0)^2$, $A^a_0A^a_0C^\alpha_0C^\alpha_0$ and $B_0^2C^\alpha_0C^\alpha_0$ from the effective theory for the heavy scale. Spatial gluons do not couple to the scalar field, while the self interactions of temporal gluons and their interactions with other adjoint fields would have a very small contribution to our quantities of interest, such as the scalar mass parameter of the effective theory for the light scale, cf.~section~\ref{sec:int_heavyscale}.

\item In the superheavy $\sigma$ case, a momentum-dependent four-point
  self-interaction of the Higgs doublet is generated through the
  $\sigma$-exchange diagram shown in Eq.~\eqref{sigma_exchange}. To
  see this, simply expand the $\sigma$ propagator in powers of
  momentum, or equivalently solve the equation of motion of $\sigma$
  including just its kinetic term and the $\mu_m$ coupling. This
  yields an induced interaction for the Higgs,
\begin{equation}
\La_\text{ind}=-\frac18\mu_m^2(\he\phi\phi)\frac1{-\Box+\mu_\sigma^2}(\he\phi\phi).
\label{indinter}
\end{equation}
From an expansion in powers of derivatives, one gets an infinite
series of interactions. Since $\mu_\sigma$ is of order $g^0$ while the
momentum in the effective theory for the heavy scale is of order
$g^1$, the expansion starts at order $g^2$. Every power of $\Box$ then
adds an extra factor of $g^2$. The first operator containing
derivatives, $(\he\phi\phi)\Box(\he\phi\phi)$, therefore comes with an
order-$g^4$ coefficient, which is safe to neglect to the order at
which we work.

\item The first non-derivative self-coupling of the Higgs doublet, not included in our effective theory, namely $(\he\phi\phi)^3$, receives a contribution proportional to $\mu_3 \mu^3_m \sim g^4$, generated by the tree-level diagram 
\begin{equation}
\parbox{30mm}{\begin{fmfgraph*}(30,30)
\fmfsurroundn{s}{6}
\fmf{double,tension=2}{w,v1}
\fmf{double,tension=2}{w,v2}
\fmf{double,tension=2}{w,v3}
\fmf{scalar}{s1,v1,s2}
\fmf{scalar}{s3,v2,s4}
\fmf{scalar}{s5,v3,s6}
\fmfdot{w,v1,v2,v3}
\end{fmfgraph*}}
\end{equation}
in the superheavy $\sigma$ case. While this is dominant over the
contributions to the same operator from the SM superheavy fields,
which only start at order $g^6$, it will be likewise neglected in our
analysis carried out below.
\end{itemize}

\end{fmffile}

\begin{fmffile}{fig6}

\section{Dimensional reduction in the superheavy $\sigma$ case}
\label{sec:dimred_superheavy}

In this section, we perform the dimensional reduction step for a
superheavy singlet scalar. This requires explicitly computing a set of
Green's functions in both the full and the effective theory, requiring
that the results agree at distances of order $1/(gT)$.  The
calculations are divided into three parts: in
Section~\ref{sec:DR_correlators}, we list the results for the
necessary two- and four-point graphs; in
Section~\ref{sec:counterterms_beta_functions}, we review the explicit
counterterms needed; and in Section~\ref{sec:matrel}, we use these to
derive results for the parameters of the effective theory.

The discussion of the present section follows closely that of the
dimensional reduction in the Standard Model performed in
Ref.~\cite{Kajantie:1995dw}. In the main text, we only highlight
explicitly contributions from Feynman diagrams that are new compared
to the Standard Model, that is, those that involve at least one
$\sigma$ propagator. For the sake of completeness, the results for all
SM Feynman diagrams contributing to the effective theory parameters
are listed in Appendix~\ref{sec:DR_diagrams}. Note that, in contrast
to Ref.~\cite{Kajantie:1995dw}, we do not make the scaling assumption
$g' \sim g^{3/2}$. Consequently, we must consider a group of SM
diagrams that were neglected in that work.

\subsection{Correlators for the dimensional reduction}
\label{sec:DR_correlators}

We start by calculating a set of correlators in the full
four-dimensional theory. The results listed below are given in terms
of a set of master sum-integrals introduced in
Appendix~\ref{sec:master_integrals}.  Special attention is paid to
subtleties related to the assumed superheavy nature of $\sigma$: apart
from having to deal with massive $\sigma$ propagators, a major
modification is that we also need to include graphs which are
one-$\sigma$-reducible. Led by practical convenience, we evaluate the
contributions to wavefunction renormalization and to the interaction
vertices of the temporal gauge fields by a direct diagrammatic
analysis. The correlators in the scalar sector, on the other hand, are
determined afterwards using the effective potential.

\subsubsection{Self-energy diagrams}
\label{subsec:debye}

We start by considering the two-point functions. In order to be able
to extract the contributions to both the kinetic terms and the mass
parameters of the fields, we expand the correlators to second order in
the external momentum $P$.

\paragraph{$\gr{SU(2)}_L$ gauge boson self-energy.}
\begin{align}
\notag
\\[-8ex]
\notag
a\mu\;&
\parbox{20mm}{%
\begin{fmfgraph}(20,20)
\fmfleft{l}
\fmfright{r}
\fmf{boson}{r,v,l}
\fmfv{d.sh=circle,d.fi=shaded,d.si=5mm}{v}
\end{fmfgraph}}
\; b\nu\\[-2ex]
\label{eq:SU2_self_energy_00}
&=g^2\delta_{ab}\bigl[-(d-1)(2d-1)I^{4b}_1+\tfrac16(16-3d+2d^2)P^2I^{4b}_2\bigr]\\
\notag
&\phantom{{}={}}+g^2\delta_{ab}(d-1)N_f(1+N_c)\bigl[(2^{2-d}-1)I^{4b}_1-\tfrac16(2^{4-d}-1)P^2I^{4b}_2\bigr]\\
\notag
&\text{for $\mu=\nu=0$},\\[2ex]
\label{eq:SU2_self_energy_rs}
&=g^2\delta_{ab}\bigl[\tfrac16(31-2d)-\tfrac13(2^{4-d}-1)N_f(1+N_c)\bigr](\delta_{rs}P^2-P_rP_s)I^{4b}_2\\
\notag
&\text{for $\mu=r$, $\nu=s$}.
\end{align}

\paragraph{$\gr{U(1)}_Y$ gauge boson self-energy.}
\begin{align}
\notag
\\[-8ex]
\notag
\mu\;&
\parbox{20mm}{%
\begin{fmfgraph}(20,20)
\fmfleft{l}
\fmfright{r}
\fmf{zigzag}{r,v,l}
\fmfv{d.sh=circle,d.fi=shaded,d.si=5mm}{v}
\end{fmfgraph}}
\;\nu\\[-2ex]
\label{eq:U1_self_energy_00}
&=g'^2\bigl[(1-d)I^{4b}_1-\tfrac23(1-\tfrac d4)P^2I^{4b}_2\bigr]-\tfrac12g'^2(d-1)N_f\\
\notag
&\phantom{{}={}}\times[2Y_\ell^2+Y_e^2+N_c(2Y_q^2+Y_u^2+Y_d^2)]\bigl[(1-2^{2-d})I^{4b}_1+\tfrac16(2^{4-d}-1)P^2I^{4b}_2\bigr]\\
\notag
&\text{for $\mu=\nu=0$,}\\[2ex]
\label{eq:U1_self_energy_rs}
&=-\tfrac16g'^2\bigl\{1+(2^{4-d}-1)N_f[2Y_\ell^2+Y_e^2+N_c(2Y_q^2+Y_u^2+Y_d^2)]\bigr\}\\
\notag
&\phantom{{}={}}\times(\delta_{rs}P^2-P_rP_s)I^{4b}_2\\
\notag
&\text{for $\mu=r$, $\nu=s$.}
\end{align}
These two-point gauge-field correlators are not affected by $\sigma$
at one-loop order. In an analogous manner, we could determine the gluon Debye mass through the $\gr{SU(3)}_c$ gauge boson self energy, but instead we take it from the literature, cf.~Section~\ref{sec:thermal_masses}. The wave-function renormalization of the temporal gluon fields is not needed at all, as the temporal gluons do not couple to the Higgs field at tree level like the $\gr{SU(2)}_L$ and $\gr{U(1)}_Y$ gauge fields.

\paragraph{Higgs doublet self-energy.} Here we only consider the
contributions to wavefunction renormalization, i.e.~the $P^2$ part of
the correlator. (The corrections to the mass parameter will be
extracted below from the effective potential.) In the Standard Model
alone, there are three different one-loop diagrams that contribute to
wavefunction renormalization, corresponding to the exchange of
$A^a_\mu$ and $B_\mu$ and to the fermion loop,
respectively. Altogether, they give
\begin{equation}
\label{eq:higg_self_energy}
\delta_{ij}\bigl\{\tfrac94g^2+\tfrac34g'^2-(2^{4-d}-1)\tr\bigl[h^{(e)}h^{(e)\dagger}+N_ch^{(u)}h^{(u)\dagger}+N_ch^{(d)}h^{(d)\dagger}\bigr]\bigr\}P^2I^{4b}_2.
\end{equation}
In addition, there is one diagram containing a massive $\sigma$ propagator:
\begin{equation}
\selftwo{scalar}{scalar}{double}{scalar}
\nonumber
\end{equation}
A brief calculation shows that the $P^2$ piece of this diagram
reads\footnote{In our notation four-momenta are written
  $K=(K_0,\vek{k})$; see Appendix~\ref{sec:master_integrals}.}
\begin{equation}
\tfrac14\mu_m^2\delta_{ij}P^2\,\sumint K\frac1{K^2(K^2+\mu_\sigma^2)^2}\left(\frac4d\frac{k^2}{K^2+\mu_\sigma^2}-1\right)=\tfrac14\mu_m^2\delta_{ij}P^2\bigl[\tfrac4d\tilde J_{3/1,0,1}^{4b}(\mu_\sigma)-\tilde J_{2/1}^{4b}(\mu_\sigma)\bigr].
\label{eq:higg_self_energy_sigma_part}
\end{equation}
Note that this sum-integral is manifestly finite and thus does not
require any regularization. Also, unlike the sum-integrals with SM
propagators only, the zero mode \emph{is} included here.

\subsubsection{Correlators with gauge fields}
\label{sec:gauge_field_correlators}

We consider first the self-couplings of the temporal gauge fields. At one loop, these do not receive any contributions from the $\sigma$ field, and we therefore merely list the results.

\paragraph{The $A^a_0A^b_0A^c_0A^d_0$ correlator.}
\begin{align}
\label{eq:adjoint_cor_1}
\parbox{20mm}{%
\begin{fmfgraph}(20,20)
\fmfleftn{l}{2}
\fmfrightn{r}{2}
\fmf{boson}{l1,v}
\fmf{boson}{v,l2}
\fmf{boson}{r1,v}
\fmf{boson}{v,r2}
\fmfv{d.sh=circle,d.fi=shaded,d.si=5mm}{v}
\end{fmfgraph}}%
={}&\tfrac16(d-1)(d-3)\bigl[8d-7+(1-2^{4-d})N_f(1+N_c)\bigr]\\[-4ex]
\notag
&\times g^4(\delta_{ab}\delta_{cd}+\delta_{ac}\delta_{bd}+\delta_{ad}\delta_{bc})I^{4b}_2.
\end{align}

\paragraph{The $B_0^4$ correlator.}
\begin{align}
\label{eq:adjoint_cor_2}
\parbox{20mm}{%
\begin{fmfgraph}(20,20)
\fmfleftn{l}{2}
\fmfrightn{r}{2}
\fmf{zigzag}{l1,v}
\fmf{zigzag}{v,l2}
\fmf{zigzag}{r1,v}
\fmf{zigzag}{v,r2}
\fmfv{d.sh=circle,d.fi=shaded,d.si=5mm}{v}
\end{fmfgraph}}%
={}&\tfrac12(d-1)(d-3)\bigl\{1+\tfrac12(1-2^{4-d})N_f\\[-4ex]
\notag
&\times\bigl[2Y_\ell^4+Y_e^4+N_c(2Y_q^4+Y_u^4+Y_d^4)\bigr]\bigr\}g'^4I^{4b}_2.
\end{align}

\paragraph{The $A^a_0A^b_0B_0^2$ correlator.}
\begin{equation}
\label{eq:adjoint_cor_3}
\parbox{20mm}{%
\begin{fmfgraph}(20,20)
\fmfleftn{l}{2}
\fmfrightn{r}{2}
\fmf{boson}{l1,v}
\fmf{boson}{v,l2}
\fmf{zigzag}{r1,v}
\fmf{zigzag}{v,r2}
\fmfv{d.sh=circle,d.fi=shaded,d.si=5mm}{v}
\end{fmfgraph}}%
=\tfrac12(d-1)(d-3)\bigl[1+(1-2^{4-d})N_f(Y_\ell^2+N_cY_q^2)\bigr]\delta_{ab}g^2g'^2I^{4b}_2.
\end{equation}

Next, we consider the four-point functions with two gauge field and
two scalar legs. Knowing the wavefunction renormalization factors of
all the fields, the correlators with temporal gauge fields determine
the new couplings of these fields in the three-dimensional effective
theory, whereas the correlators with spatial gauge fields determine
the gauge couplings $g_3$ and $g_3'$. In principle, the same gauge
couplings can also be extracted from four-point gauge correlators,
which are however somewhat more difficult to evaluate. The correlators
used here, albeit simpler to calculate, have a downside: they contain
explicit contributions from $\sigma$, which in the final expressions
for $g_3$ and $g_3'$ have to cancel against similar contributions
coming from the Higgs field wavefunction
renormalization\footnote{Since neither the two-point nor the
  four-point gauge correlators contain any $\sigma$ propagators at one
  loop, the effective theory gauge couplings are manifestly
  independent of $\sigma$.}. We consider this a nontrivial test of the
correctness of our calculation.

\paragraph{The $\phi^{\dag i}\phi^jA^a_\mu A^b_\nu$ correlator.} We first put together all 1PI diagrams without any $\sigma$ propagators, getting
\begin{align}
\label{eq:phiphiAA_00}
&\delta_{ij}\delta_{ab}\bigl\{d(d-\tfrac{25}8)g^4+\tfrac d8g^2g'^2+3(d-3)\lambda_hg^2\\
\notag
&+\tfrac12(2^{4-d}-1)(2-d)g^2\tr\bigl[h^{(e)}h^{(e)\dagger}+N_ch^{(u)}h^{(u)\dagger}+N_ch^{(d)}h^{(d)\dagger}\bigr]\bigr\}I^{4b}_2\\
\notag
&\text{for $\mu=\nu=0$,}\\[2ex]
\notag
&\delta_{ij}\delta_{ab}\delta_{rs}\bigl\{-\tfrac38g^4+\tfrac38g^2g'^2-\tfrac12(2^{4-d}-1)g^2\tr\bigl[h^{(e)}h^{(e)\dagger}+N_ch^{(u)}h^{(u)\dagger}+N_ch^{(d)}h^{(d)\dagger}\bigr]\bigr\}I^{4b}_2\\
\label{eq:phiphiAA_rs}
&\text{for $\mu=r$, $\nu=s$.}
\end{align}
In addition, there are two one-$\sigma$-irreducible (1$\sigma$I) and two one-$\sigma$-reducible (1$\sigma$R) diagrams which can be grouped into two pairs according to the coupling of the external gauge legs to the loop. The first pair reads
\begin{align}
\notag
\scatthreeinv{boson}{boson}{scalar}{double}{scalar}{scalar}{scalar}+\scatexthree{scalar}{scalar}{double}{scalar}{scalar}{boson}{boson}&=-\tfrac18\delta_{ij}\delta_{ab}\delta_{\mu\nu}\mu_m^2g^2\,\sumint K\frac1{(K^2)^2}\left(\frac1{K^2+\mu_\sigma^2}+\frac2{\mu_\sigma^2}\right)\\[-2ex]
\label{eq:phiphiAA_sigma1}
&=-\tfrac18\delta_{ij}\delta_{ab}\delta_{\mu\nu}\mu_m^2g^2\left[\tilde J^{4b}_{1/2}(\mu_\sigma)+\frac2{\mu_\sigma^2}\tilde I^{4b}_2\right].
\end{align}
Note that both sum-integrals contain an infrared divergence due to the
presence of the zero Matsubara mode of a massless field. For diagrams
without a $\sigma$ propagator, such divergences cancel
straightforwardly in the matching against a contribution of the
corresponding diagram in the three-dimensional theory, and can thus be
dropped immediately. The treatment of diagrams with a $\sigma$
propagator is, however, more subtle since $\sigma$ is missing from the
dimensionally reduced theory. This is resolved thanks to the
tree-level self-interaction of the Higgs field, induced by a $\sigma$
exchange, cf.~graph~\eqref{sigma_exchange} and
Eq.~\eqref{indinter}. Its effect can be viewed as a modification of
the quartic Higgs coupling $\lambda_h$. When inserted in the
diagram~\eqref{diagram1} in the three-dimensional theory, this
correction yields
$-3\delta_{ij}\delta_{ab}\delta_{\mu\nu}\mu_m^2g^2/(8\mu_\sigma^2)\int_k\frac1{(k^2)^2}$,
which is easily seen to cancel the infrared divergence in
Eq.~\eqref{eq:phiphiAA_sigma1}. It is important to keep in mind that
as a result of nonzero $\mu_\sigma$ in the $\sigma$ propagator, the
zero mode contribution to Eq.~\eqref{eq:phiphiAA_sigma1} contains a
finite remainder even after the infrared divergence has been canceled,
which has to be taken into account.

The other pair of diagrams with a $\sigma$ propagator reads
\begin{align}
\notag
\scatfour{scalar}{scalar}{double}{scalar}{scalar}{scalar}{boson}{boson}+\scatexfour{scalar}{scalar}{double}{scalar}{scalar}{scalar}{boson}{boson}&=\tfrac12\delta_{ij}\delta_{ab}\mu_m^2g^2\,\sumint K\frac{K_\mu K_\nu}{(K^2)^3}\left(\frac1{K^2+\mu_\sigma^2}+\frac2{\mu_\sigma^2}\right)\\[-2ex]
\label{eq:phiphiAA_sigma2}
&=\tfrac12\delta_{ij}\delta_{ab}\mu_m^2g^2\left[\tilde J^{4b}_{1/3,1,0}(\mu_\sigma)+\frac2{\mu_\sigma^2}\tilde I^{4b}_{3,1}\right]\\
\notag
&\text{for $\mu=\nu=0$,}\\[2ex]
\notag
&=\tfrac1{2d}\delta_{ij}\delta_{ab}\delta_{rs}\mu_m^2g^2\left[\tilde J^{4b}_{1/3,0,1}(\mu_\sigma)+\frac2{\mu_\sigma^2}\tilde I^{4b}_{3,0,1}\right]\\
\notag
&\text{for $\mu=r$, $\nu=s$.}
\end{align}
The temporal part of this expression is infrared finite. The spatial
part, however, has an infrared divergence. This is canceled by the
mechanism described above, namely by inserting the $\sigma$-induced
correction to $\lambda_h$ into the diagram~\eqref{diagram2} in the
three-dimensional theory. Again, there is a finite leftover which must
be evaluated properly.

\paragraph{The $\phi^{\dag i}\phi^jB_\mu B_\nu$ correlator.} This
correlator is calculated following the same steps as above,
albeit with different combinatorial factors. We first present
the sum of all 1PI diagrams without a $\sigma$ propagator,
\begin{align}
\label{eq:phiphiBB_1}
&\delta_{ij}\Bigl(\tfrac38dg^2g'^2+\tfrac d8g'^4+3(d-3)\lambda_hg'^2-\tfrac12(2^{4-d}-1)g'^2\\
\notag
&\times\tr\bigl\{(d-2)\bigl[(Y_\ell^2+Y_e^2)h^{(e)}h^{(e)\dagger}+N_c(Y_q^2+Y_u^2)h^{(u)}h^{(u)\dagger}+N_c(Y_q^2+Y_d^2)h^{(d)}h^{(d)\dagger}\bigr]\\
\notag
&-2\bigl[Y_eY_\ell h^{(e)}h^{(e)\dagger}+N_cY_uY_qh^{(u)}h^{(u)\dagger}+N_cY_dY_qh^{(d)}h^{(d)\dagger}\bigr]\bigr\}\Bigr)I^{4b}_2\\
\notag
&\text{for $\mu=\nu=0$,}\\[2ex]
\label{eq:phiphiBB_2}
&\delta_{ij}\delta_{rs}\bigl\{\tfrac98g^2g'^2+\tfrac38g'^4-\tfrac12(2^{4-d}-1)g'^2\\
\notag
&\times\tr\bigl[(Y_e-Y_\ell)^2h^{(e)}h^{(e)\dagger}+N_c(Y_u-Y_q)^2h^{(u)}h^{(u)\dagger}+N_c(Y_d-Y_q)^2h^{(d)}h^{(d)\dagger}\bigr]\bigr\}I^{4b}_2\\
\notag
&\text{for $\mu=r$, $\nu=s$.}
\end{align}
In addition to these, there are again two pairs of diagrams with a $\sigma$ propagator, which differ from those with $\gr{SU(2)}_L$ gauge boson lines just by replacing $g$ with $g'$ and removing the overall factor $\delta_{ab}$. We collect the results here for completeness:
\begin{align}
\label{eq:phiphiBB_3}
\scatthreeinv{zigzag}{zigzag}{scalar}{double}{scalar}{scalar}{scalar}+\scatexthree{scalar}{scalar}{double}{scalar}{scalar}{zigzag}{zigzag}&=-\tfrac18\delta_{ij}\delta_{\mu\nu}\mu_m^2g'^2\left[\tilde J^{4b}_{1/2}(\mu_\sigma)+\frac2{\mu_\sigma^2}\tilde I^{4b}_2\right],\\
\notag
\scatfour{scalar}{scalar}{double}{scalar}{scalar}{scalar}{zigzag}{zigzag}+\scatexfour{scalar}{scalar}{double}{scalar}{scalar}{scalar}{zigzag}{zigzag}&=\tfrac12\delta_{ij}\mu_m^2g'^2\left[\tilde J^{4b}_{1/3,1,0}(\mu_\sigma)+\frac2{\mu_\sigma^2}\tilde I^{4b}_{3,1}\right]\\[-2ex]
\notag
&\text{for $\mu=\nu=0$,}\\[2ex]
\notag
&=\tfrac1{2d}\delta_{ij}\delta_{rs}\mu_m^2g'^2\left[\tilde J^{4b}_{1/3,0,1}(\mu_\sigma)+\frac2{\mu_\sigma^2}\tilde I^{4b}_{3,0,1}\right]\\
\notag
&\text{for $\mu=r$, $\nu=s$.}
\end{align}
Upon subtracting the contribution of the three-dimensional theory, there is again a finite leftover which must be carefully accounted for, and which expresses the contribution of the zero mode of $\sigma$ to the effective theory coupling.

\paragraph{The $\phi^{\dag i}\phi^jA^a_\mu B_\nu$ correlator.} Since the information about the gauge couplings in the three-dimensional theory can be extracted from the above correlators with two $A^a_\mu$ or two $B_\mu$ fields, we only need the temporal correlators, $\mu=\nu=0$, here. Putting first together all the diagrams without any $\sigma$ propagators gives
\begin{align}
\label{eq:phiphiAB_1}
&(\tau_a)^{ij}I^{4b}_2\Bigl(\bigl[\tfrac d8g^3g'+\tfrac d8gg'^3+(d-3)\lambda_hgg'\bigr]+\tfrac12(1-2^{4-d})gg'\\
\notag
&\times\tr\bigl\{[(d-2)Y_\ell-Y_e]h^{(e)}h^{(e)\dagger}-N_c[(d-2)Y_q-Y_u]h^{(u)}h^{(u)\dagger}+N_c[(d-2)Y_q-Y_d]h^{(d)}h^{(d)\dagger}\bigr\}\Bigr).
\end{align}
Next, we have to consider diagrams containing a $\sigma$ propagator, and there is again a one-to-one correspondence between the 1$\sigma$I and 1$\sigma$R graphs. First, we find
\begin{equation}
\label{eq:phiphiAB_2}
\scatthreeinv{boson}{zigzag}{scalar}{double}{scalar}{scalar}{scalar}+\scatexthree{scalar}{scalar}{double}{scalar}{scalar}{boson}{zigzag}=-\tfrac18\mu_m^2gg'(\tau_a)^{ij}\tilde J^{4b}_{1/2}(\mu_\sigma);
\end{equation}
the latter diagram vanishes thanks to the trace in the scalar doublet loop. The same is true for the other 1$\sigma$R diagrams, and we therefore only show here the 1$\sigma$I graphs which give a nontrivial contribution, namely
\begin{equation}
\label{eq:phiphiAB_3}
\scatfour{scalar}{scalar}{double}{scalar}{scalar}{scalar}{zigzag}{boson}+\scatfour{scalar}{scalar}{double}{scalar}{scalar}{scalar}{boson}{zigzag}=\tfrac12\mu_m^2gg'(\tau_a)^{ij}\tilde J^{4b}_{1/3,1,0}(\mu_\sigma).
\end{equation}

\paragraph{The $\phi^{\dag i}\phi^jC^\alpha_0 C^\beta_0$ correlator.}
\begin{equation}
\label{eq:adjoint_gluon_cor}
\parbox{20mm}{%
\begin{fmfgraph}(20,20)
\fmfleftn{l}{2}
\fmfrightn{r}{2}
\fmf{scalar}{l1,v}
\fmf{scalar}{v,l2}
\fmf{gluon}{r1,v}
\fmf{gluon}{v,r2}
\fmfv{d.sh=circle,d.fi=shaded,d.si=5mm}{v}
\end{fmfgraph}}%
=2(2^{4-d}-1)(3-d)g^2_s \tr\bigl[h^{(u)}h^{(u)\dagger} + h^{(d)}h^{(d)\dagger} ] \delta_{ij}\delta_{\alpha\beta} I^{4b}_2.
\end{equation}


\subsubsection{Effective potential for the scalars}
\label{sec:effective_potential}

The correlators in the scalar sector are comprised of a large number of diagrams even at the one-loop level. It is therefore advantageous to obtain the corresponding operators in the effective theory using the effective potential method. To that end, we shift the scalar fields by their assumed expectation values,
\begin{equation}
\langle\phi^i\rangle\equiv\varphi^i,\qquad
\langle\sigma\rangle\equiv\rho.
\end{equation}
The shift affects the $\La_\text{scalar}$ and $\La_\text{Yukawa}$
parts of the Lagrangian~\eqref{classlag}. For $\La_\text{Yukawa}$, the
shift simply results in a number of mass terms for the fermions. On
the other hand the shift of the scalar fields has a twofold effect on
$\La_\text{scalar}$. First, it leads to a modification of some of the
couplings in the form $\mu_i \to \tilde{\mu}_i$, with\footnote{The
  modified linear coupling $\tilde\mu_1$ is not needed for the
  calculation of the effective potential, and thus is not given here
  explicitly.}
\begin{equation}
\begin{aligned}
\tilde\mu_h^2&=\mu_h^2-2\lambda_h\he\varphi\varphi-\frac12\mu_m\rho-\frac12\lambda_m\rho^2,\qquad
&
\tilde\mu_3&=\mu_3+3\lambda_\sigma\rho,\\
\tilde\mu_\sigma^2&=\mu_\sigma^2+2\mu_3\rho+3\lambda_\sigma\rho^2+\lambda_m\he\varphi\varphi,\qquad
&
\tilde\mu_m&=\mu_m+2\lambda_m\rho.
\end{aligned}
\end{equation}
Second, it produces a number of new operators that do not appear in the original Lagrangian. There are several new interaction vertices, encoded in
\begin{equation}
\begin{split}
\La^\text{new}_\text{int}={}&\frac14(g^2\vec A_\mu\cdot\vec A_\mu+g'^2B_\mu^2)(\he\varphi\phi+\he\phi\varphi)+\frac12gg'B_\mu\vec A_\mu\cdot(\he\varphi\vec\tau\phi+\he\phi\vec\tau\varphi)\\
&+2\lambda_h\he\phi\phi(\he\phi\varphi+\he\varphi\phi)+\frac12\lambda_m\sigma^2(\he\phi\varphi+\he\varphi\phi),
\end{split}
\end{equation}
and in addition a number of new bilinear terms which introduce mixings between the fields. Together with the already existing bilinear terms for the gauge fields and the scalars, these can be written in the form
\begin{align}
\La_\text{bilin}={}&\frac14G^a_{\mu\nu}G^a_{\mu\nu}+\frac14F_{\mu\nu}F_{\mu\nu}+\frac14(g^2\vec A_\mu\cdot\vec A_\mu+g'^2B_\mu^2)\he\varphi\varphi+\frac12gg'B_\mu\vec A_\mu\cdot\he\varphi\vec\tau\varphi\\
\notag
&+\partial_\mu\he\phi\partial_\mu\phi-\tilde\mu_h^2\he\phi\phi+\frac12(\partial_\mu\sigma)^2+\frac12\tilde\mu_\sigma^2\sigma^2+\left(\frac12\mu_m+\lambda_m\rho\right)\sigma(\he\phi\varphi+\he\varphi\phi)\\
\notag
&+\lambda_h(\he\phi\varphi+\he\varphi\phi)^2+\frac{\imag g}2\vec A_\mu\cdot(\he\varphi\vec\tau\partial_\mu\phi-\partial_\mu\he\phi\vec\tau\varphi)+\frac{\imag g'}2B_\mu(\he\varphi\partial_\mu\phi-\partial_\mu\he\phi\varphi).
\end{align}
Note that the last two operators are irrelevant as they vanish when
contracted with a gauge boson propagator in the Landau gauge. The
bilinear part of the Lagrangian, $\La_\text{bilin}$ -- together with a
similar bilinear Lagrangian for the fermions -- completely determines
the effective potential at the one-loop level. Hence we only
need to know the eigenvalues of the mass matrix for all the
fields. Clearly, the masses of gluons as well as of all the ghosts are
independent of $\varphi$ and $\rho$, and these fields therefore do not
contribute to the effective potential (apart from its constant part,
which does not play a role in the matching to the three-dimensional
effective theory). In the electroweak gauge boson sector, the squared
masses read
\begin{equation}
0,\qquad
m_W^2=\frac12g^2\he\varphi\varphi\quad[2\times],\qquad
m_Z^2=\frac12(g^2+g'^2)\he\varphi\varphi.
\end{equation}
In the scalar sector, we find three modes with mass squared
$-\tilde\mu_h^2$. The remaining component of the Higgs doublet mixes
with $\sigma$, and the mass eigenvalues have to be found by explicit
diagonalization,
\begin{equation}
m^2_\pm=\frac{\tilde\mu_\sigma^2-\tilde\mu_h^2}2+2\lambda_h\he\varphi\varphi\pm\sqrt{\left(\frac{\tilde\mu_\sigma^2+\tilde\mu_h^2}2-2\lambda_h\he\varphi\varphi\right)^2+2\he\varphi\varphi\left(\frac12\mu_m+\lambda_m\rho\right)^2}.
\end{equation}
The full one-loop effective potential of the four-dimensional theory then reads
\begin{equation}
V_\text{eff}=d\tilde K^{4b}(m_Z)+2d\tilde K^{4b}(m_W)+3\tilde K^{4b}(\imag\tilde\mu_h)+\tilde K^{4b}(m_+)+\tilde K^{4b}(m_-)-4\sum_i\tilde K^{4f}(h_i\sqrt{\he\varphi\varphi}),
\label{effpot}
\end{equation}
where the sum runs over all eigenvalues of the Yukawa coupling matrices,
\begin{equation}
h_i\in\text{spectrum}\Bigl(\sqrt{h^{(e)}h^{(e)\dagger}},\sqrt{h^{(u)}h^{(u)\dagger}},\sqrt{h^{(d)}h^{(d)\dagger}}\Bigr)
\end{equation}
including the $N_c$-fold degeneracy due to different colors.

\subsubsection{Scalar correlators from the effective potential}

The 1PI correlators at zero momentum can be determined from the
effective potential~\eqref{effpot}. As this is still an exact
expression, we merely have to determine the scaling of individual
couplings. Together with the tree-level potential, the result can be
written using the generic notation
\begin{equation}
V_\text{eff}=V_{0,0}+V_{2,0}\he\varphi\varphi+V_{4,0}(\he\varphi\varphi)^2+V_{0,1}\rho+V_{0,2}\rho^2+V_{0,3}\rho^3+V_{0,4}\rho^4+V_{2,1}\rho\he\varphi\varphi+\dotsb,
\end{equation}
where the relevant coefficients read
\begin{align}
\notag
V_{2,0}={}&-\mu_h^2+\bigl(\tfrac34dg^2+\tfrac14 dg'^2+6\lambda_h\bigr)I^{4b}_1-2\sum_ih_i^2I^{4f}_1-\tfrac14\mu_m^2\tilde J^{4b}_{1/1}(\mu_\sigma)+\tfrac12\lambda_m\tilde J^{4b}_1(\mu_\sigma),\\
\notag
V_{4,0}={}&\lambda_h-\bigl(\tfrac3{16}dg^4+\tfrac1{16}dg'^4+\tfrac18dg^2g'^2+12\lambda_h^2\bigr)I^{4b}_2+\sum_ih_i^4I^{4f}_2\\
\notag
&+\tfrac32\lambda_h\mu_m^2\tilde J^{4b}_{1/2}(\mu_\sigma)-\tfrac1{16}\mu_m^4\tilde J^{4b}_{2/2}(\mu_\sigma)+\tfrac14\lambda_m\mu_m^2\tilde J^{4b}_{2/1}(\mu_\sigma)-\tfrac14\lambda_m^2\tilde J^{4b}_2(\mu_\sigma),\\
V_{0,1}={}& \mu_1 + \mu_mI^{4b}_1+\mu_3\tilde J^{4b}_1(\mu_\sigma),\\
\notag
V_{0,2}={}&\tfrac12\mu_\sigma^2+\lambda_mI^{4b}_1-\tfrac14\mu_m^2I^{4b}_2-\mu_3^2\tilde J^{4b}_2(\mu_\sigma)+\tfrac32\lambda_\sigma\tilde J^{4b}_1(\mu_\sigma),\\
\notag
V_{2,1}={}&\tfrac12\mu_m-3\lambda_h\mu_mI^{4b}_2+\tfrac12\mu_3\mu_m^2\tilde J^{4b}_{2/1}(\mu_\sigma)+\tfrac18\mu_m^3\tilde J^{4b}_{1/2}(\mu_\sigma)-\lambda_m\mu_3\tilde J^{4b}_2(\mu_\sigma)-\lambda_m\mu_m\tilde J^{4b}_{1/1}(\mu_\sigma).
\end{align}
This is, however, not quite enough. In order to match to the three-dimensional theory, we have to determine all one-$\phi$-irreducible contributions to the Higgs correlators. These include in particular diagrams that are 1$\sigma$R.

We first consider the Higgs two-point function. The corresponding
wavefunction renormalization was already found above, see
Eqs.~\eqref{eq:higg_self_energy}
and~\eqref{eq:higg_self_energy_sigma_part}. The static two-point
function, on the other hand, consists of contributions of the
forms\\[1ex]
\begin{equation}
\parbox{20mm}{%
\begin{fmfgraph}(20,20)
\fmfleft{l}
\fmfright{r}
\fmf{scalar}{r,v,l}
\fmfv{d.sh=circle,d.fi=shaded,d.si=5mm}{v}
\end{fmfgraph}}
\,+
\parbox{20mm}{%
\begin{fmfgraph}(20,20)
\fmfleft{l}
\fmfright{r}
\fmftop{t}
\fmf{scalar}{r,v,l}
\fmfv{d.sh=circle,d.fi=shaded,d.si=5mm}{v}
\fmffreeze
\fmf{double}{t,v}
\fmfv{d.sh=circle,d.fi=full,d.si=5mm}{t}
\end{fmfgraph}}
\,+\dotsb,\\[-3ex]
\label{higgsmass}
\end{equation}
where the ellipsis stands for 1$\sigma$R diagrams carrying contributing of orders $g^n$, $n>2$. Note that the full circle denotes connected Green's functions, whereas the shaded circle the 1PI ones. To order $g^2$, we therefore obtain
\begin{equation}
\label{eq:phi2_cor}
\begin{split}
\Pi_2 ={}& -V_{2,0}+\frac{\mu_m}{2\mu_\sigma^2}V_{0,1}\\
={}&\mu_h^2 + \frac{\mu_1 \mu_m}{2\mu^2_\sigma} -\bigl(\tfrac34dg^2+\tfrac14 dg'^2+6\lambda_h\bigr)I^{4b}_1+2\sum_ih_i^2I^{4f}_1\\
&+\frac{\mu_3\mu_m}{2\mu_\sigma^2}\tilde J^{4b}_1(\mu_\sigma)+\tfrac14\mu_m^2\left[\tilde J^{4b}_{1/1}(\mu_\sigma)+\frac{2I^{4b}_1}{\mu_\sigma^2}\right]-\tfrac12\lambda_m\tilde J^{4b}_1(\mu_\sigma).
\end{split}
\end{equation}

The one-$\phi$-irreducible static four-point correlator is evaluated in a similar fashion. Symbolically, it is given by
\begin{equation}
\parbox{20mm}{%
\begin{fmfgraph}(20,20)
\fmfleftn{l}{2}
\fmfrightn{r}{2}
\fmf{scalar}{l1,v}
\fmf{scalar}{v,l2}
\fmf{scalar}{r1,v}
\fmf{scalar}{v,r2}
\fmfv{d.sh=circle,d.fi=shaded,d.si=5mm}{v}
\end{fmfgraph}}%
+
\parbox{35mm}{%
\begin{fmfgraph}(35,20)
\fmfleftn{l}{2}
\fmfrightn{r}{2}
\fmf{scalar,tension=1}{l1,vl}
\fmf{scalar,tension=1}{vl,l2}
\fmf{scalar,tension=1}{r1,vr}
\fmf{scalar,tension=1}{vr,r2}
\fmf{double,tension=1}{vl,v,vr}
\fmfv{d.sh=circle,d.fi=shaded,d.si=5mm}{vl,vr}
\fmfv{d.sh=circle,d.fi=full,d.si=5mm}{v}
\end{fmfgraph}}
+\quad
\parbox{25mm}{%
\begin{fmfgraph}(25,20)
\fmfleftn{l}{3}
\fmfrightn{r}{2}
\fmf{scalar,tension=1}{l1,vl}
\fmf{scalar,tension=1}{vl,l3}
\fmf{scalar,tension=1}{r1,vr}
\fmf{scalar,tension=1}{vr,r2}
\fmf{double,tension=1}{vl,vr}
\fmfv{d.sh=circle,d.fi=shaded,d.si=5mm}{vl,vr}
\fmffreeze
\fmf{double}{l2,vl}
\fmfv{d.sh=circle,d.fi=full,d.si=5mm}{l2}
\end{fmfgraph}}
+
\parbox{25mm}{%
\begin{fmfgraph}(25,20)
\fmfleftn{l}{2}
\fmfrightn{r}{3}
\fmf{scalar,tension=1}{l1,vl}
\fmf{scalar,tension=1}{vl,l2}
\fmf{scalar,tension=1}{r1,vr}
\fmf{scalar,tension=1}{vr,r3}
\fmf{double,tension=1}{vl,vr}
\fmfv{d.sh=circle,d.fi=shaded,d.si=5mm}{vl,vr}
\fmffreeze
\fmf{double}{r2,vr}
\fmfv{d.sh=circle,d.fi=full,d.si=5mm}{r2}
\end{fmfgraph}}
\quad+\dotsb,
\label{fourpoint}
\end{equation}
where the ellipsis now denotes contributions beyond order $g^4$. To
compute this, we need to know the two-point function of
$\sigma$. Fortunately, all the $\sigma$ propagators in these diagrams
carry zero momentum so only the mass of $\sigma$ is needed. This is
given by\\[1ex]
\begin{equation}
\parbox{20mm}{%
\begin{fmfgraph}(20,20)
\fmfleft{l}
\fmfright{r}
\fmf{double}{r,v,l}
\fmfv{d.sh=circle,d.fi=shaded,d.si=5mm}{v}
\end{fmfgraph}}
\,+\,
\parbox{20mm}{%
\begin{fmfgraph}(20,20)
\fmfleft{l}
\fmfright{r}
\fmftop{t}
\fmf{double}{r,v,l}
\fmfv{d.sh=circle,d.fi=shaded,d.si=5mm}{v}
\fmffreeze
\fmf{double}{t,v}
\fmfv{d.sh=circle,d.fi=full,d.si=5mm}{t}
\end{fmfgraph}}
\,+\dotsb,\\[-3ex]
\end{equation}
in a similar manner to Eq.~\eqref{higgsmass}. As we are calculating
the Higgs four-point function to order $g^4$ and the
$\sigma\he\phi\phi$ vertex begins at order $g$, we need to know the
$\sigma$ mass to order $g^2$. Hence the one-point function of $\sigma$
is only needed to order $g$ and is given by $V_{0,1}$. Putting all the
pieces together, we find the ``renormalized squared $\sigma$ mass'' to
be
\begin{equation}
\mu_{\sigma,\text{ren}}^2=2\left(V_{0,2}-\frac{\mu_3}{\mu_\sigma^2}V_{0,1}\right).
\end{equation}
The full static four-point correlator of the Higgs field, indicated in Eq.~\eqref{fourpoint}, then becomes $(\delta_{ik}\delta_{j\ell}+\delta_{i\ell}\delta_{jk})\Pi_4$, where
\begin{align}
\notag
\Pi_4={}&-2V_{4,0}+\frac{V_{2,1}^2}{\mu_{\sigma,\text{ren}}^2}-\frac4{\mu_\sigma^4}V_{0,1}V_{2,1}V_{2,2}\\
\notag
={}&-2\lambda_h+\frac{\mu_m^2}{4\mu_\sigma^2}+\frac{\mu_1 \mu_3 \mu^2_m}{2 \mu^6_\sigma} - \frac{\mu_1 \mu_m \lambda_m}{\mu^4_\sigma}+\bigl(\tfrac38dg^4+\tfrac18dg'^4+\tfrac14dg^2g'^2+24\lambda_h^2\bigr)I^{4b}_2\\
\notag
&-2\sum_ih_i^4I^{4f}_2-\frac{3\lambda_h\mu_m^2}{\mu_\sigma^2}\bigl[I^{4b}_2+\mu_\sigma^2\tilde J^{4b}_{1/2}(\mu_\sigma)\bigr]+\frac{\mu_3\mu_m^3}{2\mu_\sigma^6}\bigl[I^{4b}_1+\mu_\sigma^4\tilde J^{4b}_{2/1}(\mu_\sigma)\bigr]\\
\label{eq:phi4_cor}
&+\frac{\mu_3^2\mu_m^2}{2\mu_\sigma^6}\bigl[\tilde J^{4b}_1(\mu_\sigma)+\mu_\sigma^2\tilde J^{4b}_2(\mu_\sigma)\bigr]-\frac{\lambda_m\mu_3\mu_m}{\mu_\sigma^4}\bigl[\tilde J^{4b}_1(\mu_\sigma)+\mu_\sigma^2\tilde J^{4b}_2(\mu_\sigma)\bigr]\\
\notag
&-\frac{3\lambda_\sigma\mu_m^2}{4\mu_\sigma^4}\tilde J^{4b}_1(\mu_\sigma)+\frac{\mu_m^4}{8\mu_\sigma^4}\bigl[I^{4b}_2+\mu_\sigma^2\tilde J^{4b}_{1/2}(\mu_\sigma)+\mu_\sigma^4\tilde J^{4b}_{2/2}(\mu_\sigma)\bigr]\\
\notag
&-\frac{\lambda_m\mu_m^2}{2\mu_\sigma^4}\bigl[3I^{4b}_1+2\mu_\sigma^2\tilde J^{4b}_{1/1}(\mu_\sigma^2)+\mu_\sigma^4\tilde J^{4b}_{2/1}(\mu_\sigma)\bigr]+\tfrac12\lambda_m^2\tilde J^{4b}_2(\mu_\sigma).
\end{align}

\subsection{Counterterms and $\beta$-functions}
\label{sec:counterterms_beta_functions}

All the counterterms of the SSM are defined in
Section~\ref{sec:renormalization}. We use the modified minimal
subtraction (\MSbar) scheme. When implemented in combination with
dimensional regularization and the definition of momentum integrals
according to Eq.~\eqref{MSint}, finding the counterterms then amounts
to extracting the pole part of the corresponding
correlators\footnote{We note that the ultraviolet divergences are
  independent of temperature, hence the counterterms can be extracted
  from correlators computed either at nonzero temperature or in the
  vacuum. We can, and will, therefore make use of the previously
  calculated thermal correlators.}. From this point on, we make the
substitution $N_c=3$ in the main text. The field renormalization
counterterms can be extracted from the two-point correlators, obtained
in Section~\ref{subsec:debye}, giving
\begin{equation}
\begin{split}
\delta Z_A &= \frac{g^2}{16\pi^2\epsilon}\biggl[\frac{25}{6}-\frac{4}{3}N_f \biggr],\\
\delta Z_B &=-\frac{g'^2}{96\pi^2\epsilon}\bigl\{1+N_f \bigl[2Y^2_\ell + Y^2_e + 3 ( 2Y^2_q + Y^2_u + Y^2_d)\bigr] \bigr\},\\
\delta Z_\phi &= \frac{1}{16\pi^2\epsilon}\biggl\{\frac{9}{4}g^2 + \frac{3}{4}g'^2 - \tr\bigl[h^{(e)}h^{(e)\dagger}+3 h^{(u)}h^{(u)\dagger}+3 h^{(d)}h^{(d)\dagger}\bigr] \biggr\}.
\label{eq:field_renormalisation}
\end{split}
\end{equation}
The $\sigma$ field receives no divergent contribution to wavefunction
renormalization at one loop, and hence $Z_\sigma=1$. The counterterms
for the gauge and Yukawa couplings are not affected by $\sigma$ at one
loop, and thus agree with the SM results. We list here the results,
valid in Landau gauge and assuming that only the top quark Yukawa
coupling is nonzero~\cite{Kajantie:1995dw},
\begin{equation}
\begin{split}
\delta g &= -\frac{g^3}{32\pi^2\epsilon}\biggl(\frac{43}{6}-\frac{4}{3}N_f \biggr),\\
\delta g' &= \frac{g'^3}{192\pi^2\epsilon}\biggl(1+ \frac{40}{3}N_f \biggr),\\
\delta g_Y &= \frac{g_Y}{16\pi^2\epsilon}\biggl(\frac{9}{4}g^2_Y - \frac{9}{8}{g}^2 - \frac{17}{24}g'^2 - 4 g^2_s \biggr).
\end{split}
\end{equation}
The counterterms for the couplings in the scalar sector can be found using the one-loop effective potential, calculated in Section~\ref{sec:effective_potential}. We take Eq.~\eqref{effpot} and expand the integrals around the actual mass parameters of the fields in powers of the classical fields $\varphi$ and $\rho$. The individual counterterms are then readily identified as
\begin{align}
\notag
\delta \mu^2_h &= \frac{1}{64\pi^2\epsilon}\bigl(24 \lambda_h \mu^2_h - \mu^2_m - 2\lambda_m \mu^2_\sigma  \bigr),\\
\notag
\delta \lambda_h &= \frac{1}{256\pi^2\epsilon}\bigl(9g^4 + 3g'^4 + 6 g^2 g'^2 - 48 g^4_Y + 192 \lambda^2_h + 4 \lambda^2_m  \bigr),\\
\notag
\delta \mu^2_\sigma &= \frac{1}{32\pi^2\epsilon}\bigl(4 \mu^2_3 + \mu^2_m + 6\lambda_\sigma \mu^2_\sigma - 4 \lambda_m \mu^2_h \bigr),\\
\delta \mu_1 &= -\frac{1}{16\pi^2\epsilon}\bigl(\mu^2_h \mu_m - \mu^2_\sigma \mu_3 \bigr),\\
\notag
\delta \mu_3 &= \frac{3}{32\pi^2\epsilon}\bigl(6 \lambda_\sigma \mu_3 + \lambda_m \mu_m \bigr),\\
\notag
\delta \lambda_\sigma &= \frac{1}{16\pi^2\epsilon}\bigl(\lambda^2_m + 9 \lambda^2_\sigma \bigr),\\
\notag
\delta \mu_m &= \frac{1}{8\pi^2\epsilon}\bigl[3 \lambda_h \mu_m + \lambda_m(\mu_3 + \mu_m) \bigr],\\
\notag
\delta \lambda_m &= \frac{1}{16\pi^2\epsilon} \lambda_m \bigl( 6 \lambda_h + 2 \lambda_m + 3 \lambda_\sigma \bigr).
\end{align}
The wavefunction renormalization factors together with the
coupling counterterms determine, in the usual manner, the running of
the couplings with renormalization scale $\Lambda$. From the
one-loop counterterms listed above, we obtain the one-loop
$\beta$-functions of all the couplings of the SSM:
\begin{align}
\label{running_g}
\Lambda \frac{d}{d\Lambda}g^2 ={}& - \frac{g^4}{8 \pi^2} \bigg( \frac{43}{6}-\frac{4}{3}N_f \bigg), \\
\Lambda \frac{d}{d\Lambda}g'^2 ={}& \frac{g'^4}{8 \pi^2} \bigg( \frac{1}{6} + \frac{20}{9}N_f \bigg), \\
\Lambda \frac{d}{d\Lambda}g^2_Y ={}& \frac{1}{8 \pi^2} \bigg( \frac{9}{2}g^4_Y - 8 g^2_s g^2_Y - \frac{9}{4}g^2 g^2_Y - \frac{17}{12}g'^2 g^2_Y \bigg), \\
\Lambda \frac{d}{d\Lambda}\mu^2_h ={}& \frac{1}{8\pi^2} \bigg[- \frac{\mu^2_m}{4} - \frac{1}{2}\lambda_m \mu^2_\sigma  + \mu^2_h \bigg( -\frac{9}{4}g^2 - \frac{3}{4}g'^2 + 6 \lambda_h + g^2_{Y,1} \bigg) \bigg], \\
\Lambda \frac{d}{d\Lambda} \lambda_h ={}& \frac{1}{8\pi^2}\biggl[ 12
  \lambda^2_h + \frac{1}{4} \lambda^2_m + \frac{9}{16}g^4  +
  \frac{3}{8}g^2 g'^2 + \frac{3}{16}g'^4 - \sum_i h^4_i \notag \\
& - \lambda_h \frac{3}{2} \big(3g^2 + g'^2\big)  + 2 \lambda_h g^2_{Y,1}   \biggr], \\ 
\Lambda \frac{d}{d\Lambda}\mu^2_\sigma ={}& \frac{1}{8\pi^2} \bigg( 2 \mu^2_3  + \frac{1}{2} \mu^2_m + 3 \lambda_\sigma \mu^2_\sigma - 2 \lambda_m \mu^2_h \bigg), \\
\Lambda \frac{d}{d\Lambda}\mu_1 ={}& \frac{1}{8\pi^2} \big(\mu_3 \mu^2_\sigma - \mu^2_h \mu_m \big), \\
\Lambda \frac{d}{d\Lambda}\mu_3 ={}& \frac{3}{8\pi^2}\bigg( 3 \lambda_\sigma \mu_3 + \frac{1}{2} \lambda_m \mu_m \bigg),\\
\Lambda \frac{d}{d\Lambda}\lambda_\sigma ={}& \frac{1}{8\pi^2}\big(\lambda^2_m + 9 \lambda^2_\sigma \big) ,\\ 
\Lambda \frac{d}{d\Lambda}\mu_m ={}& \frac{1}{8\pi^2} \biggl[ 2
  \lambda_m \mu_3 + \mu_m \bigg( -\frac{9}{4}g^2 - \frac{3}{4}g'^2 + 6
  \lambda_h + 2 \lambda_m + g^2_{Y,1} \bigg)  \biggr], \quad \text{and} \\
\Lambda \frac{d}{d\Lambda}\lambda_m ={}& \frac{\lambda_m}{8\pi^2} \biggl( - \frac{9}{4}g^2 - \frac{3}{4}g'^2 + g^2_{Y,1}+2 \lambda_m + 6 \lambda_h + 3 \lambda_\sigma \biggr) .
\label{running_end}
\end{align}
Note that the running of the strong coupling $g_s$ is not included here, as it is not needed at the order of our calculation due to the fact that the parameter does not appear in any tree-level results.

\subsection{Matching relations}
\label{sec:matrel}

Having considered above both the correlators needed for dimensional
reduction as well as all the required counterterms, we are ready to
move to the explicit derivation of the effective theory
parameters. For the heavy scale effective theory, these are obtained
by matching the long-distance behavior of various static Green's
functions with the full theory. The final step to the effective theory
for the light scale is described below in
Section~\ref{sec:int_heavyscale}.

\subsubsection{Thermal masses and normalization of fields}
\label{sec:thermal_masses}

We start with the mass parameters for the temporal gauge fields. These
are forbidden in the four-dimensional theory by gauge invariance, and
therefore arise solely from integration of the superheavy modes. To
leading order in powers of the gauge couplings, they can simply
  be read off the static limits of the two-point correlators,
Eqs.~\eqref{eq:SU2_self_energy_00} and \eqref{eq:U1_self_energy_00},
\begin{align}
\notag
m_D^2&= g^2\big[(d-1)(2d-1)-4N_f(d-1)(2^{2-d}-1)\big]I^{4b}_1\\ 
&= g^2T^2\bigg(\frac{5}{6}+\frac{N_f}{3}\bigg), \\
\notag
m'^2_D&= g'^2(d-1)\bigg\{1-\frac{1}{2}N_f\big[2Y_l^2+Y_e^2+3(2Y_q^2+Y_u^2+Y_d^2)\big]\big(2^{2-d}-1\big)\bigg\}I^{4b}_1\\ 
&= g'^2T^2\bigg(\frac{1}{6}+\frac{5N_f}{9}\bigg).
\end{align}
The gluon Debye mass $m''_D$ can be taken from the literature, see e.g. Ref.~\cite{Braaten:1994pk}
\begin{align}
m''^2_D ={}&g^2_s T^2 \bigg(1+\frac{N_f}{6}\bigg).
\end{align}

For the evaluation of (most of) the other couplings of the effective
theory, we need to know the relation between the three-dimensional and
the four-dimensional fields. Within this section, these will be
distinguished by the lower indices $\td$ and $\fd$, respectively. For
a generic field, the relation between the fields reads
\begin{equation}
\psi_\td^2 = \frac{1}{T}\big[1 + \Pi_{\psi}'(0) - \delta Z_\psi \big] \psi_\fd^2,
\label{fieldmatching}
\end{equation}
where $\Pi_{\psi}(P)$ is the self-energy of the field, a prime denotes
a derivative with respect to $P^2$, and $\delta Z_\psi$ is the field
renormalization counterterm. We will now consider all the
fields of the three-dimensional effective theory one by one.

\paragraph{The $\gr{SU(2)}_L$ gauge fields.}
Using the momentum-dependent parts of
Eqs.~\eqref{eq:SU2_self_energy_00} and \eqref{eq:SU2_self_energy_rs}
as well as the counterterm from Eq.~\eqref{eq:field_renormalisation},
the general relation~\eqref{fieldmatching} immediately leads to
\begin{align}
A_{\td,0}^2 &=\frac{A_{\fd,0}^2}{T}\bigg\{1+\frac{g^2}{(4\pi)^2}\bigg[-\frac{25L_b}{6}+3+\frac{4N_f}{3}(L_f-1)\bigg]\bigg\},\\
A_{\td,r}^2 &=\frac{A_{\fd,r}^2}{T}\bigg[1+\frac{g^2}{(4\pi)^2}\bigg(-\frac{25L_b}{6}-\frac{2}{3}+\frac{4N_f}{3}L_f\bigg)\bigg], 
\end{align}
where we have followed Ref.~\cite{Kajantie:1995dw} in defining
\begin{equation}
L_b\equiv2\log\bigg(\frac{\Lambda}{4\pi T}\bigg)+2\gamma,\qquad
L_f\equiv L_b+4\log2.
\end{equation}
Note that the divergences coming from the two-point correlators and
the wavefunction renormalization factors have to cancel each
other in the final matching relations for the fields. This is another
nontrivial check that our calculation is correct.

\paragraph{The $\gr{U(1)}_Y$ gauge fields.}
Here we analogously use the momentum-dependent parts of
Eqs.~\eqref{eq:U1_self_energy_00} and \eqref{eq:U1_self_energy_rs} in
combination with the counterterm from
Eq.~\eqref{eq:field_renormalisation} to get
\begin{align}
B_{\td,0}^2
&=\frac{B_{\fd,0}^2}{T}\bigg\{1+\frac{g'^2}{(4\pi)^2}\bigg[\frac{L_b}{6}+\frac{1}{3}+\frac{20N_f}{9}(L_f-1)\bigg]
\bigg\},\\ B_{\td,r}^2 &=\frac{B_{\fd,r}^2}{T}\bigg[1+
  \frac{g'^2}{(4\pi)^2}\bigg(\frac{L_b}{6}+\frac{20N_f}{9}L_f\bigg)\bigg].
\end{align}

\paragraph{The Higgs field.}
This is the first case where the effects of the new scalar $\sigma$
contribute. Following the same steps as for the gauge fields, we
combine Eqs.~\eqref{eq:higg_self_energy} and
\eqref{eq:higg_self_energy_sigma_part} with the counterterm from
Eq.~\eqref{eq:field_renormalisation} to get
\begin{equation}
\big(\phi^{\dagger}\phi\big)_\td =\frac{\big(\phi^{\dagger}\phi\big)_\fd}{T}\bigg[1-\frac3{4(4\pi)^2}(3g^2+g'^2)L_b+\frac{g_{Y,1}^2}{(4\pi)^2}L_f-\frac{\mu_m^2}{4}\tilde{J}_\Phi\bigg],
\end{equation}
where we have defined
\begin{align}
\tilde{J}_\Phi&\equiv\frac{4}{d}\tilde{J}^{4b}_{3/1,0,1}(\mu_\sigma)-\tilde{J}^{4b}_{2/1}(\mu_\sigma) = \frac{4}{3}\tilde{J}^{4b}_{3/1,0,1}(\mu_\sigma)-\tilde{J}^{4b}_{2/1}(\mu_\sigma), \\
g_{Y,1}^2&\equiv \tr\bigl[h^{(e)}h^{(e)\dagger}+3 h^{(u)}h^{(u)^\dagger} + 3 h^{(d)}h^{(d)\dagger}\bigr] \approx 3 g_{Y}^2.
\end{align}
Note that $\tilde J_\Phi$ is finite and can be equivalently expressed as
\begin{equation}
\tilde{J}_\Phi = -\frac{1}{32 \pi^2 \ms^2 } - \frac{T^2}{12 \ms^4} + \frac{2\pi^2 T^4}{45 \ms^6} + \frac{J_1(\ms)}{\ms^4} + \frac{J_2(\ms)}{\ms^2} - \frac{4}{3}\biggl[ \frac{ J_{1,0,1}(\ms)}{\ms^6} + \frac{ J_{2,0,1}(\ms)}{\ms^4} + \frac{ J_{3,0,1}(\ms)}{\ms^2} \biggr].
\end{equation}

\subsubsection{Coupling constants}

The couplings of the effective theory are obtained by matching the
correlators computed in the three-dimensional and the four-dimensional
theory. The calculation involves the correlators evaluated in
Section~\ref{sec:DR_correlators} as well the wavefunction
renormalization and coupling counterterms listed in
Section~\ref{sec:counterterms_beta_functions}. Note that the
correlators computed in Section~\ref{sec:DR_correlators} do not
include the effects of wavefunction renormalization, and hence are to
be treated as correlators of the \emph{bare} four-dimensional fields.

\paragraph{The gauge couplings $g_3$, $g_3'$.}
Let us illustrate the procedure by considering the $\gr{SU(2)}_L$
coupling $g_3$. We focus on the correlator with two Higgs legs and two
spatial gauge field legs. Putting together the tree-level vertices
with the results of Eqs.~\eqref{eq:phiphiAA_rs},
\eqref{eq:phiphiAA_sigma1} and \eqref{eq:phiphiAA_sigma2}, equating
the correlators in the three- and four-dimensional theories amounts to
setting
\begin{align}
\notag
&\phi^{\dag i}_\td\phi^j_\td A^a_{\td,r}A^b_{\td,s}\left(-\frac12g_3^2\delta_{ij}\delta_{ab}\delta_{rs}\right)\\
&=\frac1T\phi^{\dag i}_{\fd(b)}\phi^j_{\fd(b)} A^a_{\fd,r(b)}A^b_{\fd,s(b)}\delta_{ij}\delta_{ab}\delta_{rs}\\
\notag
&\times\biggl\{-\frac12(g^2+\delta g^2)+\biggl[-\frac38g^4+\frac38g^2g'^2-\frac12\bigl(2^{4-d}-1\bigr)g^2g_{Y,1}^2\biggr]I^{4b}_2
+\mu_m^2g^2\bigl[\tilde{J}^{(1)}_{\Phi A} + \tilde{J}^{(3)}_{\Phi A}\bigr]\biggr\},
\end{align}
where we have defined
\begin{align}
\tilde{J}^{(1)}_{\Phi A} &\equiv -\frac{1}{8}\bigg[\tilde{J}^{4b}_{1/2}(\mu_\sigma)+\frac{2}{\mu_\sigma^2}I^{4b}_2\bigg], \\
\tilde{J}^{(2)}_{\Phi A}&\equiv \frac{1}{2}\bigg[\tilde{J}^{4b}_{1/3,1,0}(\mu_\sigma)+\frac{2}{\mu_\sigma^2}\left(1-\frac{d}{4}\right) I^{4b}_{2}\bigg], \\
\tilde{J}^{(3)}_{\Phi A}&\equiv \frac{1}{2d}\bigg[\tilde{J}^{4b}_{1/3,0,1}(\mu_\sigma)+\frac{2}{\mu_\sigma^2} \frac{d}{4} I^{4b}_{2}\bigg].
\end{align}
Note that the combination $\tilde{J}^{(1)}_{\Phi A} +
\tilde{J}^{(2)}_{\Phi A}$ entering the above matching relation is
ultraviolet finite. It is now a matter of simple algebra to put
together the definitions of the bare fields in terms of the
wavefunction renormalization factors, the coupling counterterm and the
above-derived expressions for relations between the renormalized
three-dimensional and four-dimensional fields. One arrives then at the
final result for the three-dimensional coupling,
\begin{equation}
g_3^2= g^2(\Lambda)T \bigg\{1 +\frac{g^2}{(4\pi)^2}\bigg(\frac{43}{6}L_b+\frac{2}{3}-\frac{4N_f}{3}L_f\bigg)+\frac{\mu_m^2}{4}\tilde{J}_\Phi- 2\mu_m^2 \bigl[\tilde{J}^{(1)}_{\Phi A} + \tilde{J}^{(3)}_{\Phi A}\bigr] \bigg\}. 
\end{equation}
Here we have indicated explicitly the dependence of the
four-dimensional coupling $g$ on the renormalization scale
$\Lambda$. However, it is easy confirm with the
renormalization flow equation~\eqref{running_g} that $g_3$ is
independent of $\Lambda$. The same comment applies to all the other
three-dimensional couplings discussed below. Again, renormalization
group independence represents a nontrivial check of the correctness of
our calculation.

The evaluation of the coupling $g_3'$ proceeds along the same
lines, and we therefore just quote the final result,
\begin{equation}
g'^2_3=g'^2(\Lambda)T \bigg\{1 +\frac{g'^2}{(4\pi)^2}\bigg(-\frac{1}{6}L_b -  \frac{20N_f}{9}L_f\bigg)+\frac{\mu_m^2}{4}\tilde{J}_\Phi- 2\mu_m^2\bigl[\tilde{J}^{(1)}_{\Phi A} + \tilde{J}^{(3)}_{\Phi A}\bigr] \bigg\}.
\end{equation}
An interested reader can easily check this expression themselves, using Eqs.~\eqref{eq:phiphiBB_2} and \eqref{eq:phiphiBB_3}. We note that both $g_3$ and $g_3'$ depend on the same combination of massive master integrals, which can be seen to vanish by an explicit manipulation,
\begin{equation}
\frac{1}{4}\tilde{J}_\Phi - 2\bigl[\tilde{J}^{(1)}_{\Phi A} + \tilde{J}^{(3)}_{\Phi A}\bigr] =\frac{1}{\mu^2_\sigma} \bigg\{ \frac{J_2(\ms)}{4} - \frac{J_{3,0,1}(\ms)}{3} + \frac{1}{\ms^2} \bigg[\frac{J_1(\ms)}{2} - \frac{J_{2,0,1}(\ms)}{3} \bigg] \bigg\}=0.
\end{equation}
This is to be expected; we could have instead
determined $g_3$ and $g_3'$ using four-point correlators of the gauge
fields, to which $\sigma$ does not contribute at the one-loop level.

\paragraph{The temporal gauge field self-couplings $\lambda_3$, $\lambda_3'$, $\lambda_3''$.}
These couplings are forbidden by gauge invariance in the
four-dimensional theory. They are generated at nonzero temperature by
loop effects, and hence all appear at order $g^4$. Since we do not
require higher precision in our setup, wavefunction renormalization
does not contribute. Hence all the couplings can be straightforwardly
obtained from the correlators of
Eqs.~(\ref{eq:adjoint_cor_1}-\ref{eq:adjoint_cor_3}),
\begin{align}
\lambda_3 &= T \frac{g^4}{16 \pi^2} \frac{17 - 4 N_f}{3}, \\ 
\lambda_3' &= T \frac{g'^4}{16 \pi^2}\bigg[ \frac{1}{3} - \frac{N_f}6 \big(3 Y^4_d + Y^4_e + 2 Y^4_\ell + 6 Y^4_q + 3 Y^4_u \big) \big] \notag \\
&=  T \frac{g'^4}{16 \pi^2} \Big(\frac{1}{3} - \frac{380}{81} N_f \Big), \\
\lambda_3''& = T \frac{g^2g'^2}{16 \pi^2}\big[2 - 2N_f \big(Y^2_\ell + 3 Y^2_q \big) \big] = T \frac{g^2g'^2}{16 \pi^2} \bigg( 2 - \frac{8}{3} N_f \bigg).
\end{align}

\paragraph{The Higgs-gauge field couplings $h_3$, $h_3'$, $h_3'', \delta_3$.} These couplings are extracted from the correlators with two Higgs legs and two temporal gauge field legs. Putting together the results of Eqs.~\eqref{eq:phiphiAA_00}, \eqref{eq:phiphiAA_sigma1} and \eqref{eq:phiphiAA_sigma2}, equating the three-dimensional and the four-dimensional correlators amounts to setting
\begin{align}
\notag
&\phi^{\dag i}_\td\phi^j_\td A^a_{\td,0}A^b_{\td,0}\left(-2h_3\delta_{ij}\delta_{ab}\right)\\
&=\frac1T\phi^{\dag i}_{\fd(b)}\phi^j_{\fd(b)} A^a_{\fd,0(b)}A^b_{\fd,0(b)}\delta_{ij}\delta_{ab}\biggl\{-\frac12(g^2+\delta g^2)+\mu_m^2g^2\bigl[\tilde{J}^{(1)}_{\Phi A} + \tilde{J}^{(2)}_{\Phi A}\bigr]\\
\notag
&+\biggl[d\left(d-\frac{25}8\right)g^4+\frac d8g^2g'^2+3(d-3)\lambda_hg^2+\frac12\bigl(2^{4-d}-1\bigr)(2-d)g^2g_{Y,1}^2\biggr]I^{4b}_2\biggr\}.
\end{align}
A straightforward calculation then leads to
\begin{equation}
\begin{split}
h_3={}& \frac{g^2(\Lambda)T}{4} \Bigg(1 +\frac{1}{(4\pi)^2}\bigg\{\bigg[\frac{43}{6}L_b+\frac{17}{2}-\frac{4N_f}{3}(L_f-1)\bigg]g^2+\frac{g'^2}{2}-2g_{Y,1}^2+12\lambda_h\bigg\} \\
&+\frac{\mu_m^2}{4}\tilde{J}_\Phi - 2 \mu_m^2 \bigl[\tilde{J}^{(1)}_{\Phi A} + \tilde{J}^{(2)}_{\Phi A}\bigr] \Bigg).
\end{split}
\end{equation}
The coupling $h_3'$ is obtained from the correlator with two external
$B_0$ legs, and its evaluation proceeds in exactly the same
fashion. Using Eqs.~\eqref{eq:phiphiBB_1} and \eqref{eq:phiphiBB_3}
yields
\begin{align}
\notag
h'_3={}& \frac{g'^2(\Lambda)T}{4} \Bigg(1 +\frac{1}{(4\pi)^2}\bigg\{\frac{3g^2}{2}+\bigg[ -\frac{1}{6}(L_b-1) - \frac{20N_f}{9}(L_f-1)\bigg]g'^2+ G_{Y,1}^2 +12\lambda_h\bigg\}\\
&+\frac{\mu_m^2}{4}\tilde{J}_\Phi - 2\mu_m^2 \bigl[\tilde{J}^{(1)}_{\Phi A} + \tilde{J}^{(2)}_{\Phi A}\bigr] \Bigg),
\end{align}
where
\begin{equation}
G^2_{Y,1} \equiv \frac{2}{3}\tr\bigl[15 h^{(e)}h^{(e)\dagger}+ 17 h^{(u)}h^{(u)\dagger} + 5 h^{(d)}h^{(d)\dagger}\bigr].
\end{equation}
Finally, the $h_3''$ coupling is obtained from the correlator with one $A^a_0$ and $B_0$ external leg,
\begin{align}
h''_3={}& \frac{g'g T}{2} \bigg\{ 1 +  \frac{1}{(4 \pi)^2} \bigg[ -g^2 + \frac{1}{3}g'^2 + L_b \bigg(\frac{43}{12}g^2  - \frac{1}{12}g'^2 \bigg) - N_f (L_f-1) \bigg(\frac{2}{3}g^2 + \frac{10}{9}g'^2 \bigg) \nonumber \\
&+ 4 \lambda_h + G_{Y,2}^2 \bigg] + \mu^2_m \bigg( \frac{1}{4} \tilde{J}_\Phi  -2 \tilde{J}_{\Phi A B} \bigg)  \bigg\},
\end{align}
where we have defined
\begin{align}
\tilde{J}_{\Phi A B}& \equiv -\frac{1}{8} \tilde{J}^{4b}_{1/2} + \frac{1}{2} \tilde{J}^{4b}_{1/3,1,0},\\
G_{Y,2}^2 &\equiv 2 \tr [ h^{(e)}h^{(e)\dagger} + h^{(u)}h^{(u)\dagger} - h^{(d)}h^{(d)\dagger}].
\end{align}
Note that this coupling has a different sign compared to that of
Ref.~\cite{Kajantie:1995dw} due to our different convention for the
covariant derivative of the $B_\mu$-field.

The combinations of massive master integrals that enter the above
expressions for $h_3$, $h_3'$ and $h_3''$ can be further simplified,
as was the case for the $g_3$ and $g_3'$ couplings. A short
manipulation shows that
\begin{equation}
\begin{split}
\frac{1}{4}\tilde{J}_\Phi - 2\bigl[\tilde{J}^{(1)}_{\Phi A} + \tilde{J}^{(2)}_{\Phi A}\bigr] &=H(\mu_\sigma),\\
\frac{1}{4}\tilde{J}_\Phi - 2 \tilde{J}_{\Phi A B}  &= H(\mu_\sigma) - \frac{1}{16\pi^2 \mu^2_\sigma},
\end{split}
\end{equation}
where the integral $H(\mu_\sigma)$ is defined by Eq.~(\ref{Hintegral}).

Finally, the coupling $\delta_3$ is obtained from the correlator with two external $C^\alpha_0$ legs, Eq.~(\ref{eq:adjoint_gluon_cor})
\begin{align}
\delta_3= -\frac{2 g^2_s G^2_{Y,3} T}{(4\pi)^2},
\end{align}
where we defined
\begin{equation}
G_{Y,3}^2 \equiv \tr [h^{(u)}h^{(u)\dagger} + h^{(d)}h^{(d)\dagger}].
\end{equation}

\paragraph{The scalar couplings $\mu_{h,3}$, $\lambda_{h,3}$.}
The mass parameter of the Higgs doublet is assumed to be heavy, and
one-loop corrections contribute to it at the same order. At this
order, $g^2$, wavefunction renormalization of the Higgs field does not
play a role and the squared mass parameter in the three-dimensional
theory can be extracted directly from Eq.~(\ref{eq:phi2_cor}), to which
we add the counterterm contributions
$\delta\mu_h^2+\delta\mu_1\mu_m/(2\mu^2_\sigma)$, the result being
\begin{align} \mu^2_{h,3}={}&\mu^2_h(\Lambda) +
    \frac{\mu_m(\Lambda) \mu_1(\Lambda)}{2\mu^2_\sigma(\Lambda)}
    -T^2\bigg(\frac{3}{16}g^2+\frac{1}{16}g'^2+\frac{\lambda_h}{2}+\frac{1}{12}\sum_i
    h^2_i\bigg) \nonumber \\ &+\frac{1}{64
      \pi^2}(2\ms^2\lambda_m-2\mu_3\mu_m+\mu^2_m)\left[1+\log\left(\frac{\Lambda^2}{\ms^2}\right)\right]+\frac{T^2}{16}\frac{\mu^2_m}{\ms^2}
    \nonumber\\ &+J_1(\ms)\bigg(-\frac{\lambda_m}{2}+\frac{\mu_3\mu_m}{2
      \ms^2}-\frac{\mu^2_m}{4 \ms^2}\bigg).
\end{align}
The scalar self-coupling $\lambda_{h,3}$ can be obtained in a similar
fashion from Eq.~(\ref{eq:phi4_cor}). This results in a lengthy
expressions that is displayed in full in the overview of the matching
relations below.

\subsubsection{Collected matching relations at one-loop order}
\label{sec:matching_1loop}

For the reader's convenience, we collect here all one-loop results for
the parameters of the effective theory for the heavy scale. The mass
parameters are given to order $g^2$, while the couplings are given to
order $g^4$. 
\begin{align}
m_D^2={}&g^2T^2\bigg(\frac{5}{6}+\frac{N_f}{3}\bigg),\\
m'^2_D={}&g'^2T^2\bigg(\frac{1}{6}+\frac{5N_f}{9}\bigg),\\
m''^2_D ={}&g^2_s T^2 \bigg(1+\frac{N_f}{6}\bigg),\\
g_3^2={}&g^2(\Lambda)T\bigg[1 +\frac{g^2}{(4\pi)^2}\bigg(\frac{43}{6}L_b+\frac{2}{3}-\frac{4N_f}{3}L_f\bigg)\bigg],\\
g'^2_3={}&g'^2(\Lambda)T\bigg[1 +\frac{g'^2}{(4\pi)^2}\bigg(-\frac{1}{6}L_b-\frac{20N_f}{9}L_f\bigg)\bigg],\\
\lambda_3={}&T\frac{g^4}{16 \pi^2} \frac{17-4N_f}{3},\\
\lambda_3'={}&T\frac{g'^4}{16\pi^2} \bigg(\frac{1}{3}-\frac{380}{81} N_f\bigg),\\
\lambda_3''={}&T\frac{g^2g'^2}{16\pi^2}\bigg(2-\frac{8}{3}N_f\bigg),\\
\notag
h_3={}&\frac{g^2(\Lambda)T}{4}\bigg(1+\frac{1}{(4\pi)^2}\bigg\{\bigg[\frac{43}{6}L_b+\frac{17}{2}-\frac{4N_f}{3}(L_f-1)\bigg]g^2+\frac{g'^2}{2}-2g_{Y,1}^2+12\lambda_h\bigg\}\\
&+\mu^2_m H(\ms)\bigg),\\
\notag
h'_3={}&\frac{g'^2(\Lambda)T}{4}\bigg(1 +\frac{1}{(4\pi)^2}\bigg\{\frac{3g^2}{2}+\bigg[-\frac{1}{6}(L_b-1)  -\frac{20N_f}{9}(L_f-1)\bigg]g'^2-G_{Y,1}^2+12\lambda_h\bigg\}\\
&+\mu^2_m H(\ms)\bigg),\\
\notag
h''_3={}&\frac{g(\Lambda)g'(\Lambda)T}{2}\bigg\{1+\frac{1}{(4\pi)^2}\bigg[-g^2+ \frac{1}{3}g'^2+L_b\bigg(\frac{43}{12}g^2 -\frac{1}{12}g'^2\bigg)\\
&-N_f(L_f-1)\bigg(\frac{2}{3}g^2+\frac{10}{9}g'^2\bigg)+4\lambda_h+G_{Y,2}^2\bigg]+\mu^2_m\bigg[H(\ms)-\frac{1}{16\pi^2\ms^2}\bigg]\bigg\},\\
\notag
\delta_3={}& -\frac{2 g^2_s G^2_{Y,3} T}{(4\pi)^2},\\ \notag
\mu^2_{h,3}={}&\mu^2_h(\Lambda) + \frac{\mu_m(\Lambda) \mu_1(\Lambda)}{2\mu^2_\sigma(\Lambda)} -T^2\bigg(\frac{3}{16}g^2+\frac{1}{16}g'^2+\frac{\lambda_h}{2}+\frac{1}{12}\sum_i h^2_i\bigg)\\
\notag
&+\frac{1}{64 \pi^2}(2\ms^2\lambda_m-2\mu_3\mu_m+\mu^2_m)\left[1+\log\left(\frac{\Lambda^2}{\ms^2}\right)\right]\\
&+\frac{T^2}{16}\frac{\mu^2_m}{\ms^2}+J_1(\ms)\bigg(-\frac{\lambda_m}{2}+\frac{\mu_3\mu_m}{2 \ms^2}-\frac{\mu^2_m}{4 \ms^2}\bigg),\\
\notag
\lambda_{h,3}={}&T\Bigg\{\lambda_h(\Lambda)-\frac{1}{8}\frac{\mu^2_m(\Lambda)}{\mu^2_\sigma(\Lambda)}  - \frac{1}{4} \frac{\mu^2_m \mu_3 \mu_1(\Lambda)}{\mu^6_\sigma} + \frac{1}{2} \frac{\mu_m \lambda_m \mu_1(\Lambda)}{\mu^4_\sigma}  \\
\notag 
&+\frac{1}{16 \pi^2} \frac{1}{16} \bigg\{ 6 g^4 + 4 g^2 g'^2 + 2 g'^4 + L_f \Big( 16 \sum_i h^4_i - 32 \lambda_h g_{Y,1}^2 \Big) \\
\notag
&-3 L_b \Big[ 3 g^4 + g'^4 + 2 g'^2 (g^2 - 4 \lambda_h) + 8 \lambda_h (-3g^2 + 8 \lambda_h) \Big] \bigg\} \\
\notag
&-\frac{1}{16\pi^2} \frac{\mu^2_m}{\mu^2_\sigma}  \frac{1}{2} \bigg[ L_b \bigg( \frac{9}{8}g^2 + \frac{3}{8}g'^2 \bigg) - \frac{1}{2} L_f g_{Y,1}^2 \bigg] + \bigg( \frac{1}{2} \lambda_h \mu^2_m - \frac{1}{16} \frac{\mu^4_m}{\mu^2_\sigma} \bigg) \bigg[H(\ms)+\frac1{16\pi^2\ms^2}\bigg]\\
\notag
&-\frac{1}{2 \ms^4} \Bigg[ \frac{1}{128 \pi^2} \bigg( \ms^2 \Big[ \mu _m^2 \left(24 \lambda _h+6 \lambda _{\sigma }-12 \lambda _m\right)+8 \mu _3 \lambda _m \mu _m \Big ]+\mu _m^2 (-4 \mu _3^2+4\mu _3 \mu_m-3 \mu _m^2)\\
\notag
&+L_b \mu _m^2 (3 \mu _m^2-48 \ms^2 \lambda _h ) + \log \bigg (\frac{\Lambda ^2}{\ms^2}\bigg) \Big\{\mu _m^2 \left[24 \ms^2 \lambda _h+2 \ms^2 \left(3 \lambda _{\sigma }-4 \lambda _m\right)-2 \mu _m^2\right]+4 \ms^4 \lambda _m^2 \Big\} \bigg)\\
\notag
&+T^2 \mu _m^2 \left(\frac{\lambda _h }{4}-\frac{\mu_m^2}{32 \ms^2}+\frac{\mu _3 \mu _m}{12 \ms^2}-\frac{\lambda _m }{4}\right)\\ 
\notag
&+J_1(\ms) \left[\mu _m^2 \left(-3 \lambda _h +\frac{3 \mu _m^2}{8 \ms^2}-\frac{\mu _3 \mu _m}{2 \ms^2}+\frac{\mu _3^2 }{2  \ms^2}-\frac{3 \lambda_{\sigma } }{4}+\frac{3 \lambda _m }{2}\right) - \mu _3 \lambda _m \mu _m \right]\\
&+J_2(\ms) \left[ \mu _m^2\left(\frac{\mu _m^2}{8}-\frac{\mu _3 \mu _m}{2}+\frac{\mu _3^2 }{2}+\frac{ \ms^2 \lambda _m }{2}\right)- \ms^2 \mu _3 \lambda _m \mu _m + \ms^4\frac{\lambda _m^2}{2}\right] \Bigg] \Bigg\}.
\end{align}
In these expressions, we have used the following notation
  introduced earlier in this section:
\begin{align}
L_b&\equiv2\ln\Big(\frac{\Lambda}{T}\Big)-2[\ln(4\pi)-\gamma], \\ 
L_f&\equiv L_b+4\ln2,\\
g_{Y,1}^2&\equiv\tr\bigl[h^{(e)}h^{(e)\dagger}+3h^{(u)}h^{(u)\dagger}+3h^{(d)}h^{(d)\dagger}\bigr]\approx3g_{Y}^2,\\
G^2_{Y,1}&\equiv\frac{2}{3}\tr\bigl[15h^{(e)}h^{(e)\dagger}+17h^{(u)}h^{(u)\dagger}+5h^{(d)}h^{(d)\dagger}\bigr]\approx\frac{34}3g_Y^2,\\
G_{Y,2}^2&\equiv2\tr\big[h^{(e)}h^{(e)\dagger}+h^{(u)}h^{(u)\dagger}-h^{(d)}h^{(d)\dagger}\big]\approx2g_Y^2,\\
G_{Y,3}^2&\equiv \tr\big[h^{(u)}h^{(u)\dagger}+h^{(d)}h^{(d)\dagger}\big]\approx g_Y^2.
\end{align}
The given approximate values apply when only the top quark Yukawa
coupling $g_Y$ is nonzero. We also have $\sum_i h_i^n\approx 3g_Y^n$,
since here the sum runs over the spectrum of Gram matrices of the
Yukawa couplings, including the three-fold degeneracy due to different
colors,
\begin{equation}
h_i\in\text{spectrum}\Bigl(\sqrt{h^{(e)}h^{(e)\dagger}},\sqrt{h^{(u)}h^{(u)\dagger}},\sqrt{h^{(d)}h^{(d)\dagger}}\Bigr).
\end{equation}
The massive master integrals $J_1(m),J_2(m)$ and $H(m)$ are defined in
Appendix~\ref{sec:master_integrals}.

\subsection{Integration over the heavy scale}
\label{sec:int_heavyscale}

The effective theory for the heavy scale is already identical to the
SM case, only differing by the contributions of the new scalar
$\sigma$ to the effective couplings. The next step, in which
  heavy degrees of freedom are integrated out to leave an effective
  theory for the light scale alone, therefore goes through without
any modification. In particular, the matching conditions that relate
the couplings of the effective theories for the heavy and light
scales can be taken from Ref.~\cite{Kajantie:1995dw}. However,  in addition we include here the leading contribution of the temporal gluons to the scalar mass parameter in the last term in Eq.~(\ref{eq:bar_mu_h3}).  We
list these matching conditions here for the reader's convenience,
\begin{align}
\bar g^2_3 &= g^2_3 \bigg( 1 - \frac{g^2_3}{24\pi m_D} \bigg), \\
\bar g'^2_3 &= g'^2_3, \\
\bar\mu^2_{h,3} &= \mu^2_{h,3} + \frac{1}{4\pi}(3 h_3 m_D + h'_3 m'_D + 8\delta_3 m''_D), \label{eq:bar_mu_h3} \\
\bar{\lambda}_{h,3} &= \lambda_{h,3} - \frac{1}{8\pi}\bigg( \frac{3h^2_3}{m_D} + \frac{h'^2_3}{m'_D} + \frac{h''^2_3}{m_D+m'_D} \bigg).
\end{align}

\subsection{Relations to physical parameters}
\label{sec:rel_physical}

In this section, we have derived expressions for the
parameters of the three-dimensional effective theories of the
SSM in terms of the running \MSbar{} parameters of the
original four-dimensional theory. The ultimate aim is to
translate the behavior of the effective theory into physical
insights concerning, amongst other things, the order of the
electroweak phase transition in the full theory. To do
  this, however, we need to express the \MSbar{} parameters in terms
of measurable quantities such as pole masses and the Fermi
constant. In this article, we have worked only up to one-loop order in
the scalar mass parameters ($\sim g^2$) so it suffices to
perform this translation at tree level. However, if our  matching results are eventually generalised to two-loop order,
providing $g^4$ accuracy, then one would need the \MSbar{}
parameters to be related to the physical ones at the same $g^4$ order,
requiring a one-loop renormalization of the
theory~\cite{Kajantie:1995dw}. This rather tedious exercise is left to
a forthcoming paper.

For the gauge couplings, we use the Standard Model results of
Ref.~\cite{Kajantie:1995dw},
\begin{equation}
\begin{split}
g^2 &= g^2_0,\\
g'^2 &= \frac{g^2_0}{m^2_W}(m^2_Z - m^2_W),
\end{split}
\end{equation}
where we have denoted $g^2_0 \equiv 4\sqrt{2} G_f m^2_W$, with $G_f$ being the Fermi constant. By inverting the mass eigenvalues (cf.~Section~\ref{sec:effective_potential}) for both of the physical scalars, the W boson and the top quark, one can on the other hand obtain the desired tree-level relations for the corresponding parameters. An important simplification can be achieved by fixing the parameter $\mu_1$ as given in Eq.~(\ref{eq:fix_mu_1}) such that singlet Vacuum Expectation Value vanishes, $\rho=0$, while the doublet VEV is the same as in the SM, i.e.~$\nu=\mu_h/\sqrt{\lambda_h}$. As a result, we obtain
\begin{align}
\mu^2_h &= \frac{1}{4} \left[m_-^2+m_+^2 \pm \frac{\sqrt{g^4_0 m_W^4 \left(m_-^2-m_+^2\right)^2-4 g^2_0 \mu _m^2 m_W^6}}{g^2_0  m_W^2}\right] ,\\
\lambda_h &=\frac{g^2_0 m_W^2 \left(m_-^2+m_+^2\right) \pm \sqrt{g^4_0 m_W^4 \left(m_-^2-m_+^2\right)^2-4 g^2_0 \mu _m^2 m_W^6}}{16 m_W^4}  ,\\
\mu^2_\sigma &= \frac{1}{2} \left[m_-^2+m_+^2-\frac{4 \lambda _m m_W^4 \pm \sqrt{g^4_0 m_W^4 \left(m_-^2-m_+^2\right)^2-4 g^2_0 \mu _m^2 m_W^6}}{g^2_0 m_W^2}\right],
\end{align}
where one must consistently take the same sign in all three
equations. Identifying the Higgs mass with $m_-$ requires us to take
the positive sign for each.

It is easy to see that in the decoupling limit where the portal couplings vanish, the above relations reduce to those of the SM. Here, it would have been possible to  eliminate one of the portal couplings in favor of the mixing angle of the two physical scalars, but for practical reasons we have kept both portal couplings as input parameters. Furthermore, the Yukawa coupling obeys the relation
\begin{equation}
g^2_Y = \frac{g^2_0}{2}\frac{m^2_t}{m^2_W},
\end{equation}
which is same as in the SM. 

To obtain the \MSbar{} and effective theory parameters as
functions of the renormalization scale, the above relations are used
as initial conditions at the scale $\Lambda=m_Z$. We emphasize that
fixing $\mu_1$ in terms of $\mu_m$ and the doublet VEV at the initial
scale -- such that $\rho=0$ there -- does not make this
  parameter vanish, in general. However, by solving $\rho$ and $\nu$
in terms of the coupling constants by requiring that they minimize the
tree-level scalar potential, and allowing these expressions to run
with the renormalization scale, the changes in the VEVs remain
numerically small.


\section{Discussion}
\label{sec:discussion}

In the present work, we have performed a high-temperature dimensional
reduction of the Standard Model augmented by a singlet scalar field
coupled in the most general way to the Higgs field. For our
purposes, the singlet is treated as a superheavy degree of
freedom and integrated out of the theory altogether; the only
light fields remaining in the 3D theory correspond to the Higgs, SU(2)
and U(1) zero modes. As a consequence, the presence of the singlet in
the 4D theory appears through the enlarged RG-system of couplings
(Eq. \ref{running_g}-\ref{running_end}); through the multiple
occurrences of the non-SM couplings $\mu_\sigma^2$, $\mu_m$, $\mu_1$,
$\mu_3$, $\lambda_m$ and $\lambda_\sigma$ in the matching relations in
Section~\ref{sec:matching_1loop} (and after integrating
  out the heavy scale in Section \ref{sec:int_heavyscale});
and through the matching to physical parameters as described in
Section \ref{sec:rel_physical}. The SM limit (taking
$\lambda_m$ and $\mu_m$ to zero) stands out clearly in the expressions
of Section \ref{sec:matching_1loop}, and we note that the
singlet addition is a highly non-trivial generalisation of these
expressions.

\begin{figure}
\begin{center}
\includegraphics[width=12cm]{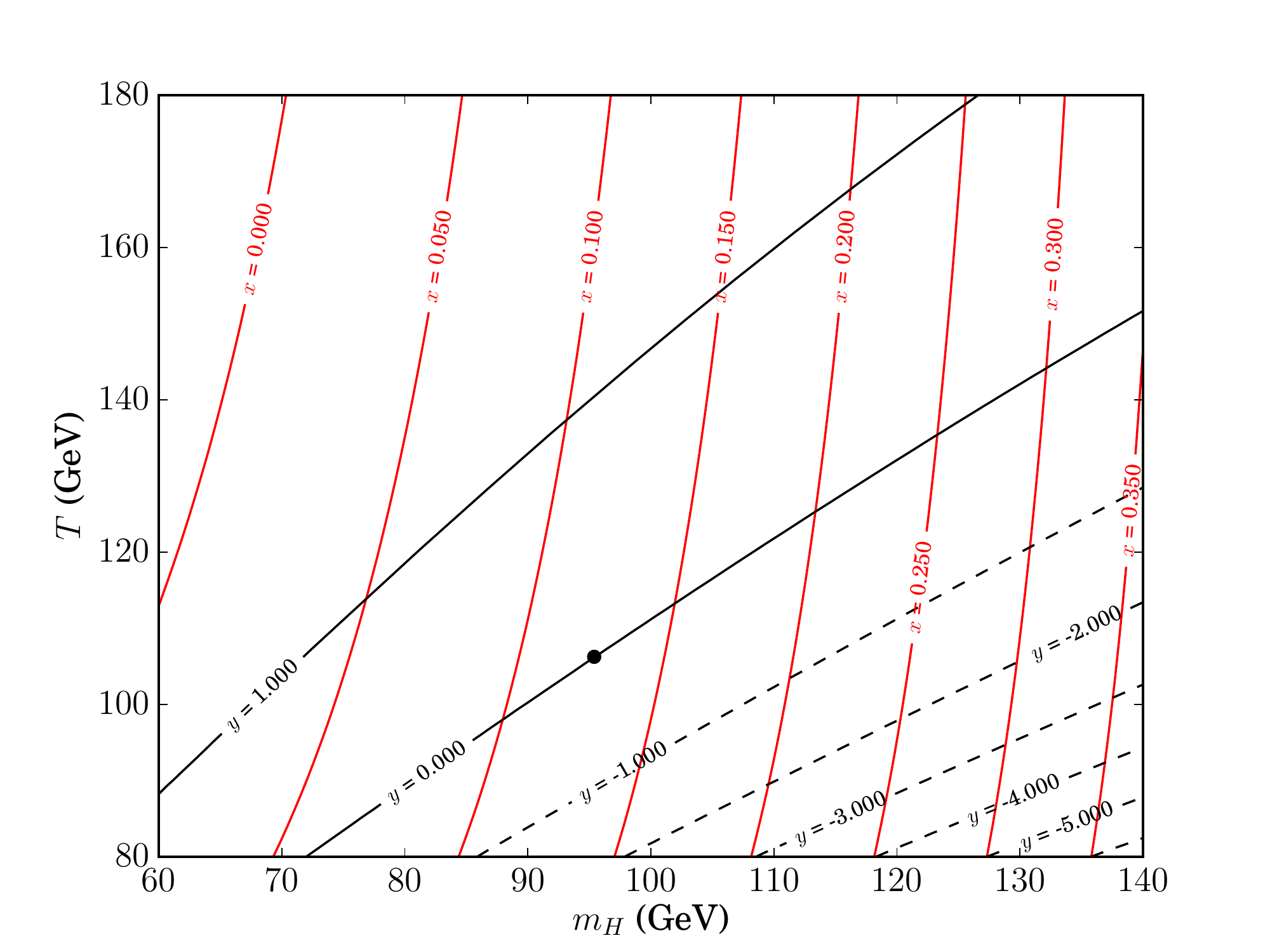}
\end{center}
\caption{The $m_H$-$T$ plane in the Standard Model. Overlaid, curves
  of constant $x$ and $y$ as defined in the main text. The black dot
  denotes the critical point in the 3D theory. This can be
    compared with Fig.~8 of Ref.~\cite{Kajantie:1995dw}, but note that
    we have chosen a different set of approximations in carrying out
    the dimensional reduction, as explained in the main text.}
\label{fig:SMxy}
\end{figure}

We match to the exact same 3D theory as in the seminal
papers~\cite{Kajantie:1995kf,Kajantie:1996mn},
%
where the nonperturbative lattice simulations are phrased in terms
  of the dimensionless combinations
\begin{eqnarray}
\frac{\bar{g}_3^2}{T},\qquad x = \frac{\bar{\lambda}_3}{\bar{g}_3^2},\qquad y = \frac{\bar{m}_3^2}{\bar{g}_3^4}.
\end{eqnarray}
It turns out that $\bar{g}_3^2$ varies very little for the parameter
range considered, and one is left with finding the position of the
phase transition in $x$-$y$-space. This computation involves only the
3D theory, and the result applies to any 4D theory that is matched
to it. As one might expect, the phase transition happens near $y=0$,
where the mass parameter $\bar{m}_3^2$ changes sign. The central
result of \cite{Kajantie:1996mn,Kajantie:1996qd} is that there is a
line $0<x<x_c$, $y\simeq 0$, where the phase transition is first
order. This line ends at a critical point $x_c\simeq 0.125$, beyond
which the transition is a crossover.

In Fig.~\ref{fig:SMxy} we show the Higgs mass-temperature ($T-m_H$)
plane for the Standard Model. Overlaid are curves of constant
$x$ and $y$ as defined by the matching relations. We have marked the
point $(x_c,y=0)$ with a black dot, and we see that it corresponds to
a value of $m_H$ much below the measured value of $\simeq 125$ GeV. The Fig.~\ref{fig:SMxy} can be compared with Fig.~8 of Ref.~\cite{Kajantie:1995dw}. The difference between these two is due to a different set of approximations in carrying out
    the dimensional reduction: While we have included order ${g'}^4$ effect of $\gr{U(1)}_Y$ gauge field and the effect of temporal gluons, we have not included two-loop contributions to the mass parameter of Higgs, nor the one-loop relations between \MSbar{} parameters and physical quantities. Major difference comes from the omission of two-loop contributions. 

The familiar conclusion is that in the Standard Model, the transition
is a crossover. One may approximately recover the whole first
order range by following the $y=0$ line from the black dot towards
$x=0$. Nonperturbative simulations give a slight deviation
from $y=0$, but the conclusion is the same. We see that the physical
point has $x_{SM}\simeq 0.25$ at $y=0$. This is one way of quantizing
``how far'' the Standard Model is from the first order range.

\begin{figure}
\begin{center}
\includegraphics[width=12cm]{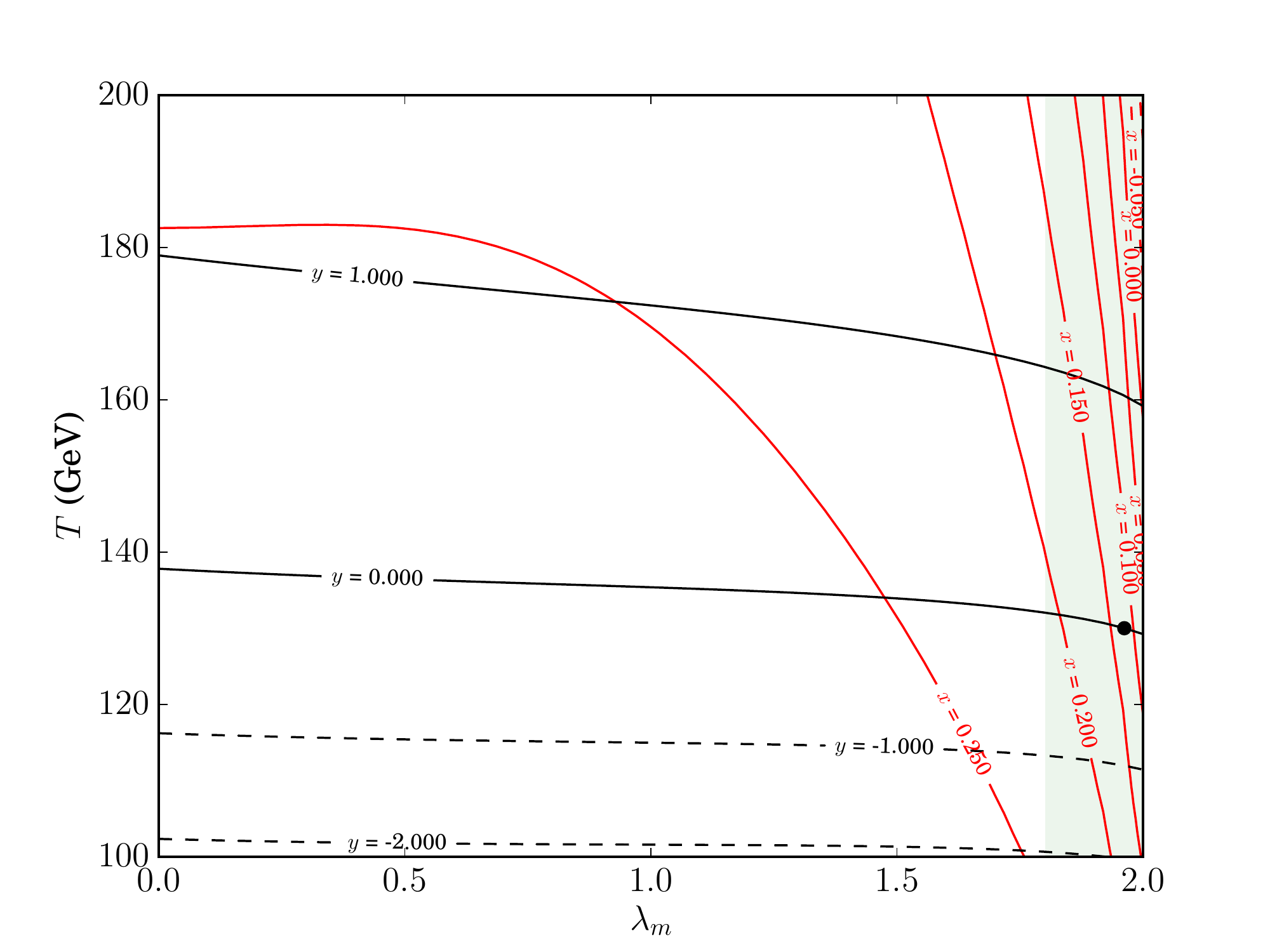}
\end{center}
\caption{The $\lambda_m$-$T$ plane in the $Z_2$-symmetric
  singlet-extended Standard Model, when $m_\sigma=250$ GeV and
  $\lambda_\sigma=1/4$. Overlaid, curves of constant $x$ and $y$ as
  defined in the main text. The black dot denotes the critical point
  in the 3D theory. In the shaded region the computation is
  unreliable due to $\mu_h>\mu_\sigma$, which violates our assumption about scale hierarchy for mass parameters. }
\label{fig:Z2xy}
\end{figure}

The object of this work is to investigate whether adding a singlet allows for a first order transition while insisting that $m_H=125$ GeV. 
The
difference that the singlet makes is that 
because the matching
relations have changed, 
for a given set of 4D parameters in
the 5-dimensional SM-singlet parameter space, varying the temperature
over the electroweak transition results in a different trajectory in
$\{ \bar{g}^2_3,
\bar{g}'^{2}_3,\bar{\mu}^2_{h,3},\bar{\lambda}^2_{h,3} \}$-space,
and, in turn, in $x$-$y$-space.  The task is therefore to
identify these trajectories and perform similar multicanonical
simulations \cite{Kajantie:1995kf,Kajantie:1996mn}. For a complete
scan of the singlet model, this is a challenge, but not
impossible. Fortunately, comprehensive scans already exist employing
perturbation theory computations of the 4D effective potential
which, together with experimental constraints, may be used to guide
non-perturbative searches
\cite{Huber:2000mg,O'Connell:2006wi,Ahriche:2007jp,Profumo:2007wc,Espinosa:2011ax,Cline:2012hg,Damgaard:2013kva,Profumo:2014opa,Kozaczuk:2015owa,Damgaard:2015con}.
A detailed numerical investigation of this theory with the matchings
presented here is underway~\cite{TBD}.

For the present, we show in Fig.~\ref{fig:Z2xy} a pencil in the
5-dimensional parameter space, where we impose $Z_2$-symmetry
($\mu_3=\mu_m=0$). Guided by existing perturbative results, we choose
$\lambda_{\sigma}=1/4$ and $m_\sigma=250$ GeV, and scan over the
remaining parameter, the quartic portal coupling $\lambda_m$. Overlaid
are again curves of constant $x$ and $y$, and we have again placed a
black dot at the point $ (x_c,y=0)$ corresponding to the critical
point. Following the $y=0$ line towards smaller $x$ gives the first
order range.

We see that the critical point requires $\lambda_m\simeq 2$. This is
rather large, which jeopardizes the validity of our perturbative
matching relations. Another, more serious issue is that the whole
first order line is located in a region where $\mu_h>\mu_\sigma$,
explicitly violating one of the assumptions of we made about how the
mass parameters scale, namely that the $\sigma$ field is
superheavy. For more details, see Section~\ref{sec:param_relations}.
Hence although the new matching relations numerically allow us to
approach the first order region, we cannot go closer than $x=0.2$ and
still trust our computation. That large couplings are necessary in the
$Z_2$-symmetric case is also true perturbatively (see for instance
\cite{Damgaard:2015con}).

\begin{figure}
\begin{center}
\includegraphics[width=12cm]{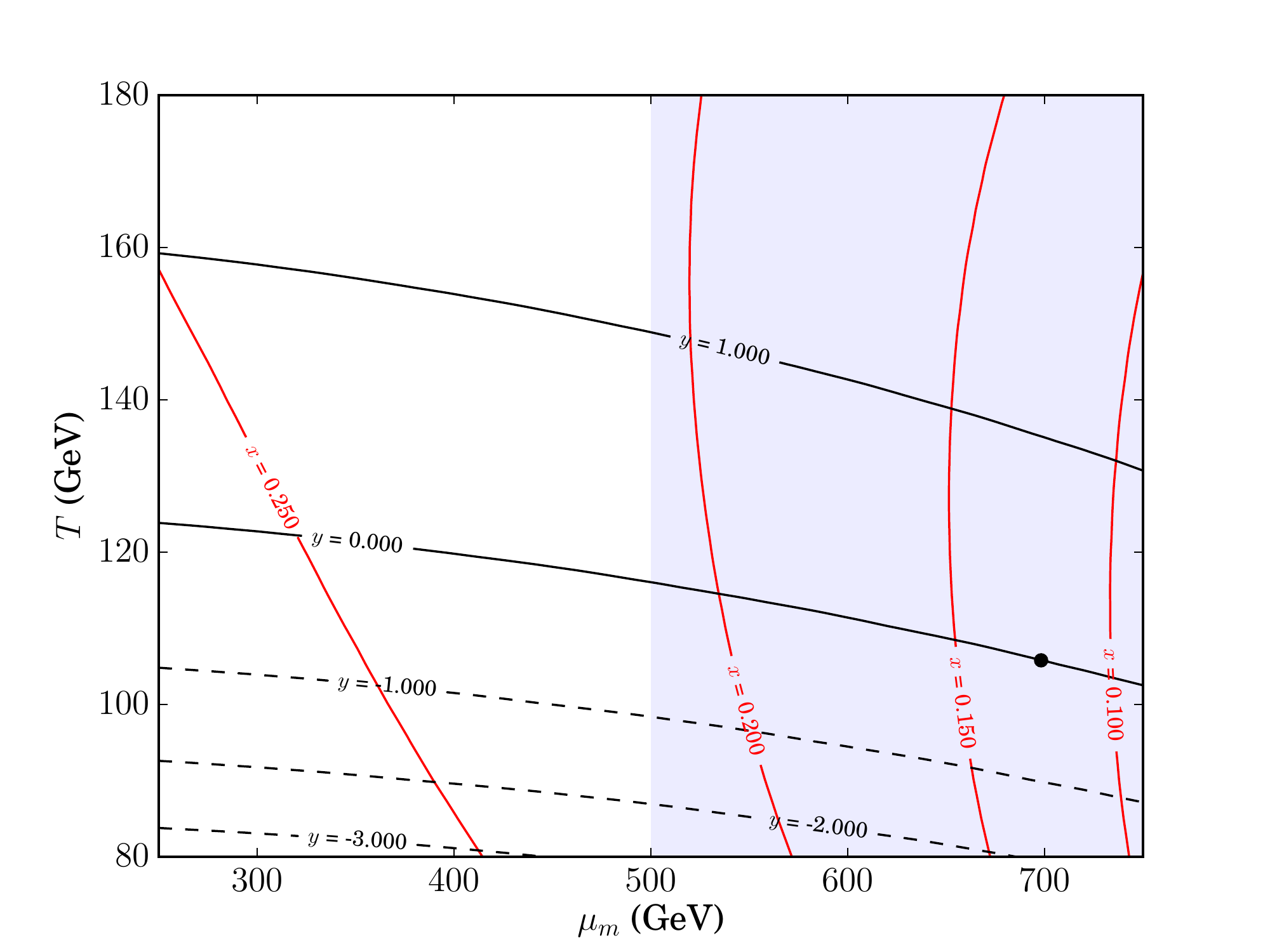}
\end{center}
\caption{The $\mu_m$-$T$ plane in the singlet-extended Standard Model, when $m_\sigma=500$ GeV, $\lambda_m=0$, $\mu_3=0$ and $\lambda_\sigma=1/4$. Overlaid, curves of constant $x$ and $y$ as defined in the main text. The black dot denotes the critical point in the 3D theory. In the shaded region the computation is
  unreliable due to $\mu_m>\mu_\sigma$, which violates our scaling assumptions. }
\label{fig:nonZ2xy}
\end{figure}

However $Z_2$-symmetry need not be imposed, and in
Fig.~\ref{fig:nonZ2xy} we show the case where $\lambda_m=0$,
$\lambda_{\sigma}=1/4$, $\mu_3=0$ and where we have chosen
$m_\sigma=500$ GeV, while varying the cubic portal coupling $\mu_m$.
We see that the critical point is within the scope of the matching
relations, but it turns out that another of our scaling
assumptions $\mu_m<\mu_\sigma$ is not fulfilled.
The closest we can go seems to again be $x\simeq 0.2$, and we have
been unable to find a parameter set that obeys all the
assumptions of Section~\ref{sec:param_relations} while
providing a first order transition. We have however not systematically
scanned the full parameter space in the present paper.

We conclude that, within the limits of our current approximation, we
cannot argue that adding a singlet provides a first order electroweak
phase transition. We do believe that at least in the
non-$Z_2$-symmetric case, effort would be well spent on improving on
this approximation with the view of confirming the picture in
Fig.~\ref{fig:nonZ2xy}. In the case of superheavy $\sigma$, our
current approximation can be improved by adding two-loop contributions
to the mass parameters in the dimensional reduction step. Another
improvement would be to give the relations between \MSbar{} parameters
and physical quantities to one-loop accuracy.  With these
improvements, one would obtain full $g^4$ accuracy, analogous to
Ref.~\cite{Kajantie:1995dw} for the Standard Model. This work is
already underway.

There are many other generalisations of our results that deserve
  further consideration. One could treat the singlet as a heavy rather
  than a superheavy field, which would still allow us to integrate it
  out, and the theory would still reduce to the same 3D
  theory. Another possibility is to include higher order
operators (dimension 6 and above) in the 3D effective
theory. It is also possible to treat the singlet as a light
field, permitting the (expectation value of the) singlet to
play an active role in the phase transition, with different values in
the high- and low-temperature phases. This is in contrast to the
present case, where the super-heavy singlet only acts as an additional
spectator degree of freedom, impinging on the Higgs effective
potential through modified effective couplings. If the singlet
  was light, a 3D singlet-Higgs potential would come into play. Then
the numerics would involve a whole new 3D theory, with some additional
work required for a consistent lattice implementation.

Other 4D theories also deserve investigation. A strong candidate for
future study is the Two-Higgs Doublet Model. This consists of two
equivalent Higgs fields coupled to each other and to gauge fields, and
with one (Type I) or both (Type II) of the Higgs fields coupled to
fermions. Work on this is already underway.  Several attempts have
been made on computing the strength of the 2HDM phase transition
perturbatively
\cite{Cline:1996mga,Fromme:2006cm,Cline:2011mm,Curtin:2012aa,Chung:2012vg},
most recently in \cite{Haarr:2016qzq}. However, the full 2HDM
parameters space is 10-dimensional, posing an even bigger numerical
challenge than the singlet model addressed here. On the other hand,
the 2HDM readily allows for the inclusion of CP-violation, which the
singlet model itself does not\footnote{One may remedy this by adding
  CP-violating higher dimensional operators.}. From the point of view
of baryogenesis, this is appealing, whereas it has no relevance for
sourcing observable gravitational waves.

\acknowledgments TT has been supported by the Vilho, Yrj\"{o} and
Kalle V\"{a}is\"{a}l\"{a} Foundation and AV was supported by the
Academy of Finland grant nos.~1273545 and 1303622.  DJW was supported
by the People Programme (Marie Sk{\l}odowska-Curie actions) of the
European Union Seventh Framework Programme (FP7/2007-2013) under grant
agreement number PIEF-GA-2013-629425. This research was supported
  by the Munich Institute for Astro- and Particle Physics (MIAPP) of
  the DFG cluster of excellence ``Origin and Structure of the
  Universe''. The authors would like to thank Keijo Kajantie, Mikko
Laine and Kari Rummukainen for enlightening discussions.

\appendix

\end{fmffile}
\begin{fmffile}{fig7}

\section{Feynman rules in the unbroken phase}
\label{sec:4d_feynrules_unbroken}

In this appendix we list the Feynman rules valid in the high-temperature phase where the expectation value of the Higgs field vanishes. Note that our list is not complete in that we leave out the gluon sector, which is not needed for dimensional reduction at the order considered here.

\paragraph{Projectors to specific polarization states.}
\begin{equation}
\begin{split}
\text{Transverse projector:}&\quad\mathcal P_T(K)_{\mu\nu}\equiv\delta_{\mu\nu}-\frac{K_\mu K_\nu}{K^2}\\
\text{Chiral projectors:}&\quad\mathcal P_R\equiv\tfrac12(1+\gamma_5),\quad\mathcal P_L\equiv\tfrac12(1-\gamma_5)
\end{split}
\end{equation}

\subsection{Propagators in the Landau gauge}

\vspace{-2ex}
\begin{equation}
\begin{split}
\text{$\gr{SU(2)}_L$ gauge bosons:}&\quad\selfzero{boson}{}\,=\delta^{ab}\frac{\mathcal P_T(K)_{\mu\nu}}{K^2}\\[-6ex]
\text{$\gr{U(1)}_Y$ gauge boson:}&\quad\selfzero{zigzag}{}\,=\frac{\mathcal P_T(K)_{\mu\nu}}{K^2}\\[-6ex]
\text{$\gr{SU(3)}_c$ gauge bosons:}&\quad\selfzero{gluon}{}\,=\delta^{\alpha\beta}\frac{\mathcal P_T(K)_{\mu\nu}}{K^2}\\[-6ex]
\text{$\gr{SU(2)}_L$ ghosts:}&\quad\selfzero{ghost}{}\,=\delta^{ab}\frac1{K^2}\\[-6ex]
\text{fermions:}&\quad\selfzero{fermion}{}\,=\mathcal P_{L/R}\frac\imag{\slashed K}\qquad\text{(left/right-handed)}\\[-6ex]
\text{Higgs doublet:}&\quad\selfzero{scalar}{}\,=\delta^{ij}\frac1{K^2}\\[-6ex]
\text{neutral scalar:}&\quad\selfzero{double}{}\,=\frac1{K^2+\mu_\sigma^2}
\end{split}
\end{equation}
In the case of heavy $\sigma$, its propagator is to be expanded in powers of $\mu_\sigma^2$.

\subsection{Interaction vertices}

For oriented lines, momentum is understood to flow in the given direction. For unoriented lines, momentum flows into the interaction vertex.

\paragraph{Gauge self-interactions.}

\begin{equation}
\vertone{boson}{boson}{boson}{b\nu}{a\mu}{c\lambda}{P}{K}{Q}\hspace{3ex}=-\imag g\epsilon^{abc}[(P-Q)_\mu\delta_{\nu\lambda}+(Q-K)_\nu\delta_{\lambda\mu}+(K-P)_\lambda\delta_{\mu\nu}]\\[2ex]
\end{equation}
When $c\lambda$ is an external line with $Q=0$, this vertex reduces in the Landau gauge to
\begin{equation}
2\imag g\epsilon^{abc}P_\lambda\delta_{\mu\nu}=-2\imag g\epsilon^{abc}K_\lambda\delta_{\mu\nu}.\\[2ex]
\end{equation}
\begin{align}
\scatone{boson}{boson}{boson}{boson}{b\nu}{a\mu}{c\kappa}{d\lambda}\hspace{2ex}={}&g^2[\delta_{ab}\delta_{cd}(\delta_{\mu\kappa}\delta_{\nu\lambda}+\delta_{\mu\lambda}\delta_{\nu\kappa}-2\delta_{\mu\nu}\delta_{\kappa\lambda})\\[-5ex]
\notag
&+\delta_{ac}\delta_{bd}(\delta_{\mu\nu}\delta_{\kappa\lambda}+\delta_{\mu\lambda}\delta_{\nu\kappa}-2\delta_{\mu\kappa}\delta_{\nu\lambda})\\
\notag
&+\delta_{ad}\delta_{bc}(\delta_{\mu\nu}\delta_{\kappa\lambda}+\delta_{\mu\kappa}\delta_{\nu\lambda}-2\delta_{\mu\lambda}\delta_{\nu\kappa})]\\[4ex]
\vertone{ghost}{ghost}{boson}{c}{b}{a\mu}{}{P}{}\hspace{3ex}={}&\imag g\epsilon^{abc}P_\mu\\
\notag
\end{align}

\paragraph{Gauge-matter interactions.}

\begin{align}
\vertone{fermion}{fermion}{boson}{j}{i}{a\mu}{}{}{}\hspace{3ex}&=\frac\imag 2g(\tau_a)^{ij}\gamma^E_\mu\qquad\text{(left-handed fermions)}\\[4ex]
\vertone{fermion}{fermion}{zigzag}{}{}{\mu}{}{}{}\hspace{3ex}&=\frac\imag 2g'Y\gamma^E_\mu\qquad\text{(all fermions)}\\[4ex]
\vertone{fermion}{fermion}{gluon}{\rho}{\beta}{\alpha \mu}{}{}{}\hspace{3ex}&=\frac\imag 2g_s(\lambda_\alpha)^{\beta\rho}\gamma^E_\mu\qquad\text{(all quarks)}
\end{align}
\begin{align}
\vertone{scalar}{scalar}{boson}{j}{i}{a\mu}{K}{P}{}\hspace{3ex}&=-\frac12g(\tau_a)^{ij}(K+P)_\mu
\qquad 
&\scatone{scalar}{scalar}{boson}{boson}{j}{i}{b\nu}{a\mu}\hspace{2ex}&=-\frac12g^2\delta^{ij}\delta^{ab}\delta_{\mu\nu}\\[4ex]
\vertone{scalar}{scalar}{zigzag}{j}{i}{\mu}{K}{P}{}\hspace{3ex}&=-\frac12g'\delta^{ij}(K+P)_\mu
&\scatone{scalar}{scalar}{zigzag}{zigzag}{j}{i}{\nu}{\mu}\hspace{2ex}&=-\frac12g'^2\delta^{ij}\delta_{\mu\nu}
\end{align}
\begin{equation}
\scatone{scalar}{scalar}{boson}{zigzag}{j}{i}{a\mu}{\nu}\hspace{2ex}=-\frac12gg'(\tau_a)^{ij}\delta_{\mu\nu}.\\[4ex]
\end{equation}

\paragraph{Scalar self-interactions.}

\begin{equation}
\scatone{scalar}{scalar}{scalar}{scalar}{k}{i}{\ell}{j}\hspace{2ex}=-2\lambda_h(\delta_{ik}\delta_{j\ell}+\delta_{i\ell}\delta_{jk})\\[4ex]
\end{equation}
\begin{equation}
\vertone{scalar}{scalar}{double}{j}{i}{}{}{}{}=-\frac12\mu_m\delta^{ij}
\qquad
\scatone{scalar}{scalar}{double}{double}{j}{i}{}{}=-\lambda_m\delta^{ij}\\[4ex]
\end{equation}
\begin{equation}
\parbox{12mm}{%
\begin{fmfgraph*}(12,20)
\fmfleft{l}
\fmfright{r}
\fmf{double}{l,r}
\fmfdot{l}
\end{fmfgraph*}}=-\mu_1
\qquad
\vertone{double}{double}{double}{}{}{}{}{}{}=-2\mu_3
\qquad
\scatone{double}{double}{double}{double}{}{}{}{}=-6\lambda_\sigma\\[3ex]
\end{equation}

\paragraph{Yukawa interactions.} The family indices are indicated explicitly.

\begin{align}
\vertone{fermion}{fermion}{scalar}{B}{iA}{j}{e}{\ell}{}\hspace{2ex}&=-\delta^{ij}h^{(e)}_{AB}
\qquad
&\vertoneinv{fermion}{fermion}{scalar}{iA}{B}{j}{\ell}{e}{}\hspace{2ex}&=-\delta^{ij}h^{(e)*}_{AB}\\[6ex]
\vertoneinv{fermion}{fermion}{scalar}{B}{iA}{j}{u}{q}{}\hspace{2ex}&=-\imag(\tau_2)^{ij}h^{(u)}_{AB}
\qquad
&\vertone{fermion}{fermion}{scalar}{B}{iA}{j}{d}{q}{}\hspace{2ex}&=-\delta^{ij}h^{(d)}_{AB}\\[6ex]
\vertoneinv{fermion}{fermion}{scalar}{iA}{B}{j}{q}{d}{}\hspace{2ex}&=-\delta^{ij}h^{(d)*}_{AB}
\qquad
&\vertone{fermion}{fermion}{scalar}{iA}{B}{j}{q}{u}{}\hspace{2ex}&=-\imag(\tau_2)^{ij}h^{(u)*}_{AB}\\[2ex]
\notag
\end{align}

\section{Integrals for the dimensional reduction step}
\label{sec:master_integrals}

For spatial momentum integration, we use the shorthand notation
\begin{equation}
\int_p\equiv\left(\frac{e^\gamma\Lambda^2}{4\pi}\right)^\epsilon\int\frac{\dd^{d}\vek p}{(2\pi)^d},
\label{MSint}
\end{equation}
where $d\equiv3-2\epsilon$. The Euclidean four-momentum is denoted as
$P=(\omega_n,\vek p)$ for bosons, where $\omega_n\equiv2n\pi T$, and
as $P=(\nu_n,\vek p)$ for fermions, where $\nu_n\equiv(2n+1)\pi
T$. For the combined Matsubara sum and spatial momentum
  integration, we use the following shorthand:
\begin{align}
\notag
\text{bosons:}&&\sumint P&\equiv T\sum_{\omega_n}\int_p,\\
&&\sumint P'&\equiv T\sum_{\omega_n\neq0}\int_p\qquad\qquad\text{(sum over nonzero modes)},\\
\notag
\text{fermions:}&&\sumint{\braced P}&\equiv T\sum_{\nu_n}\int_p.
\end{align}


\subsection{Massless bosonic sum-integrals}
\begin{align}
I^{4b}_{\alpha,\beta,\delta}\equiv\,\sumint P'\frac{(P_0^2)^\beta(\vek p^2)^\delta}{(P^2)^\alpha}={}&\frac{(e^\gamma\Lambda^2)^\epsilon}{8\pi^2}\frac{\Gamma\left(\alpha-\tfrac d2-\delta\right)\Gamma\left(\tfrac d2+\delta\right)\zeta(2\alpha-2\beta-2\delta-d)}{\Gamma\left(\tfrac12\right)\Gamma(\alpha)\Gamma\left(\tfrac d2\right)}\\
&\times(2\pi T)^{1+d-2\alpha+2\beta+2\delta}, \notag \\
I^{4b}_{\alpha,\beta}\equiv I^{4b}_{\alpha,\beta,0}=\,\sumint P'\frac{(P_0^2)^\beta}{(P^2)^\alpha}={}&\frac{(e^\gamma\Lambda^2)^\epsilon}{8\pi^2}\frac{\Gamma\left(\alpha-\tfrac d2\right)\zeta(2\alpha-2\beta-d)}{\Gamma\left(\tfrac12\right)\Gamma(\alpha)}(2\pi T)^{1+d-2\alpha+2\beta},\\
I^{4b}_\alpha\equiv I^{4b}_{\alpha,0}=\,\sumint{P}'\frac{1}{(P^2)^\alpha}={}&\frac{(e^\gamma\Lambda^2)^\epsilon}{8\pi^2}\frac{\Gamma\left(\alpha-\tfrac d2\right)\zeta(2\alpha-d)}{\Gamma\left(\tfrac12\right)\Gamma(\alpha)}(2\pi T)^{1+d-2\alpha},\\
I^{4b}_1=\,\sumint P'\frac1{\phantom(P^2\phantom{)^\alpha}}={}&\frac{T^2}{12}\mufourpiT2\left\{1+2\left[\log2\pi+\gamma-\frac{\zeta'(2)}{\zeta(2)}\right]\epsilon+\OO(\epsilon^2)\right\},\\
I^{4b}_2=\,\sumint P'\frac1{(P^2)^2}={}&\frac1{16\pi^2}\mufourpiT2\left[\frac1\epsilon+2\gamma+\OO(\epsilon)\right].
\end{align}
Useful recursive relations among the sum-integrals:
\begin{align}
I^{4b}_{\alpha+1,\beta+1}&=\left(1-\frac d{2\alpha}\right)I^{4b}_{\alpha,\beta},\\
I^{4b}_{\alpha+1,\beta,\delta+1}&=\frac{\frac d2+\delta}\alpha I^{4b}_{\alpha,\beta,\delta},\\
I^{4b}_{\alpha,\beta-1,\delta+1}&=-\frac{\frac d2+\delta}{1+\frac d2-\alpha+\delta} I^{4b}_{\alpha,\beta,\delta}.
\end{align}
Occasionally, we need analogous sum-integrals including the zero
Matsubara mode; these are denoted by a tilde, e.g.~$\tilde
I^{4b}_{\alpha,\beta,\delta}$. Explicit expressions for these
sum-integrals are not needed, we merely note that they do not satisfy
the above recursive relations.


\subsection{Massless fermionic sum-integrals}
\begin{align}
I^{4f}_{\alpha,\beta,\delta}\equiv\,\sumint{\braced P}\frac{(P_0^2)^\beta(\vek p^2)^\delta}{(P^2)^\alpha}&=\Bigl(2^{2\alpha-2\beta-2\delta-d}-1\Bigr)I^{4b}_{\alpha,\beta,\delta},\\
I^{4f}_{\alpha,\beta}\equiv I^{4f}_{\alpha,\beta,0}=\,\sumint{\braced P}\frac{(P_0^2)^\beta}{(P^2)^\alpha}&,\\
I^{4f}_\alpha\equiv I^{4f}_{\alpha,0}=\,\sumint{\braced P}\frac{1}{(P^2)^\alpha}&,\\
I^{4f}_1=\,\sumint{\braced P}\frac1{\phantom(P^2\phantom{)^\alpha}}&=-\frac{T^2}{24}\mufourpiT2\left\{1+2\left[\log\pi+\gamma-\frac{\zeta'(2)}{\zeta(2)}\right]\epsilon+\OO(\epsilon^2)\right\},\\
I^{4f}_2=\,\sumint{\braced P}\frac1{(P^2)^2}&=\frac1{16\pi^2}\mufourpiT2\left[\frac1\epsilon+2\gamma+4\log2+\OO(\epsilon)\right].
\end{align}
Due to the first of the above relations, the fermionic sum-integrals satisfy the same recursive identities as their bosonic counterparts.


\subsection{Massive sum-integrals}
\label{sec:massive_sumints}
\begin{align}
\tilde K^{4b}(m)&\equiv\frac12\,\,\sumint P\log(P^2+m^2),\\
\tilde K^{4f}(m)&\equiv\frac12\,\,\sumint{\braced P}\log(P^2+m^2),\\
\tilde J^{4b}_{\kappa/\alpha,\beta,\delta}(m)&\equiv\,\sumint P\frac{(P_0^2)^\beta(\vek p^2)^\delta}{(P^2)^\alpha(P^2+m^2)^\kappa},\\
\tilde J^{4b}_{\kappa/\alpha}(m)&\equiv\tilde J^{4b}_{\kappa/\alpha,0,0}=\,\sumint P\frac{1}{(P^2)^\alpha(P^2+m^2)^\kappa},\\
\tilde J^{4b}_{\kappa}(m)&\equiv\tilde J^{4b}_{\kappa/0}=\,\sumint P\frac{1}{(P^2+m^2)^\kappa},
\end{align}
and likewise for the version without the zero mode,
$J^{4b}_{\kappa/\alpha,\beta,\delta}(m)$. For $\beta>0$, the two
integrals -- with and without the zero mode -- coincide. The
two-index integrals satisfy the recursive relation
\begin{equation}
\tilde J^{4b}_{\kappa/\alpha}(m)=\tilde J^{4b}_{\kappa-1/\alpha+1}(m)-m^2\tilde J^{4b}_{\kappa/\alpha+1}(m).
\end{equation}
It is straightforward to verify that the following relations hold:
\begin{align}
\tilde{J}^{4b}_{1/1}(m) &= \frac{1}{m^2}\big[I^{4b}_1-\tilde{J}^{4b}_{1}(m)\big],\\
\tilde{J}^{4b}_{1/2}(m) &= \frac{1}{m^2}I^{4b}_2-\frac{1}{m^4}\big[I^{4b}_1-\tilde{J}^{4b}_{1}(m)\big],\\
\tilde{J}^{4b}_{2/1}(m) &= -\frac{1}{m^2}\tilde{J}^{4b}_{2}(m)+\frac{1}{m^4}\big[I^{4b}_1-\tilde{J}^{4b}_{1}(m)\big],\\
\tilde{J}^{4b}_{2/2}(m) &= \frac{1}{m^4} \big[I^{4b}_2 + \tilde{J}^{4b}_{2}(m)\big] -\frac{2}{m^6}\big[I^{4b}_1-\tilde{J}^{4b}_{1}(m)\big],\\
\tilde{J}^{4b}_{1/3,0,1}(m) &= \frac{1}{m^2}I^{4b}_{3,0,1}-\frac{1}{m^4}I^{4b}_{2,0,1}+\frac{1}{m^6}I^{4b}_{1,0,1}-\frac{1}{m^6}\tilde{J}^{4b}_{1/0,0,1}(m),\\
\tilde{J}^{4b}_{1/3,1,0}(m) &= \tilde{J}^{4b}_{1/2}(m) - \tilde{J}^{4b}_{1/3,0,1}(m),\\
\tilde{J}^{4b}_{3/1,0,1}(m) &= -\frac{1}{m^2}\tilde{J}^{4b}_{3/0,0,1}(m)-\frac{1}{m^4}\tilde{J}^{4b}_{2/0,0,1}(m)-\frac{1}{m^6}\tilde{J}^{4b}_{1/0,0,1}(m)+\frac{1}{m^6}I^{4b}_{1,0,1}.
\end{align}
Furthermore
\begin{align}
\tilde{J}^{4b}_{1}(m)&=I^4_1(m)+J_1(m),\label{Jt1}\\
\tilde{J}^{4b}_{2}(m)&=I^4_2(m)+J_2(m),\\
\tilde{J}^{4b}_{3/0,0,1}(m)&=\frac{3-2\epsilon}{4-2\epsilon}\big[I^4_2(m)-m^2I^4_3(m)\big]+J_{3,0,1}(m),\\
\tilde{J}^{4b}_{2/0,0,1}(m)&=\frac{3-2\epsilon}{4-2\epsilon}\big[I^4_1(m)-m^2I^4_2(m)\big]+J_{2,0,1}(m),\\
\tilde{J}^{4b}_{1/0,0,1}(m)&=-\frac{3-2\epsilon}{4-2\epsilon}m^2I^4_1(m)+J_{1,0,1}(m) \label{Jt1001},
\end{align}
where we have defined 
\begin{equation}
I^4_\alpha(m)\equiv \bigg(\frac{e^\gamma \Lambda^2}{4\pi}\bigg)^\epsilon \int \frac{\text{d}^np}{(2\pi)^n} \frac{1}{(p^2+m^2)^\alpha} =  \bigg(\frac{e^\gamma \Lambda^2}{4\pi}\bigg)^\epsilon \frac{(m^2)^{\frac{n}{2}-\alpha}}{(4\pi)^\frac{n}{2}}\frac{\Gamma(\alpha-\frac{n}{2})}{\Gamma(\alpha)} ,
\end{equation}
where $n=4-2\epsilon$ and
\begin{align}
J_1(m)&\equiv\int\frac{\dd^3\vek p}{(2\pi)^3}\frac{n_B(E_p)}{E_p},\qquad
&J_{1,0,1}(m)&\equiv\int\frac{\dd^3\vek p}{(2\pi)^3}\frac{p^2n_B(E_p)}{E_p},\\
J_2(m)&\equiv\int\frac{\dd^3\vek p}{(2\pi)^3}\frac{n_B(E_p)}{2p^2E_p},\qquad
&J_{2,0,1}(m)&\equiv\int\frac{\dd^3\vek p}{(2\pi)^3}\frac{3n_B(E_p)}{2E_p},\\
&&J_{3,0,1}(m)&\equiv\int\frac{\dd^3\vek p}{(2\pi)^3}\frac{3n_B(E_p)}{8p^2E_p},
\end{align}
where $n_B$ is the Bose-Einstein distribution function and
$E_p\equiv\sqrt{\vek p^2+m^2}$. These integrals satisfy the simple
relations
\begin{equation}
J_{2,0,1}(m)=\frac32J_1(m),\qquad
J_{3,0,1}(m)=\frac34J_2(m).
\end{equation}
Thus, the only master integrals needed are actually just $J_1(m)$, $J_2(m)$ and $J_{1,0,1}(m)$. The summary of the results in the case of superheavy $\sigma$ only features explicitly $J_1(m)$, $J_2(m)$ and the following particular combination of the three integrals:
\begin{equation}
H(m)\equiv{}-\frac{3}{32 \pi^2m^2}-\frac{1}{m^4}\bigg[\frac{T^2}{12}+J_1(m)\bigg]+\frac{1}{m^6}\bigg[\frac{2\pi^2T^4}{45}-\frac{4J_{1,0,1}(m)}{3}\bigg].
\label{Hintegral}
\end{equation}



\end{fmffile}
\begin{fmffile}{fig8}
	
\section{Detailed results for the SM contributions to dimensional reduction}
\label{sec:DR_diagrams}

Below we provide a list of all one-loop diagrams in the SM
that arise in the four-dimensional theory, expressed in terms of the
master sum-integrals introduced in
Appendix~\ref{sec:master_integrals}. New contributions from the neutral
scalar are discussed in Section~\ref{sec:DR_correlators}. All the
diagrams listed below are given without the zero mode contribution,
which at the one-loop level trivially drops in the matching to the
three-dimensional effective theory. Many of the diagrams have already
been calculated in Ref.~\cite{Kajantie:1995dw}, but some of the
contributions of the $\gr{U(1)}_Y$ sector included here are new. The
indicated values of the diagrams already include combinatorial factors
due to permutations of external lines.

\subsection{Self-energy diagrams}
\label{subapp:debye}

These are needed for the calculation of wave-function renormalization
and of the Debye masses of the gauge bosons. The diagrams with a
single quartic vertex only contribute to the latter. The wavefunction
renormalization factors can be read off from the parts that are
quadratic in momentum, as detailed in
Section~\ref{sec:counterterms_beta_functions}.

\paragraph{$\gr{SU(2)}_L$ gauge boson self-energy.}

\begin{align}
\selfone{boson}{boson}{boson}\,&=-dg^2\delta_{ab}I^{4b}_1\\[-4ex]
\notag
&\text{for $\mu=\nu=0$,}\\[2ex]
\notag
&=g^2(1-2d)\delta_{ab}\delta_{rs}I^{4b}_1\\
\notag
&\text{for $\mu=r$, $\nu=s$,}\\
\selftwo{boson}{boson}{boson}{boson}\,&=4g^2\delta_{ab}\bigl[d\bigl(1-\tfrac d2\bigr)I^{4b}_1+\tfrac1{24}(16-3d+2d^2)P^2I^{4b}_2\bigr]\\[-4ex]
\notag
&\text{for $\mu=\nu=0$,}\\[2ex]
\notag
&=2dg^2\delta_{ab}\delta_{rs}I^{4b}_1+g^2\delta_{ab}\bigl[\tfrac16(31-2d)\delta_{rs}P^2+\tfrac13(d-17)P_rP_s\bigr]I^{4b}_2\\
\notag
&\text{for $\mu=r$, $\nu=s$,}\\
\selftwo{boson}{ghost}{ghost}{boson}\,&=g^2\delta_{ab}\bigl[(d-2)I^{4b}_1+\tfrac16(4-d)P^2I^{4b}_2\bigr]\\[-4ex]
\notag
&\text{for $\mu=\nu=0$,}\\[2ex]
\notag
&=-g^2\delta_{ab}\delta_{rs}I^{4b}_1+\tfrac16g^2\delta_{ab}(\delta_{rs}P^2+2P_rP_s)I^{4b}_2\\
\notag
&\text{for $\mu=r$, $\nu=s$,}\\
\selftwo{boson}{fermion}{fermion}{boson}\,&=g^2(d-1)\delta_{ab}N_f(1+N_c)\bigl[(2^{2-d}-1)I^{4b}_1-\tfrac16(2^{4-d}-1)P^2I^{4b}_2\bigr]\\[-4ex]
\notag
&\text{for $\mu=\nu=0$,}\\[2ex]
\notag
&=\tfrac13g^2(2^{4-d}-1)\delta_{ab}N_f(1+N_c)(P_rP_s-\delta_{rs}P^2)I^{4b}_2\\
\notag
&\text{for $\mu=r$, $\nu=s$,}\\[4ex]
\selfone{boson}{scalar}{boson}\,&=-g^2\delta_{ab}\delta_{\mu\nu}I^{4b}_1,\\[-4ex]
\selftwo{boson}{scalar}{scalar}{boson}\,&=2g^2\delta_{ab}\bigl[(1-\tfrac d2)I^{4b}_1-\tfrac13(1-\tfrac d4)P^2I^{4b}_2\bigr]\\[-4ex]
\notag
&\text{for $\mu=\nu=0$,}\\[2ex]
\notag
&=g^2\delta_{ab}\bigl[\delta_{rs}I^{4b}_1+\tfrac16(P_rP_s-\delta_{rs}P^2)I^{4b}_2\bigr]\\
\notag
&\text{for $\mu=r$, $\nu=s$.}
\end{align}

\paragraph{$\gr{U(1)}_Y$ gauge boson self-energy.}
\begin{align}
\selftwo{zigzag}{fermion}{fermion}{zigzag}\,&=-\tfrac12(d-1)g'^2N_f[2Y_\ell^2+Y_e^2+N_c(2Y_q^2+Y_u^2+Y_d^2)]\\[-4ex]
\notag
&\phantom{{}={}}\times\bigl[(1-2^{2-d})I^{4b}_1+\tfrac16(2^{4-d}-1)P^2I^{4b}_2\bigr]\\
\notag
&\text{for $\mu=\nu=0$,}\\[2ex]
\notag
&=\tfrac16(2^{4-d}-1)g'^2N_f[2Y_\ell^2+Y_e^2+N_c(2Y_q^2+Y_u^2+Y_d^2)]\\
\notag
&\phantom{{}={}}\times(P_rP_s-\delta_{rs}P^2)I^{4b}_2\\
\notag
&\text{for $\mu=r$, $\nu=s$,}\\[4ex]
\selfone{zigzag}{scalar}{zigzag}\,&=-g'^2\delta_{\mu\nu}I^{4b}_1,\\[-4ex]
\selftwo{zigzag}{scalar}{scalar}{zigzag}\,&=2g'^2\bigl[(1-\tfrac d2)I^{4b}_1-\tfrac13(1-\tfrac d4)P^2I^{4b}_2\bigr]\\[-4ex]
\notag
&\text{for $\mu=\nu=0$,}\\[2ex]
\notag
&=g'^2\bigl[\delta_{rs}I^{4b}_1+\tfrac16(P_rP_s-\delta_{rs}P^2)I^{4b}_2\bigr]\\
\notag
&\text{for $\mu=r$, $\nu=s$.}
\end{align}

\paragraph{Higgs doublet self-energy.} Only diagrams contributing to wavefunction renormalization are shown here; the mass parameter can be extracted from the effective potential.
\begin{align}
\selftwo{scalar}{scalar}{boson}{scalar}\,={}&\tfrac94g^2\delta_{ij}P^2I^{4b}_2,\\
\selftwo{scalar}{scalar}{zigzag}{scalar}\,={}&\tfrac34g'^2\delta_{ij}P^2I^{4b}_2,\\
\selftwo{scalar}{fermion}{fermion}{scalar}\,={}&2\delta_{ij}\tr\bigl[h^{(e)}h^{(e)\dag}+N_ch^{(u)}h^{(u)\dag}+N_ch^{(d)}h^{(d)\dag}\bigr]\\[-4ex]
\notag
&\times\bigl[(2^{2-d}-1)I^{4b}_1-\tfrac12(2^{4-d}-1)P^2I^{4b}_2\bigr].
\end{align}

\subsection{Correlators for gauge fields}
\label{subsec:rencoupling}

The various four-point correlators with two or four gauge field
external legs are listed below in the same order as in
Section~\ref{sec:gauge_field_correlators}.


\paragraph{The $A^a_0A^b_0A^c_0A^d_0$ correlator.}
\begin{align}
\scattwo{boson}{boson}{boson}{boson}{boson}{boson}\,&=\tfrac16d(14+d)g^4(\delta_{ab}\delta_{cd}+\delta_{ac}\delta_{bd}+\delta_{ad}\delta_{bc})I^{4b}_2,\\
\scattwo{boson}{boson}{scalar}{scalar}{boson}{boson}\,&=\tfrac12g^4(\delta_{ab}\delta_{cd}+\delta_{ac}\delta_{bd}+\delta_{ad}\delta_{bc})I^{4b}_2,\\
\scatthree{boson}{boson}{boson}{boson}{boson}{boson}{boson}\,&=\tfrac{20}3d(d-4)g^4(\delta_{ab}\delta_{cd}+\delta_{ac}\delta_{bd}+\delta_{ad}\delta_{bc})I^{4b}_2,\\
\scatthree{boson}{boson}{scalar}{scalar}{scalar}{boson}{boson}\,&=(d-4)g^4(\delta_{ab}\delta_{cd}+\delta_{ac}\delta_{bd}+\delta_{ad}\delta_{bc})I^{4b}_2,\\
\scatfour{boson}{boson}{boson}{boson}{boson}{boson}{boson}{boson}\,&=\tfrac43d(4-d)(6-d)g^4(\delta_{ab}\delta_{cd}+\delta_{ac}\delta_{bd}+\delta_{ad}\delta_{bc})I^{4b}_2,\\
\scatfour{boson}{boson}{ghost}{ghost}{ghost}{ghost}{boson}{boson}\,&=-\tfrac16(4-d)(6-d)g^4(\delta_{ab}\delta_{cd}+\delta_{ac}\delta_{bd}+\delta_{ad}\delta_{bc})I^{4b}_2,\\
\scatfour{boson}{boson}{fermion}{fermion}{fermion}{fermion}{boson}{boson}\,&=\tfrac16\bigl(1-2^{4-d}\bigr)(d-1)(d-3)N_f(1+N_c)g^4\\[-4ex]
\notag
&\phantom{{}={}}\times(\delta_{ab}\delta_{cd}+\delta_{ac}\delta_{bd}+\delta_{ad}\delta_{bc})I^{4b}_2,\\
\scatfour{boson}{boson}{scalar}{scalar}{scalar}{scalar}{boson}{boson}\,&=\tfrac16(4-d)(6-d)g^4(\delta_{ab}\delta_{cd}+\delta_{ac}\delta_{bd}+\delta_{ad}\delta_{bc})I^{4b}_2.
\end{align}


\paragraph{The $B_0^4$ correlator.}
\begin{align}
\scattwo{zigzag}{zigzag}{scalar}{scalar}{zigzag}{zigzag}\,&=\tfrac32g'^4I^{4b}_2,\\
\scatthree{zigzag}{zigzag}{scalar}{scalar}{scalar}{zigzag}{zigzag}\,&=3(d-4)g'^4I^{4b}_2,\\
\scatfour{zigzag}{zigzag}{fermion}{fermion}{fermion}{fermion}{zigzag}{zigzag}\,&=\tfrac14\bigl(1-2^{4-d}\bigr)(d-1)(d-3)N_f\\[-4ex]
\notag
&\phantom{{}={}}\times[2Y_\ell^4+Y_e^4+N_c(2Y_q^4+Y_u^4+Y_d^4)]g'^4I^{4b}_2,\\
\scatfour{zigzag}{zigzag}{scalar}{scalar}{scalar}{scalar}{zigzag}{zigzag}\,&=\tfrac12(4-d)(6-d)g'^4I^{4b}_2.
\end{align}


\paragraph{The $A^a_0A^b_0B_0^2$ correlator.}
\begin{align}
\scattwo{boson}{boson}{scalar}{scalar}{zigzag}{zigzag}\,&=\tfrac12g^2g'^2\delta_{ab}I^{4b}_2,\\
\scattwo{boson}{zigzag}{scalar}{scalar}{boson}{zigzag}\,&=g^2g'^2\delta_{ab}I^{4b}_2,\\
\scatthree{boson}{boson}{scalar}{scalar}{scalar}{zigzag}{zigzag}\,=\,\scatthree{zigzag}{zigzag}{scalar}{scalar}{scalar}{boson}{boson}\,&=-2\bigl(1-\tfrac d4\bigr)g^2g'^2\delta_{ab}I^{4b}_2,\\
\scatthree{boson}{zigzag}{scalar}{scalar}{scalar}{boson}{zigzag}\,=\,\scatthree{boson}{zigzag}{scalar}{scalar}{scalar}{zigzag}{boson}\,&=(d-4)g^2g'^2\delta_{ab}I^{4b}_2,\\
\scatfour{boson}{boson}{fermion}{fermion}{fermion}{fermion}{zigzag}{zigzag}\,+\scatfour{boson}{zigzag}{fermion}{fermion}{fermion}{fermion}{zigzag}{boson}\,&=\tfrac12\bigl(1-2^{4-d}\bigr)(d-1)(d-3)N_f(Y_\ell^2+N_cY_q^2)\\[-4ex]
\notag
&\phantom{{}={}}\times g^2g'^2\delta_{ab}I^{4b}_2,\\
\scatfour{boson}{boson}{scalar}{scalar}{scalar}{scalar}{zigzag}{zigzag}\,+\,\scatfour{boson}{zigzag}{scalar}{scalar}{scalar}{scalar}{zigzag}{boson}\,&=\tfrac12(4-d)(6-d)g^2g'^2\delta_{ab}I^{4b}_2.
\end{align}


\paragraph{The $\phi^{\dag i}\phi^jA^a_\mu A^b_\nu$ correlator.}

\begin{align}
\scattwo{scalar}{scalar}{boson}{boson}{boson}{boson}\,&=\tfrac34dg^4\delta_{ij}\delta_{ab}I^{4b}_2\\[-4ex]
\notag
&\text{for $\mu=\nu=0$,}\\[2ex]
\notag
&=\bigl(d-\tfrac34\bigr)g^4\delta_{ij}\delta_{ab}\delta_{rs}I^{4b}_2\\
\notag
&\text{for $\mu=r$, $\nu=s$,}\\
\label{diagram1}
\scattwo{scalar}{scalar}{scalar}{scalar}{boson}{boson}\,&=3\lambda_hg^2\delta_{ij}\delta_{ab}\delta_{\mu\nu}I^{4b}_2,\\
\scattwo{boson}{scalar}{scalar}{boson}{scalar}{boson}\,&=\tfrac d8g^4\delta_{ij}\delta_{ab}I^{4b}_2\\[-4ex]
\notag
&\text{for $\mu=\nu=0$,}\\[2ex]
\notag
&=\tfrac38g^4\delta_{ij}\delta_{ab}\delta_{rs}I^{4b}_2\\
\notag
&\text{for $\mu=r$, $\nu=s$,}\\
\scattwo{boson}{scalar}{scalar}{zigzag}{scalar}{boson}\,&=\tfrac d8g^2g'^2\delta_{ij}\delta_{ab}I^{4b}_2\\[-4ex]
\notag
&\text{for $\mu=\nu=0$,}\\[2ex]
\notag
&=\tfrac38g^2g'^2\delta_{ij}\delta_{ab}\delta_{rs}I^{4b}_2\\
\notag
&\text{for $\mu=r$, $\nu=s$,}\\
\scatthree{scalar}{scalar}{boson}{boson}{boson}{boson}{boson}\,&=d(d-4)g^4\delta_{ij}\delta_{ab}I^{4b}_2\\[-4ex]
\notag
&\text{for $\mu=\nu=0$,}\\[2ex]
\notag
&=-dg^4\delta_{ij}\delta_{ab}\delta_{rs}I^{4b}_2\\
\notag
&\text{for $\mu=r$, $\nu=s$,}\\
\label{diagram2}
\scatthree{scalar}{scalar}{scalar}{scalar}{scalar}{boson}{boson}\,&=3(d-4)\lambda_hg^2\delta_{ij}\delta_{ab}I^{4b}_2\\[-4ex]
\notag
&\text{for $\mu=\nu=0$,}\\[2ex]
\notag
&=-3\lambda_hg^2\delta_{ij}\delta_{ab}\delta_{rs}I^{4b}_2\\
\notag
&\text{for $\mu=r$, $\nu=s$,}\\
\scatfour{scalar}{scalar}{fermion}{fermion}{fermion}{fermion}{boson}{boson}\,&=\tfrac12(2^{4-d}-1)(2-d)g^2\delta_{ij}\delta_{ab}\tr\bigl[h^{(e)}h^{(e)\dag}+N_ch^{(d)}h^{(d)\dag}\bigr]I^{4b}_2\\[-4ex]
\notag
&\text{for $\mu=\nu=0$,}\\[2ex]
\notag
&=-\tfrac12(2^{4-d}-1)g^2\delta_{ij}\delta_{ab}\delta_{rs}\tr\bigl[h^{(e)}h^{(e)\dag}+N_ch^{(d)}h^{(d)\dag}\bigr]I^{4b}_2\\
\notag
&\text{for $\mu=r$, $\nu=s$,}\\
\scatfour{boson}{boson}{fermion}{fermion}{fermion}{fermion}{scalar}{scalar}\,&=\tfrac12(2^{4-d}-1)(2-d)g^2\delta_{ij}\delta_{ab}N_c\tr\bigl[h^{(u)}h^{(u)\dag}\bigr]I^{4b}_2\\[-4ex]
\notag
&\text{for $\mu=\nu=0$,}\\[2ex]
\notag
&=-\tfrac12(2^{4-d}-1)g^2\delta_{ij}\delta_{ab}\delta_{rs}N_c\tr\bigl[h^{(u)}h^{(u)\dag}\bigr]I^{4b}_2\\
\notag
&\text{for $\mu=r$, $\nu=s$.}
\end{align}


\paragraph{The $\phi^{\dag i}\phi^jB_\mu B_\nu$ correlator.}

\begin{align}
\scattwo{scalar}{scalar}{scalar}{scalar}{zigzag}{zigzag}\,&=3\lambda_hg'^2\delta_{ij}\delta_{\mu\nu}I^{4b}_2,\\
\scattwo{zigzag}{scalar}{scalar}{boson}{scalar}{zigzag}\,&=\tfrac 38dg^2g'^2\delta_{ij}I^{4b}_2\\[-4ex]
\notag
&\text{for $\mu=\nu=0$,}\\[2ex]
\notag
&=\tfrac98g^2g'^2\delta_{ij}\delta_{rs}I^{4b}_2\\
\notag
&\text{for $\mu=r$, $\nu=s$,}\\
\scattwo{zigzag}{scalar}{scalar}{zigzag}{scalar}{zigzag}\,&=\tfrac d8g'^4\delta_{ij}I^{4b}_2\\[-4ex]
\notag
&\text{for $\mu=\nu=0$,}\\[2ex]
\notag
&=\tfrac38g'^4\delta_{ij}\delta_{rs}I^{4b}_2\\
\notag
&\text{for $\mu=r$, $\nu=s$,}\\
\scatthree{scalar}{scalar}{scalar}{scalar}{scalar}{zigzag}{zigzag}\,&=-12\bigl(1-\tfrac d4\bigr)\lambda_h g'^2\delta_{ij}I^{4b}_2\\[-4ex]
\notag
&\text{for $\mu=\nu=0$,}\\[2ex]
\notag
&=-3\lambda_hg'^2\delta_{ij}\delta_{rs}I^{4b}_2\\
\notag
&\text{for $\mu=r$, $\nu=s$,}\\
\scatfour{scalar}{scalar}{fermion}{fermion}{fermion}{fermion}{zigzag}{zigzag}\,+\,\scatfour{zigzag}{zigzag}{fermion}{fermion}{fermion}{fermion}{scalar}{scalar}&=-\tfrac12(2^{4-d}-1)g'^2\delta_{ij}\tr\bigl[(Y_\ell^2+Y_e^2)h^{(e)}h^{(e)\dag}\\[-4ex]
\notag
&\phantom{{}={}}+N_c(Y_q^2+Y_u^2)h^{(u)}h^{(u)\dag}+N_c(Y_q^2+Y_d^2)h^{(d)}h^{(d)\dag}\bigr]\\
\notag
&\phantom{{}={}}\times I^{4b}_2\times
\begin{cases}
(d-2)\qquad\text{for $\mu=\nu=0$,}\\
\delta_{rs}\qquad\text{for $\mu=r$, $\nu=s$,}
\end{cases}\\
\scatfour{scalar}{zigzag}{fermion}{fermion}{fermion}{fermion}{zigzag}{scalar}\,&=(2^{4-d}-1)g'^2\delta_{ij}\delta_{\mu\nu}\tr\bigl[Y_eY_\ell h^{(e)}h^{(e)\dag}\\[-4ex]
\notag
&\phantom{{}={}}+N_cY_uY_qh^{(u)}h^{(u)\dag}+N_cY_dY_qh^{(d)}h^{(d)\dag}\bigr]I^{4b}_2.
\end{align}


\paragraph{The $\phi^{\dag i}\phi^jA^a_0B_0$ correlator.}

\begin{align}
\scattwo{scalar}{scalar}{scalar}{scalar}{boson}{zigzag}\,&=\lambda_hgg'(\tau_a)^{ij}I^{4b}_2,\\
\scattwo{boson}{scalar}{scalar}{boson}{scalar}{zigzag}\,=\,\scattwo{zigzag}{scalar}{scalar}{boson}{scalar}{boson}\,&=\tfrac d{16}g^3g'(\tau_a)^{ij}I^{4b}_2,\\
\scattwo{boson}{scalar}{scalar}{zigzag}{scalar}{zigzag}\,=\,\scattwo{zigzag}{scalar}{scalar}{zigzag}{scalar}{boson}\,&=\tfrac d{16}gg'^3(\tau_a)^{ij}I^{4b}_2,\\
\scatthree{scalar}{scalar}{scalar}{scalar}{scalar}{boson}{zigzag}\,=\,\scatthree{scalar}{scalar}{scalar}{scalar}{scalar}{zigzag}{boson}\,&=-2\bigl(1-\tfrac d4\bigr)\lambda_hgg'(\tau_a)^{ij}I^{4b}_2,\\
\scatfour{scalar}{scalar}{fermion}{fermion}{fermion}{fermion}{boson}{zigzag}\,=\,\scatfour{scalar}{scalar}{fermion}{fermion}{fermion}{fermion}{zigzag}{boson}\,&=\tfrac14\bigl(1-2^{4-d}\bigr)(d-2)gg'(\tau_a)^{ij}I^{4b}_2\\[-4ex]
\notag
&\phantom{{}={}}\times\tr\bigl[Y_\ell h^{(e)}h^{(e)\dag}+N_cY_qh^{(d)}h^{(d)\dag}\bigr],\\
\scatfour{boson}{zigzag}{fermion}{fermion}{fermion}{fermion}{scalar}{scalar}\,=\,\scatfour{zigzag}{boson}{fermion}{fermion}{fermion}{fermion}{scalar}{scalar}\,&=-\tfrac14\bigl(1-2^{4-d}\bigr)(d-2)N_cY_q\tr\bigl[h^{(u)}h^{(u)\dag}\bigr]\\[-4ex]
\notag
&\phantom{{}={}}\times gg'(\tau_a)^{ij}I^{4b}_2,\\
\scatfour{scalar}{boson}{fermion}{fermion}{fermion}{fermion}{zigzag}{scalar}\,&=-\tfrac12\bigl(2^{4-d}-1\bigr)N_cY_u\tr\bigl[h^{(u)}h^{(u)\dag}\bigr]gg'(\tau_a)^{ij}I^{4b}_2,\\
\scatfour{scalar}{zigzag}{fermion}{fermion}{fermion}{fermion}{boson}{scalar}\,&=\tfrac12\bigl(2^{4-d}-1\bigr)\tr\bigl[Y_eh^{(e)}h^{(e)\dag}+N_cY_dh^{(d)}h^{(d)\dag}\bigr]\\[-4ex]
\notag
&\phantom{{}={}}\times gg'(\tau_a)^{ij}I^{4b}_2.
\end{align}

\paragraph{The $\phi^{\dag i}\phi^jC^\alpha_0 C^\beta_0$ correlator.}

\begin{align}
\scatfour{scalar}{scalar}{fermion}{fermion}{fermion}{fermion}{gluon}{gluon}\,+\,\scatfour{gluon}{gluon}{fermion}{fermion}{fermion}{fermion}{scalar}{scalar}&= 2(2^{4-d}-1)(2-d)g^2_s \tr\bigl[h^{(u)}h^{(u)\dagger} + h^{(d)}h^{(d)\dagger} ] \delta_{ij}\delta_{\alpha\beta} I^{4b}_2  \\
\scatfour{scalar}{gluon}{fermion}{fermion}{fermion}{fermion}{gluon}{scalar}\,&= 2(2^{4-d}-1) g^2_s \tr\bigl[h^{(u)}h^{(u)\dagger} + h^{(d)}h^{(d)\dagger} ] \delta_{ij}\delta_{\alpha\beta}I^{4b}_2.
\end{align}


\bibliographystyle{JHEP}


\begin{thebibliography}{*}

\bibitem{Kuzmin:1985mm}
  V.~A.~Kuzmin, V.~A.~Rubakov and M.~E.~Shaposhnikov,
  Phys.\ Lett.\ B {\bf 155} (1985) 36.
  doi:10.1016/0370-2693(85)91028-7


\bibitem{Cohen:1993nk}
  A.~G.~Cohen, D.~B.~Kaplan and A.~E.~Nelson,
  Ann.\ Rev.\ Nucl.\ Part.\ Sci.\  {\bf 43} (1993) 27
  doi:10.1146/annurev.ns.43.120193.000331
  [hep-ph/9302210].

\bibitem{Rubakov:1996vz}
  V.~A.~Rubakov and M.~E.~Shaposhnikov,
  Usp.\ Fiz.\ Nauk {\bf 166} (1996) 493
   [Phys.\ Usp.\  {\bf 39} (1996) 461]
  doi:10.1070/PU1996v039n05ABEH000145
  [hep-ph/9603208].

\bibitem{Morrissey:2012db}
  D.~E.~Morrissey and M.~J.~Ramsey-Musolf,
  New J.\ Phys.\  {\bf 14} (2012) 125003
  doi:10.1088/1367-2630/14/12/125003
  [arXiv:1206.2942 [hep-ph]].

\bibitem{Shaposhnikov:1987pf}
  M.~E.~Shaposhnikov,
  Nucl.\ Phys.\ B {\bf 299} (1988) 797.
  doi:10.1016/0550-3213(88)90373-2
  
\bibitem{Farrar:1993sp}
  G.~R.~Farrar and M.~E.~Shaposhnikov,
  Phys.\ Rev.\ Lett.\  {\bf 70} (1993) 2833
   Erratum: [Phys.\ Rev.\ Lett.\  {\bf 71} (1993) 210]
  doi:10.1103/PhysRevLett.70.2833
  [hep-ph/9305274].
  
\bibitem{Farrar:1993hn}
  G.~R.~Farrar and M.~E.~Shaposhnikov,
  Phys.\ Rev.\ D {\bf 50} (1994) 774
  doi:10.1103/PhysRevD.50.774
  [hep-ph/9305275].
  
\bibitem{Gavela:1993ts}
  M.~B.~Gavela, P.~Hernandez, J.~Orloff and O.~Pene,
  Mod.\ Phys.\ Lett.\ A {\bf 9} (1994) 795
  doi:10.1142/S0217732394000629
  [hep-ph/9312215].
  
\bibitem{Gavela:1994ds}
  M.~B.~Gavela, M.~Lozano, J.~Orloff and O.~Pene,
  Nucl.\ Phys.\ B {\bf 430} (1994) 345
  doi:10.1016/0550-3213(94)00409-9
  [hep-ph/9406288].
  
\bibitem{Gavela:1994dt}
  M.~B.~Gavela, P.~Hernandez, J.~Orloff, O.~Pene and C.~Quimbay,
  Nucl.\ Phys.\ B {\bf 430} (1994) 382
  doi:10.1016/0550-3213(94)00410-2
  [hep-ph/9406289].
  
\bibitem{Brauner:2011vb}
  T.~Brauner, O.~Taanila, A.~Tranberg and A.~Vuorinen,
  Phys.\ Rev.\ Lett.\  {\bf 108} (2012) 041601
  doi:10.1103/PhysRevLett.108.041601
  [arXiv:1110.6818 [hep-ph]].
  
\bibitem{Brauner:2012gu}
  T.~Brauner, O.~Taanila, A.~Tranberg and A.~Vuorinen,
  JHEP {\bf 1211} (2012) 076
  doi:10.1007/JHEP11(2012)076
  [arXiv:1208.5609 [hep-ph]].

\bibitem{Kajantie:1995dw}
  K.~Kajantie, M.~Laine, K.~Rummukainen and M.~E.~Shaposhnikov,
  Nucl.\ Phys.\ B {\bf 458} (1996) 90
  doi:10.1016/0550-3213(95)00549-8
  [hep-ph/9508379].


\bibitem{Kajantie:1995kf}
  K.~Kajantie, M.~Laine, K.~Rummukainen and M.~E.~Shaposhnikov,
  Nucl.\ Phys.\ B {\bf 466} (1996) 189
  doi:10.1016/0550-3213(96)00052-1
  [hep-lat/9510020].

\bibitem{Kajantie:1996mn}
  K.~Kajantie, M.~Laine, K.~Rummukainen and M.~E.~Shaposhnikov,
  Phys.\ Rev.\ Lett.\  {\bf 77} (1996) 2887
  doi:10.1103/PhysRevLett.77.2887
  [hep-ph/9605288].
  
\bibitem{Kajantie:1996qd}
  K.~Kajantie, M.~Laine, K.~Rummukainen and M.~E.~Shaposhnikov,
  Nucl.\ Phys.\ B {\bf 493} (1997) 413
  doi:10.1016/S0550-3213(97)00164-8
  [hep-lat/9612006].
  
\bibitem{Csikor:1998ge}
  F.~Csikor, Z.~Fodor and J.~Heitger,
  Phys.\ Lett.\ B {\bf 441} (1998) 354
  doi:10.1016/S0370-2693(98)01127-7
  [hep-lat/9807021].
  
\bibitem{Csikor:1998eu}
  F.~Csikor, Z.~Fodor and J.~Heitger,
  Phys.\ Rev.\ Lett.\  {\bf 82} (1999) 21
  doi:10.1103/PhysRevLett.82.21
  [hep-ph/9809291].
  
\bibitem{Aoki:1999fi}
  Y.~Aoki, F.~Csikor, Z.~Fodor and A.~Ukawa,
  Phys.\ Rev.\ D {\bf 60} (1999) 013001
  doi:10.1103/PhysRevD.60.013001
  [hep-lat/9901021].
   
\bibitem{Fukugita:1986hr}
  M.~Fukugita and T.~Yanagida,
  Phys.\ Lett.\ B {\bf 174} (1986) 45.
  doi:10.1016/0370-2693(86)91126-3
  
\bibitem{Luty:1992un}
  M.~A.~Luty,
  Phys.\ Rev.\ D {\bf 45} (1992) 455.
  doi:10.1103/PhysRevD.45.455
  
\bibitem{Buchmuller:2004nz}
  W.~Buchmuller, P.~Di Bari and M.~Plumacher,
  Annals Phys.\  {\bf 315} (2005) 305
  doi:10.1016/j.aop.2004.02.003
  [hep-ph/0401240].
  
\bibitem{Davidson:2008bu}
  S.~Davidson, E.~Nardi and Y.~Nir,
  Phys.\ Rept.\  {\bf 466} (2008) 105
  doi:10.1016/j.physrep.2008.06.002
  [arXiv:0802.2962 [hep-ph]].
  
\bibitem{GarciaBellido:1999sv}
  J.~Garcia-Bellido, D.~Y.~Grigoriev, A.~Kusenko and M.~E.~Shaposhnikov,
  Phys.\ Rev.\ D {\bf 60} (1999) 123504
  doi:10.1103/PhysRevD.60.123504
  [hep-ph/9902449].
  
\bibitem{Krauss:1999ng}
  L.~M.~Krauss and M.~Trodden,
  Phys.\ Rev.\ Lett.\  {\bf 83} (1999) 1502
  doi:10.1103/PhysRevLett.83.1502
  [hep-ph/9902420].
  
\bibitem{Copeland:2001qw}
  E.~J.~Copeland, D.~Lyth, A.~Rajantie and M.~Trodden,
  Phys.\ Rev.\ D {\bf 64} (2001) 043506
  doi:10.1103/PhysRevD.64.043506
  [hep-ph/0103231].
  
\bibitem{Tranberg:2003gi}
  A.~Tranberg and J.~Smit,
  JHEP {\bf 0311} (2003) 016
  doi:10.1088/1126-6708/2003/11/016
  [hep-ph/0310342].
  

\bibitem{Abbott:2016blz}
  B.~P.~Abbott {\it et al.} [LIGO Scientific and Virgo Collaborations],
  Phys.\ Rev.\ Lett.\  {\bf 116} (2016) no.6,  061102
  doi:10.1103/PhysRevLett.116.061102
  [arXiv:1602.03837 [gr-qc]].


\bibitem{Grojean:2006bp}
  C.~Grojean and G.~Servant,
  Phys.\ Rev.\ D {\bf 75} (2007) 043507
  doi:10.1103/PhysRevD.75.043507
  [hep-ph/0607107].
  
\bibitem{No:2011fi}
  J.~M.~No,
  Phys.\ Rev.\ D {\bf 84} (2011) 124025
  doi:10.1103/PhysRevD.84.124025
  [arXiv:1103.2159 [hep-ph]].
  
\bibitem{Hindmarsh:2013xza}
  M.~Hindmarsh, S.~J.~Huber, K.~Rummukainen and D.~J.~Weir,
  Phys.\ Rev.\ Lett.\  {\bf 112} (2014) 041301
  doi:10.1103/PhysRevLett.112.041301
  [arXiv:1304.2433 [hep-ph]].
  
\bibitem{Caprini:2015zlo}
  C.~Caprini {\it et al.},
  JCAP {\bf 1604} (2016) no.04,  001
  doi:10.1088/1475-7516/2016/04/001
  [arXiv:1512.06239 [astro-ph.CO]].


\bibitem{Hashino:2016xoj} 
  K.~Hashino, M.~Kakizaki, S.~Kanemura, P.~Ko and T.~Matsui,
  arXiv:1609.00297 [hep-ph].


  
\bibitem{Barger:2007im}
  V.~Barger, P.~Langacker, M.~McCaskey, M.~J.~Ramsey-Musolf and
  G.~Shaughnessy,
  Phys.\ Rev.\ D {\bf 77} (2008) 035005
  doi:10.1103/PhysRevD.77.035005
  [arXiv:0706.4311 [hep-ph]].

\bibitem{Ashoorioon:2009nf}
  A.~Ashoorioon and T.~Konstandin,
  JHEP {\bf 0907} (2009) 086
  doi:10.1088/1126-6708/2009/07/086
  [arXiv:0904.0353 [hep-ph]].
  
\bibitem{Robens:2015gla}
  T.~Robens and T.~Stefaniak,
  Eur.\ Phys.\ J.\ C {\bf 75} (2015) 104
  doi:10.1140/epjc/s10052-015-3323-y
  [arXiv:1501.02234 [hep-ph]].



\bibitem{Kanemura:2016lkz}
     S.~Kanemura, M.~Kikuchi and K.~Yagyu,
     (2016)
     [arXiv:1608.01582 [hep-ph]].
     

\bibitem{Kanemura:2015fra}
      
	S.~Kanemura, M.~Kikuchi and K.~Yagyu,
  	Nucl. Phys. B907 (2016) 286-322 
	doi: 10.1016/j.nuclphysb.2016.04.005
	[arXiv:1511.06211 [hep-ph]].

\bibitem{Beniwal:2017eik}
  A.~Beniwal, M.~Lewicki, J.~D.~Wells, M.~White and A.~G.~Williams,
  arXiv:1702.06124 [hep-ph].
        
\bibitem{Enqvist:2014zqa}
      K.~Enqvist, S.~Nurmi, T.~Tenkanen and K.~Tuominen,
     JCAP 1408 (2014) 035
     doi:10.1088/1475-7516/2014/08/035
     [arXiv:1407.0659 [astro-ph.CO]].
     

\bibitem{Tenkanen:2016idg}
      T.~Tenkanen, K.~Tuominen and V.~Vaskonen,
      JCAP 1609 (2016) 037
      doi:10.1088/1475-7516/2016/09/037
      [arXiv:1606.06063 [hep-ph]].






\bibitem{Huber:2000mg} 
  S.~J.~Huber and M.~G.~Schmidt,
  Nucl.\ Phys.\ B {\bf 606}, 183 (2001)
  doi:10.1016/S0550-3213(01)00250-4
  [hep-ph/0003122].
  

\bibitem{O'Connell:2006wi}
  D.~O'Connell, M.~J.~Ramsey-Musolf and M.~B.~Wise,
  Phys.\ Rev.\ D {\bf 75} (2007) 037701
  doi:10.1103/PhysRevD.75.037701
  [hep-ph/0611014].

\bibitem{Ahriche:2007jp}
  A.~Ahriche,
  Phys.\ Rev.\ D {\bf 75} (2007) 083522
  [hep-ph/0701192].
  
\bibitem{Profumo:2007wc}
  S.~Profumo, M.~J.~Ramsey-Musolf and G.~Shaughnessy,
  JHEP {\bf 0708} (2007) 010
  doi:10.1088/1126-6708/2007/08/010
  [arXiv:0705.2425 [hep-ph]].

\bibitem{Espinosa:2011ax}
  J.~R.~Espinosa, T.~Konstandin and F.~Riva,
  Nucl.\ Phys.\ B {\bf 854} (2012) 592
  [arXiv:1107.5441 [hep-ph]].

\bibitem{Cline:2012hg}
  J.~M.~Cline and K.~Kainulainen,
  JCAP {\bf 1301} (2013) 012
  doi:10.1088/1475-7516/2013/01/012
  [arXiv:1210.4196 [hep-ph]].
  
\bibitem{Damgaard:2013kva}
  P.~H.~Damgaard, D.~O'Connell, T.~C.~Petersen and A.~Tranberg,
  Phys.\ Rev.\ Lett.\  {\bf 111} (2013) no.22,  221804
  doi:10.1103/PhysRevLett.111.221804
  [arXiv:1305.4362 [hep-ph]].
  
\bibitem{Profumo:2014opa}
  S.~Profumo, M.~J.~Ramsey-Musolf, C.~L.~Wainwright and P.~Winslow,
  Phys.\ Rev.\ D {\bf 91} (2015) no.3,  035018
  doi:10.1103/PhysRevD.91.035018
  [arXiv:1407.5342 [hep-ph]].

\bibitem{Kozaczuk:2015owa}
  J.~Kozaczuk,
  arXiv:1506.04741 [hep-ph].

\bibitem{Damgaard:2015con}
  P.~H.~Damgaard, A.~Haarr, D.~O'Connell and A.~Tranberg,
  JHEP {\bf 1602} (2016) 107
  doi:10.1007/JHEP02(2016)107
  [arXiv:1512.01963 [hep-ph]].
  
  \bibitem{TBD}
  T.~Brauner, A.~Haarr, T.~V.~I.~Tenkanen, A.~Tranberg, A.~Vuorinen, D.~J.~Weir, In preparation.
  
\bibitem{Appelquist:1974tg}
   T.~Appelquist and J.~Carazzone,
   Phys.\ Rev.\ D {\bf 11} (1975) 2856.
   doi:10.1103/PhysRevD.11.2856
 
\bibitem{Braaten:1995cm}
   E.~Braaten and A.~Nieto,
   Phys.\ Rev.\ D {\bf 51} (1995) 6990
   doi:10.1103/PhysRevD.51.6990
   [hep-ph/9501375].

\bibitem{Braaten:1994pk}

     E.~Braaten and A.~Nieto,
     Phys.\ Rev.\ Lett. 73 (1994) 2402-2404
     doi: 10.1103/PhysRevLett.73.2402
     [hep-ph/9408273].

 
   \bibitem{Cline:1996mga}
  J.~M.~Cline and P.~A.~Lemieux,
  Phys.\ Rev.\ D {\bf 55} (1997) 3873
  [hep-ph/9609240].
   
\bibitem{Fromme:2006cm}
  L.~Fromme, S.~J.~Huber and M.~Seniuch,
  JHEP {\bf 0611} (2006) 038
  doi:10.1088/1126-6708/2006/11/038
  [hep-ph/0605242].
  
\bibitem{Cline:2011mm}
  J.~M.~Cline, K.~Kainulainen and M.~Trott,
  JHEP {\bf 1111} (2011) 089
  doi:10.1007/JHEP11(2011)089
  [arXiv:1107.3559 [hep-ph]].
  
\bibitem{Curtin:2012aa}
  D.~Curtin, P.~Jaiswal and P.~Meade,
  JHEP {\bf 1208} (2012) 005
  doi:10.1007/JHEP08(2012)005
  [arXiv:1203.2932 [hep-ph]].
  
\bibitem{Chung:2012vg}
  D.~J.~H.~Chung, A.~J.~Long and L.~T.~Wang,
  Phys.\ Rev.\ D {\bf 87} (2013) 2,  023509
  doi:10.1103/PhysRevD.87.023509
  [arXiv:1209.1819 [hep-ph]].

\bibitem{Haarr:2016qzq}
  A.~Haarr, A.~Kvellestad and T.~C.~Petersen,
  arXiv:1611.05757 [hep-ph].
  
 \end{thebibliography}

\end{fmffile}
\end{document}